\documentclass[reprint,aps,prl,twocolumn,superscriptaddress]{revtex4-1}
\usepackage{graphicx}
\usepackage{mathrsfs}
\usepackage{bm}
\usepackage{hyperref}
\usepackage{dcolumn}
\usepackage{amsmath}
\usepackage{amssymb}
\usepackage{color}
\usepackage{subcaption}
\usepackage{caption}
\usepackage{indentfirst}
\usepackage{ifpdf}
\hypersetup{  colorlinks=true, linkcolor=blue, citecolor=red, urlcolor=blue  }

\newcommand{\bra}[1]{\ensuremath{\left\langle#1\right|}}
\newcommand{\ket}[1]{\ensuremath{\left|#1\right\rangle}}
\newcommand{\dprime}{\prime \prime}

\newcommand{\bk}{{\mathbf k}}
\newcommand{\bB}{{\mathbf B}}
\newcommand{\bs}{\bm{\mathrm{\sigma}}}

\begin{document}

\title{Ultrathin Films of Superconducting Metals as a \\
Platform for Topological Superconductivity}
\author{Chao Lei}
\affiliation{Department of Physics, The University of Texas at Austin, Austin, Texas 78712,USA}
\email{leichao.ph@gmail.com}
\author{Hua Chen}
\affiliation{Department of Physics, Colorado State University, Fort Collins, CO 80523, USA}
\affiliation{School of Advanced Materials Discovery, Colorado State University, Fort Collins, CO 80523, USA}
\author{Allan H. MacDonald}
\affiliation{Department of Physics, The University of Texas at Austin, Austin, Texas 78712,USA}

\begin{abstract}
The ingredients normally required to achieve topological superconductivity (TSC) are Cooper pairing, broken inversion symmetry, and broken time-reversal symmetry. We present a theoretical exploration of the possibility of using ultra-thin films of superconducting metals as a platform for TSC. Because they necessarily break inversion symmetry when prepared on a substrate and have intrinsic Cooper pairing, they can be TSCs when time-reversal symmetry is broken by an external magnetic field. Using microscopic density functional theory calculations we show that for ultrathin Pb and $\beta$-Sn superconductors the position of the Fermi level can be tuned to quasi-2D band extrema energies using strain, and that the $g$-factors of these Bloch states can be extremely large, enhancing the influence of external magnetic fields.
\end{abstract}
%\pacs{
%71.70.Ej,  % spin-orbit coupling
%73.21.Hb,   % electron states and collective excitations in quantum wires
%71.10.Pm,   % Anyons electronic structure
%74.45.+c,   % SN and SNS junctions (superconductivity)
%%71.20.Mq,   % band structure of semiconductors
%}

\maketitle

% \section{Introduction}
\textit{Introduction---}Topological superconductors (TSC)\cite{Beenakker2013,Beenakker2016,Stern2010,Jason2012,Martin2012,Masatoshi2017}
can host fault-tolerant qubit operations based on the exchange properties\cite{Ivanov2001,Nayak2008} of Majorana zero modes located either at the ends of topological superconducting quantum wires\cite{Kitaev2001}, or in the vortex cores of two-dimensional TSCs\cite{Ivanov2001, Beenakker2013, Stern2010,Read2000}. For weak Cooper pairing, topological superconductivity occurs whenever the host normal metal has an odd number of closed Fermi surfaces. TSCs were first realized\cite{Das2012,Deng2012,Mourik2012,Rokhinson2012} some years ago by combining\cite{Lutchyn2010,Sau2010} low density-of-states semiconductors, with strong spin-orbit coupling and external magnetic fields that lift band spin-degeneracies, and Cooper pairing provided by an adjacent superconductor. In recent research semiconductor-based TSCs have been further refined\cite{Albrecht2016,Deng2016}, and other possibilities have also been realized experimentally, including the TSCs based on magnetic atomic chains on superconducting substrates\cite{Nadj2014,Feldman2016,Pawlak2016} and two-dimensional (2D) TSCs based on topological insulator surface states\cite{He2017,Zeng2017}.

TSC has been proposed as a theoretical possibility in bulk superconductors that might have chiral order parameters \cite{Mohanta2014,Scheurer2015,Loder2015, Tzen2014,Scheurer2017,Scheurer2016,Samokhin2015,Wang2016,Loder2017,Wennerdal2017}, for example in noncentrosymmetric superconductors\cite{Samokhin2015,Scheurer2016,Wennerdal2017} with broken time-reversal or inversion symmetry. These intrinsic systems, including $\rm SrTiO_3/LaAlO_3$ heterostructures \cite{Mohanta2014,Scheurer2015,Loder2015,Scheurer2017}, bulk $\rm Sr_2RuO_4$, and superfluid $\rm He^3$, have some potential advantages over the artificial hybrid materials systems in which TSC has already been achieved experimentally. There is however no intrinsic system in which all the ingredients required for TSC states are fully established.
In the case of $\rm SrTiO_3/LaAlO_3$ heterostructures, for example, it seems difficult to achieve a sufficiently large Zeeman coupling strength since the g-factor in $\rm SrTiO_3/LaAlO_3$ is small \cite{Cheng2015}, much smaller than in Majorana platforms based on semiconductor quantum wires which have large g-factors up to 20-50\cite{Mourik2012,Albrecht2016,Deng2016}.

In this article we propose a different possibility, namely establishing 2D TSC directly in ultrathin films of superconducting metals \cite{Brun2016}, instead of semiconductors, thereby avoiding problems associated with establishing proximity coupling between a semiconductor and a superconductor. We are motivated by recent experimental demonstrations of strong robust superconductivity in ultra-thin metal films \cite{Nam2016,Xue_PbthinSC2010,Menard2017}, and by proposals for realizing topological superconductivity based on strong Rashba-like spin-orbit interactions in the surface-states of heavy metals TSC\cite{Potter2012,Yan2017}. We show that quasi 2D band extrema in ultra-thin
superconducting films can occur close to the Fermi level, that in the cases of Pb and $\beta$-Sn films the $g$-factors at the relevant band
edges can be extremely large, and that band positions can be tuned by strain. We predict that these ingredients will allow thin superconducting films to be tuned to TSC states when time-reversal invariance is broken by a weak magnetic field or a proximitized exchange interaction\cite{Menard2017}.

We concentrate below on lead (Pb) and $ \beta $-Sn thin films. Using {\it ab initio} density functional theory (DFT) calculations \cite{supplement}, we show that the strength of inversion symmetry breaking in $ \beta $-Sn and Pb thin films can be controlled by varying either film thickness or substrate material, that Fermi level positions relative to band extrema are more easily tuned by strain than by gate electric fields, and that typical $g$-factors\cite{gfactor} at band extrema are extremely large. Strains can be varied experimentally by placing the thin film on a piezoelectric substrate as illustrated in Fig. \ref{setup}, and adjusted to tune in topologically non-trivial states.

\ifpdf
\begin{figure}
\includegraphics[width=0.75\linewidth]{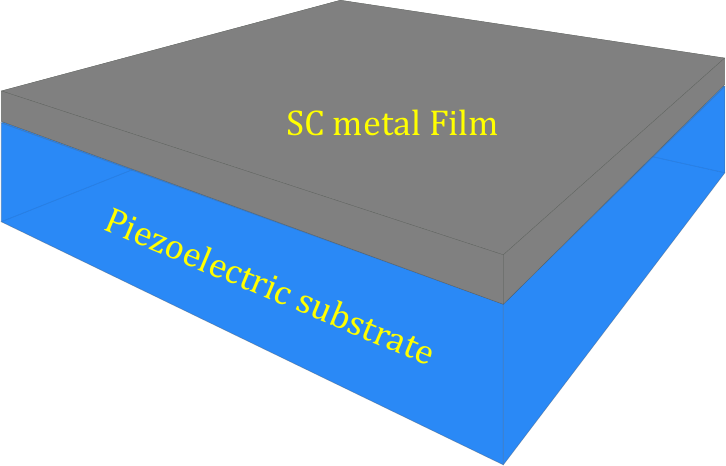}
\caption{Schematic illustration of an ultra-thin heavy superconducting(SC) metal film grown on a piezoelectric substrate. An electric field applied to the substrate can then be used to tune the film into a topological superconducting state.}
\label{setup}
\end{figure}
\fi

\textit{Superconducting metal thin film as TSCs ---}
For weak pairing TSC occurs in bands that are effectively spinless when an odd number of them cross the Fermi energy. In quasi-2D systems with strongly broken inversion symmetry, Rashba-like spin-orbit interactions lift spin-degeneracies except at the time-reversal invariant ${\bm k}$ points where Kramers theorem applies. The Kramers degeneracy can be lifted only by breaking time-reversal invariance, for example by an external magnetic field. A minimal mean field theory shows that, like the current semiconductor systems, these TSCs are class D according to the Altland-Zirnbauer (AZ) classification \cite{class_symmetry2016,supplement} which is also the main class of systems (1D or 2D) studied in present experiments. For sufficiently strong spin-orbit coupling a TSC state is realized when $\Delta_z = g \mu_B B > \sqrt{\Delta^2 + \mu^2} $, where $ \Delta_z $ is the Zeeman energy, $ \Delta $ is the pair potential, and $ \mu $ is the chemical potential measured from the zero-field band energy at the time-reversal invariant ${\bm k}$-point. Topological superconductivity in quasi-2D systems therefore requires that $\mu$ be small and that the $g$-factor that describes the Kramers degeneracy splitting be large. A quasi-2D metal film has the advantage over its bulk counterpart that it has a greater density of bands, which increases the chances for large $g$-factors
and is essential, as we shall see, if we want to find materials with small value of $\mu$. Comparing the criteria that support large $g$-factors \cite{Hota1993,Patnaik1999} with patterns in the occurrence of superconductivity\cite{SCperiodic} suggests ultra-thin films of $\beta$-Sn and Pb as promising candidates for topological superconductivity.

\begin{table}[h]
\caption{\label{g_table}
Calculated $g$-factors at the $\Gamma$ point for the band closest to the Fermi level in bulk and in thin films of $\beta$-Sn and Pb on a $\rm As_2O_3$ substrate.\cite{gfactor,supplement} For the films the average values of the g-factors of 12 subbands around fermi level is also given. The $g$-factors are obtained by evaluating the splitting between a Kramers pair at the $\Gamma$ point under a magnetic field. For the thin films the magnetic field is along the film normals [(111) for Pb and (001) for $\beta$-Sn], while for bulk Pb and $\beta$-Sn it is along the $z$-axis [(001) direction].
}
\begin{ruledtabular}
\begin{tabular}{c c c c c c}
 $\beta$-Sn & $g$-factor & avg. & Pb & $g$-factor & avg. \\ [0.5ex]
 \hline
 bulk & 681 & & bulk & 132 & \\
 7 layers & 29 & 254 & 5 layers & 57 & 140 \\
 9 layers & 572 & 243 & 7 layers & 161 & 73\\
 11 layers & 574 & 268 & 9 layers & 198 & 186\\
 \end{tabular}
 \end{ruledtabular}
\end{table}

Pb has a face-centered-cubic structure \cite{supplement} and is a widely studied superconductor with a bulk T$_c$= 7.19 K. Among the several stable phases of bulk Sn only $ \beta $-Sn, which has a tetragonal structure (A5)\cite{supplement}, is a superconductor\cite{Eisenstein1954} with T$_c$ = 3.72 K. The bands of $\beta $-Sn and Pb, illustrated in Fig. \ref{bulk_band}, reflect strong $s-p$ hybridization. Bulk $ \beta $-Sn and bulk Pb both have inversion symmetry, and therefore even degeneracies of all bands throughout the Brillouin zone.

We evaluated $g$-factors using methods informed by recent advances in the {\it ab initio} description of orbital magnetism \cite{gfactor,Mikitik2003,OrbitalMag2005a,OrbitalMag2005b}. According to our calculations\cite{gfactor}, the $ \Gamma $ point $g$-factors of bulk $ \beta $-Sn and bulk Pb are very large, as summarized in Table \ref{g_table}. In Fig. \ref{bulk_band} we see that a strongly dispersive band crosses the Fermi energy along $ \Gamma $-X in $ \beta $-Sn and along $ \Gamma $-L in Pb. Based on this observation, we expect that quasi-2D subband extrema at energies close to the Fermi energy will occur at 2D $\Gamma$ points in thin films with surface normals along the $(111)$ direction and the $(001)$ direction for Pb and $ \beta $-Sn respectively. Indeed, it is $(111)$ growth direction Pb films that are commonly studied experimentally. \cite{Nam2016}.

\ifpdf  \begin{figure}
\centering
\begin{subfigure}{0.45\linewidth}
\includegraphics[width=\linewidth,height=0.75\linewidth]{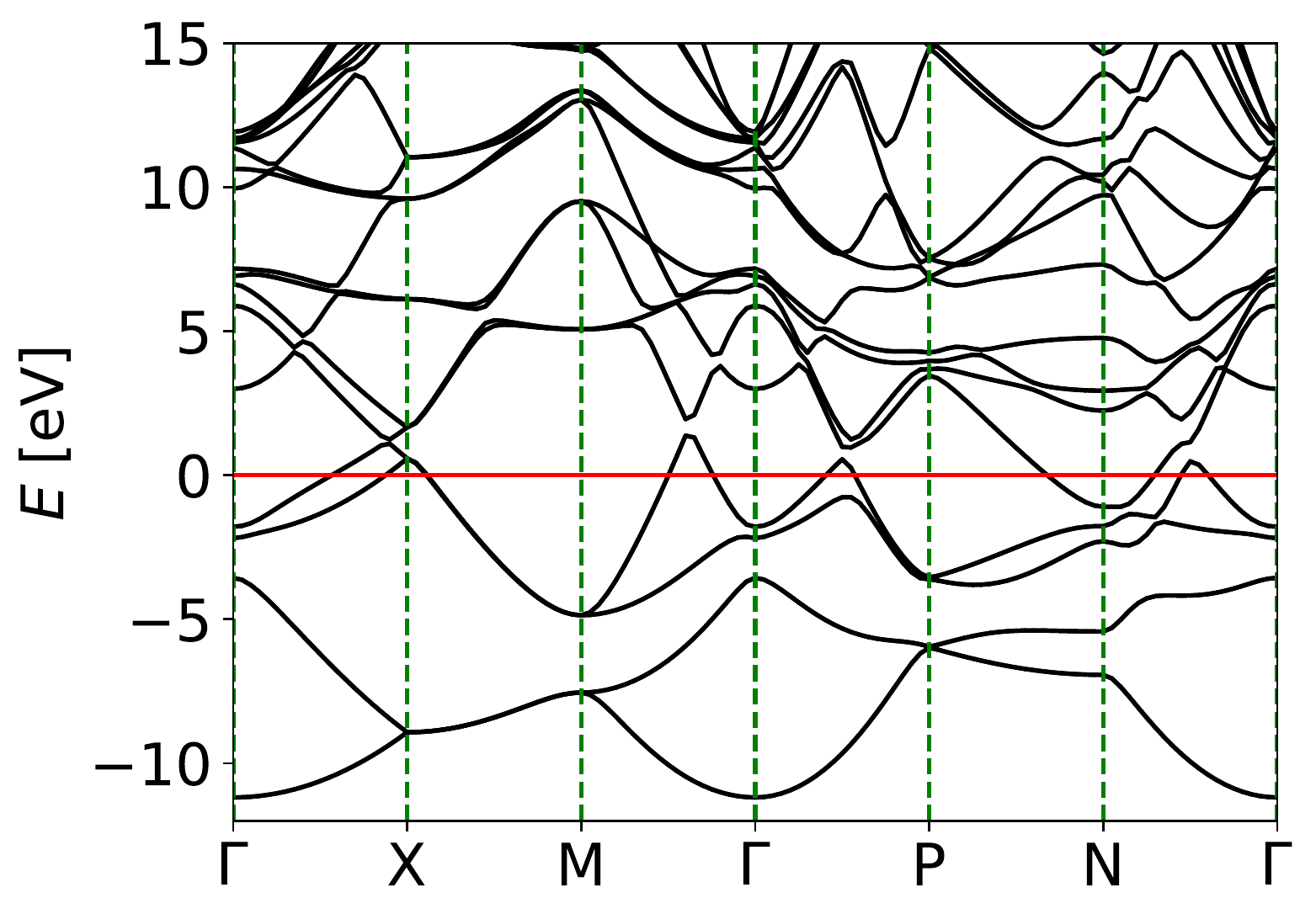}
\subcaption{Bandstructure of $ \beta $-Sn} \label{band_bulk_sn}
\end{subfigure}
\begin{subfigure}{0.45\linewidth}
\includegraphics[width=\linewidth,height=0.75\linewidth]{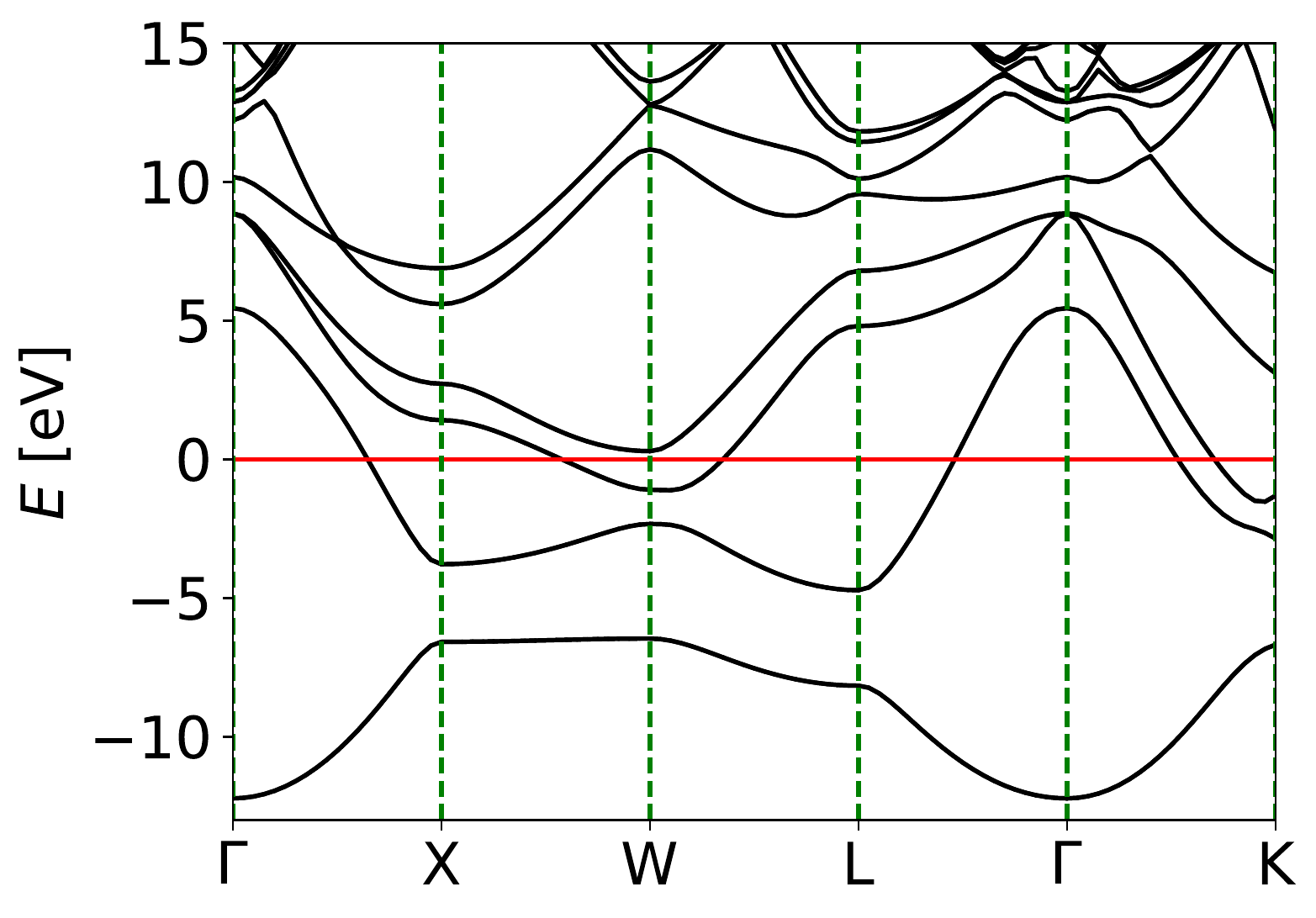}
\subcaption{Bandstructure of Pb} \label{band_bulk_pb}
\end{subfigure}
\caption{
Bandstructure of bulk $ \beta $-Sn  and Pb. The red horizontal lines mark the Fermi level. These bands are consistent with literature results for $\beta$-Sn \cite{band_tin_a,band_tin_b} and Pb \cite{band_lead}. Thin film quasi-2D bands can be crudely estimated from these bulk bands by discretizing the surface-normal momentum component.
} \label{bulk_band}
\end{figure} \fi

In thin films the inversion symmetry of a bulk structure does not survive for all surface terminations and thicknesses, even when the film
structure consists of bulk unit cells repeated in the film normal direction. For $ \beta $-Sn films grown along the $(001)$ direction, inversion symmetry is absent when the number of atomic layers is odd\cite{Sn001}. As an illustration, the band structure of single layer Sn $(001)$ is shown in Fig.~\ref{band_sn001}. The band closest to the Fermi level, which has its extremum near $\Gamma$, exhibits typical Rashba spin-orbit coupling behavior as illustrated in Fig. \ref{rashba_sn}. Similar quasi-2D bands are present for all odd-layer-number
$ \beta $-Sn $(001)$ thin films\cite{supplement}. The Rashba spin-orbit coupling strength becomes smaller with increasing number of layers. However, even at 15 layers, its value is still about 0.85 eV \AA, which is several times larger than that in semiconductor quantum wires ($\sim$0.2 eV \AA\cite{Mourik2012}).

For Pb $(111)$, on the other hand, inversion symmetry is maintained at all film thicknesses and every subband has two-fold degeneracy throughout the 2D Brillouin zone. Fig.~\ref{band_pb111} shows the example of a two-layer Pb $(111)$ thin film (see \cite{supplement} for more band structures for different number of layers). Broken inversion symmetry must then come from hybridization with a substrate. In the calculations described below we have used a single quintuple layer of As$_2$O$_3$ with the Bi$_2$Te$_3$ structure
as the substrate because it is insulating and, according to our DFT calculations, lattice-matched to Pb $(111)$. The resulting quasi-2D band structure is illustrated in Fig.~\ref{rashba_pb} (results for other thicknesses can be found in \cite{supplement}). We can see in the figure that the extremum of the lowest band at $\Gamma$ exhibits Rashba spin splitting. The Rashba spin-orbit coupling of Pb thin films on As$_2$O$_3$ is 0.15-0.4 eV\AA on average (0.01, 0.2, 0.34, and 0.05 eV\AA respectively for the four subbands around Fermi level \cite{supplement}), which is much larger than that on Si substrates\cite{Slomski2013,Dil2008} (0.03-0.04 eV \AA for 10 layers of Pb). The averaged Rashba spin-orbit coupling decreases with increasing film thickness, but even for the thicker films considered here it is still large compared with that in semiconductor quantum wires on s-wave superconductors\cite{Mourik2012}.

We also studied heavier substrates in the Pnictogen Chalcogenides family such as $\rm As_2O_3$, $\rm Sb_2S_3$, $\rm Sb_2Se_3$, $\rm Bi_2Se_3$, $\rm Bi_2Te_3$, and found an enhancement of Rashba spin-orbit coupling ranging from 0.3 $\rm eV \AA$ to around 0.7 $\rm eV \AA$ on average \cite{supplement}, and even larger for certain subbands. However, since some of Pnictogen Chalcogenides are topological insulators, those subbands may also be the surface states of topological insulators. For some Pb $(111)$ thicknesses the band extremum closest to the Fermi level lies not at a time-reversal invariant momentum, but at the $K$-point where spin-splitting is present even in the absence of a
magnetic field. In this case, valley symmetry breaking by an external magnetic field is necessary to yield a topological superconducting state.

\ifpdf  \begin{figure}
\centering
\begin{subfigure}{0.45\linewidth}
\includegraphics[width=\linewidth,height=0.75\linewidth]{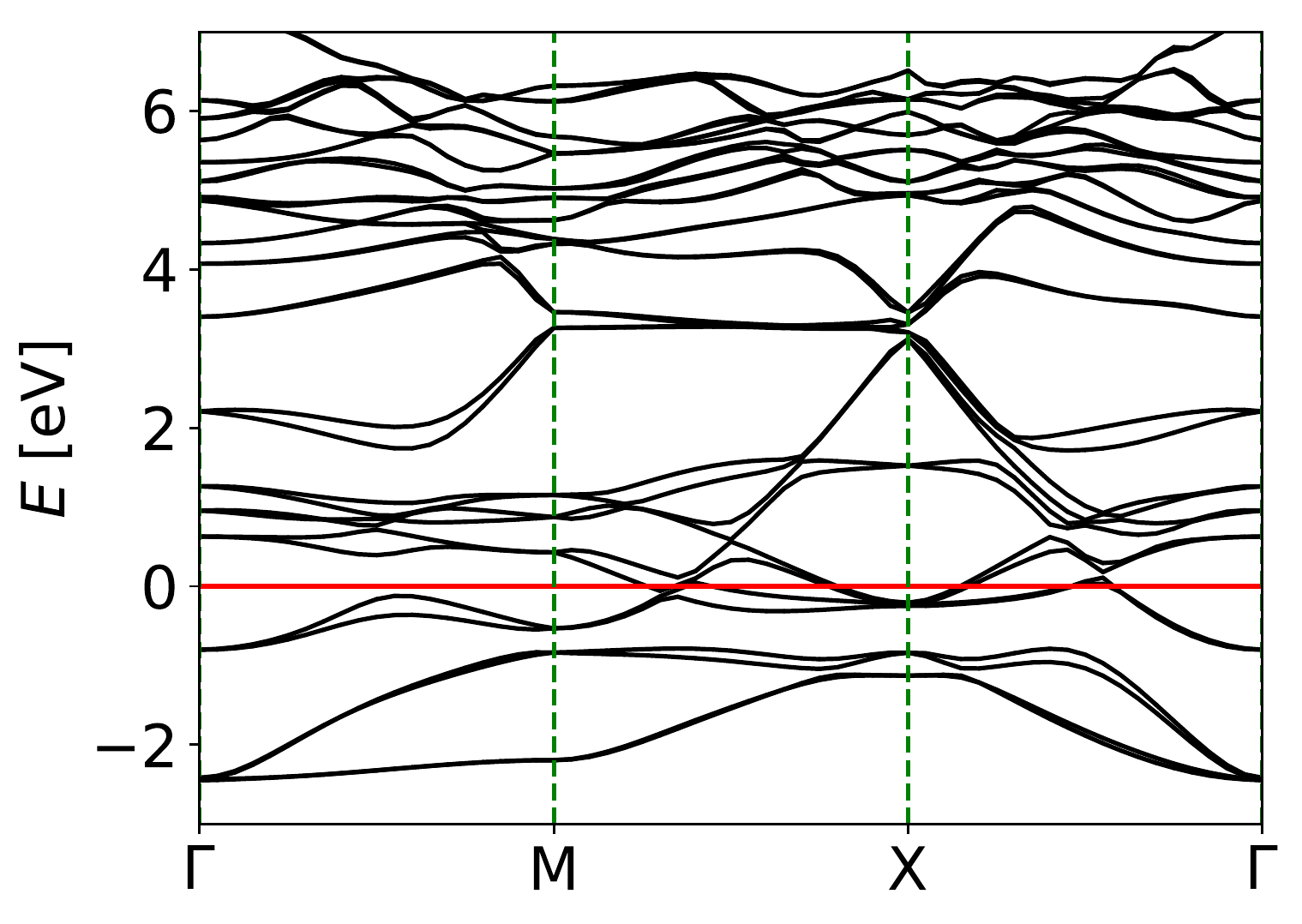}
\subcaption{1 Layer of Sn (001)} \label{band_sn001}
\end{subfigure}
\begin{subfigure}{0.45\linewidth}
\includegraphics[width=\linewidth,height=0.75\linewidth]{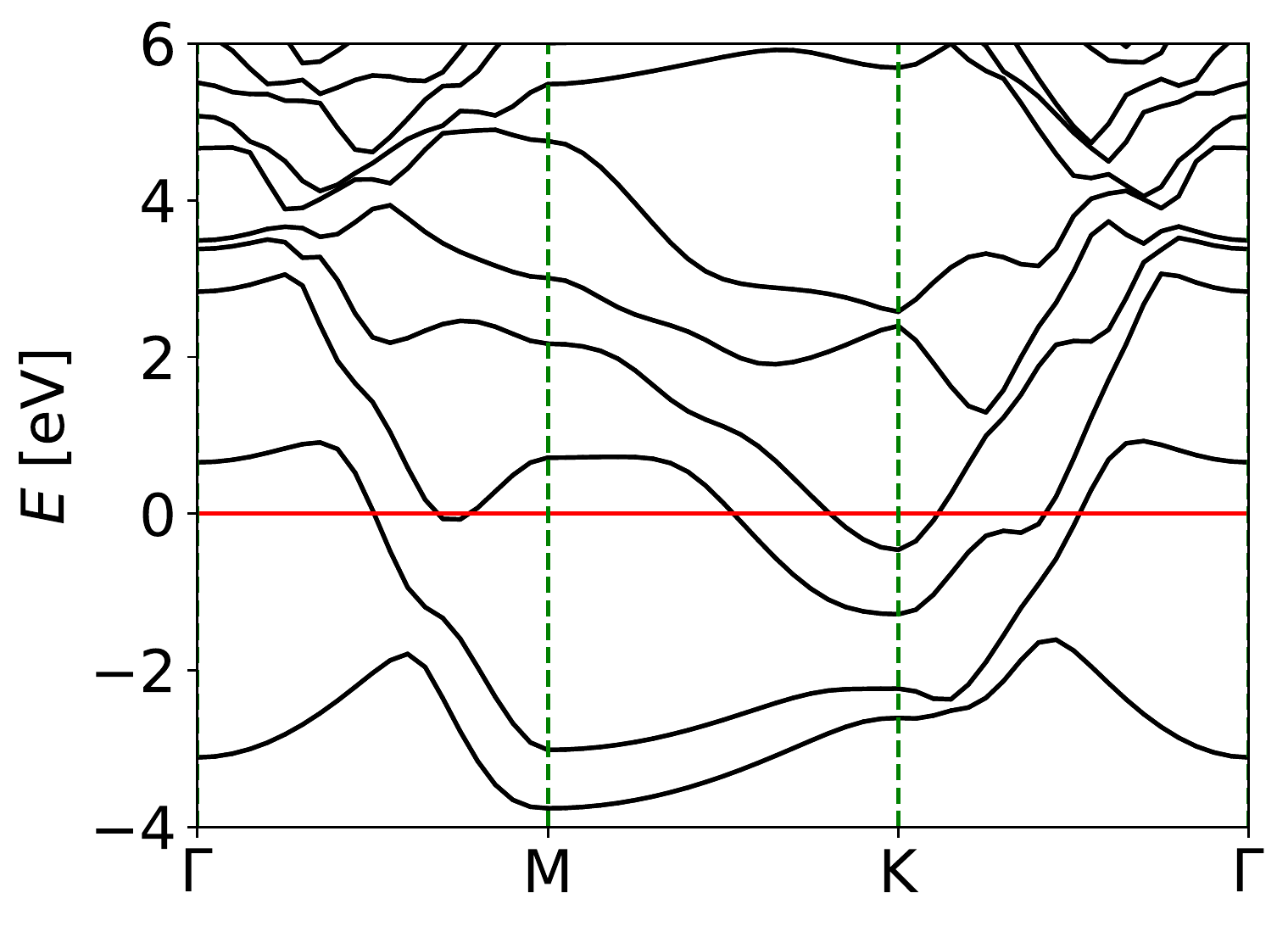}
\subcaption{Pb (111) bilayer} \label{band_pb111}
\end{subfigure}
\begin{subfigure}{0.45\linewidth}
\includegraphics[width=\linewidth,height=0.75\linewidth]{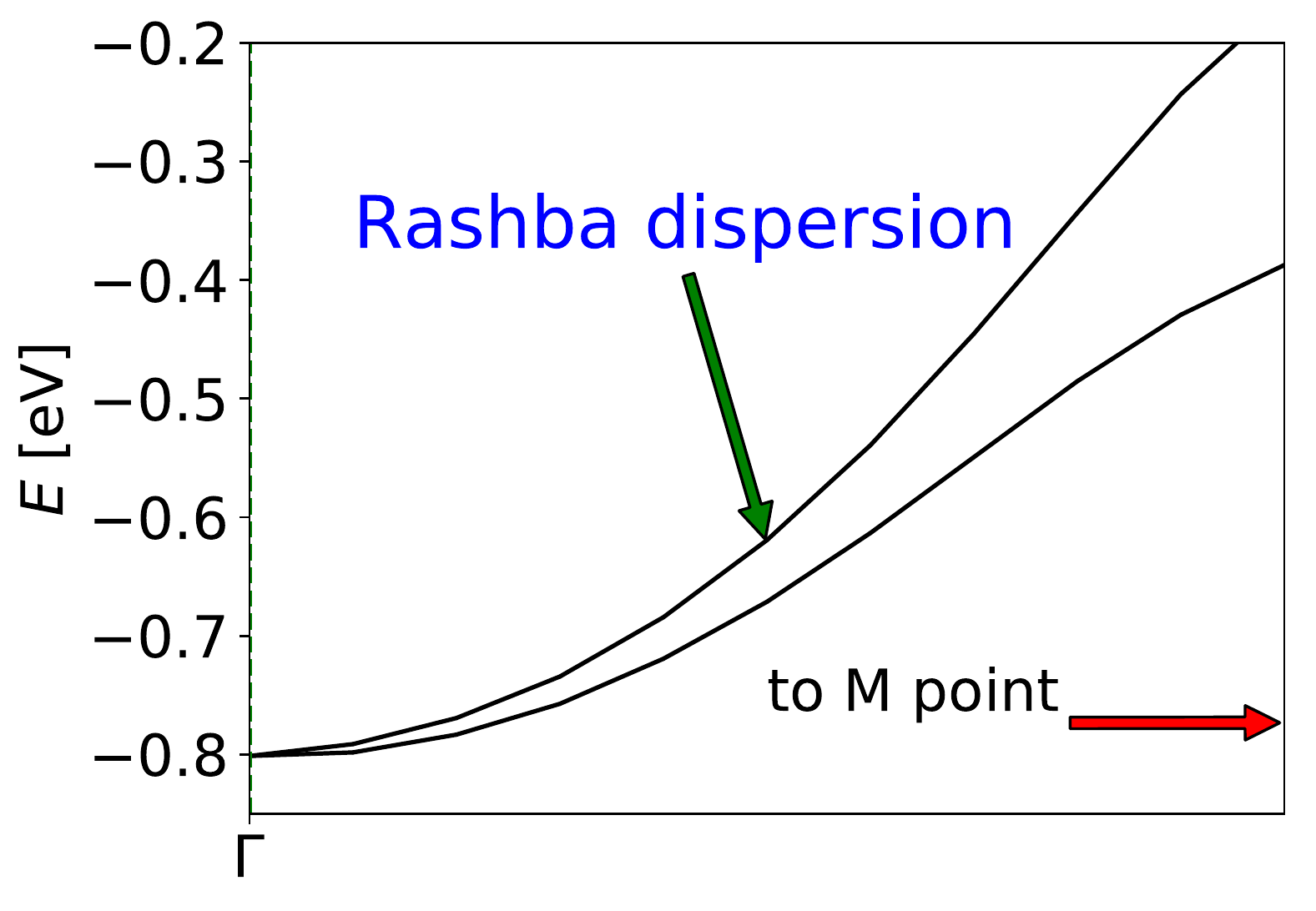}
\subcaption{Single layer Sn (001)} \label{rashba_sn}
\end{subfigure}
\begin{subfigure}{0.45\linewidth}
\includegraphics[width=\linewidth,height=0.75\linewidth]{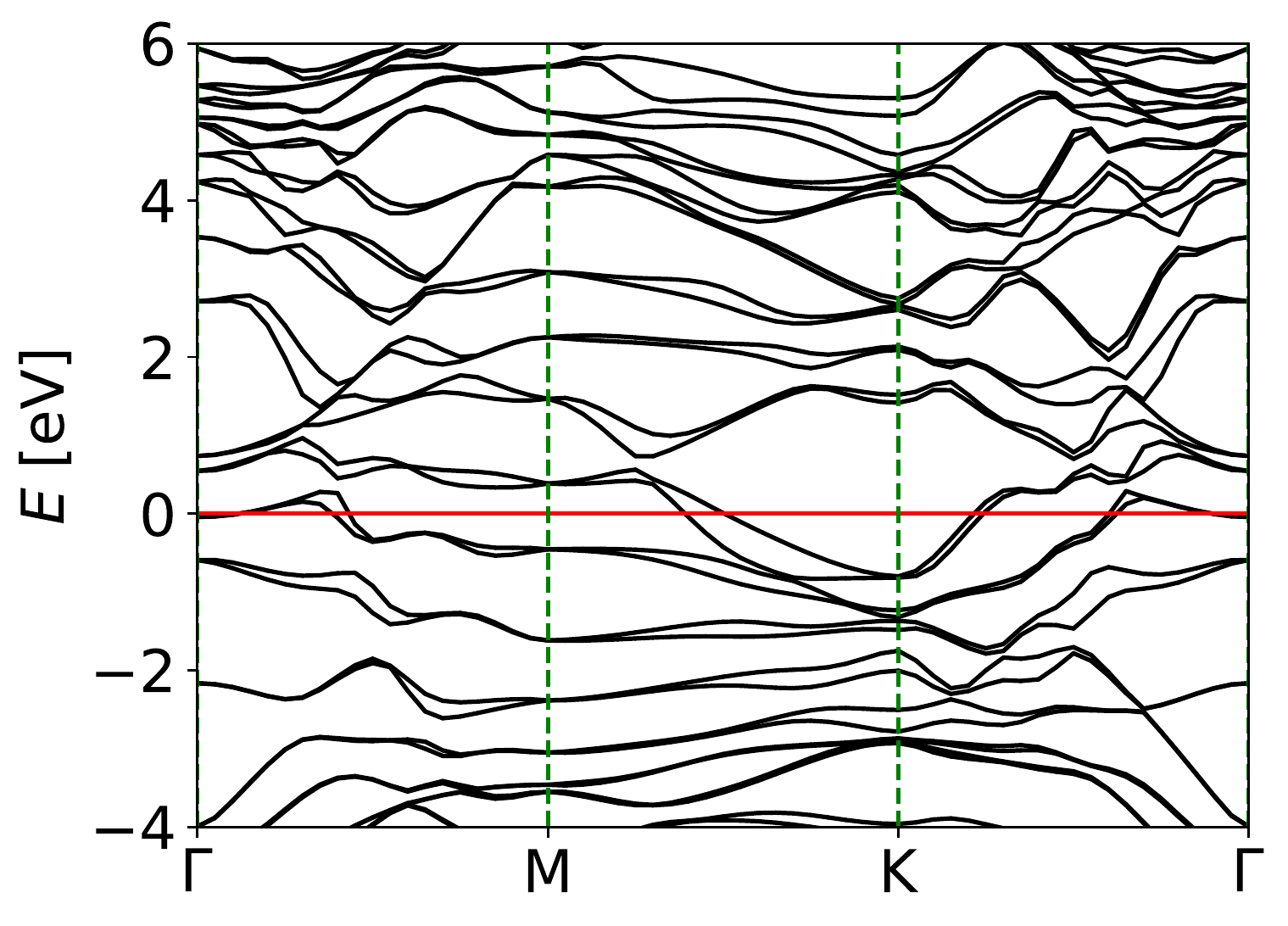}
\subcaption{Pb (111) bilayer} \label{rashba_pb}
\end{subfigure}
\caption{
Band structure of a single layer (3 atomic layers) film of $\beta $ Sn grown along the $(001)$ direction (a) and a Pb bilayer film grown along the $(111)$ direction (b). (c) Highlight of the first band below the Fermi level in (a) near the $\Gamma$-point showing Rashba-like spin-orbit splitting. (d) Pb bilayer grown along the $(111)$ direction on an As$_2$O$_3$ substrate exhibiting Rashba-like spin-orbit splitting near the $\Gamma$-point.}
\label{band_film}
\end{figure} \fi

\textit{Tuning the Fermi Level ---}
As illustrated in Figs.~\ref{rashba_sn} and \ref{rashba_pb}, the scale of the spin-orbit splitting in the metal thin films of interest is a sizable fraction of an eV and comparable to quasi-2D band widths. TSC states will therefore occur whenever the Fermi level is within $\Delta_z$ of a band extremum energy. Here $\Delta_z$ refers either to spin-splitting at a time-reversal invariant momentum, or to energetic splitting between spin-orbit split states at time-reversal partner momenta. For $g$-factors $\sim 100$, these energies are $\sim 10$ meV at the fields to which superconductivity typically survives. (The Bohr magneton is $ \sim 0.058 $meV/T. In Pb $(111)$ thin films, $H_{c\perp} = 1.56$ T for 5 monolayers and $ 0.63 $T for 13 monolayers. In-plane critical fields are much larger: $H_{c\parallel} =54.9 $T for 5 monolayers and $13.6 $T for 13 monolayers \cite{Nam2016}). It follows that TSC states should be realizable if the Fermi level can be tuned to within $\sim 10$ meV of quasi-2D band extrema, for large Rashba coupling, the most possible pairing of electrons has an intra-band form. Due to the very large $H_{c\parallel}$, the system may be driven by in-plane fields into an inter-band pairing phase with finite pairing momentum \cite{Florian2013}. %While the in-plane critical field is much larger than the perpendicular critical field, the inter-band pairing will probably dominate.

Figure~\ref{bulk_band} shows that bulk $\beta$-Sn bands cross the Fermi level along $ \Gamma $-X, and that bulk Pb bands cross the Fermi level along $ \Gamma $-L. The bandwidth of $ \beta $-Sn from $ \Gamma $ to X is about $W = 2.765 $eV. It follows that the average distance between quasi-2D subband energies at any particular 2D ${\bm k}$-point is around $W/2N$, or $\sim 150 $meV for a 10 layer thick film. In Fig.~\ref{EnLayerSn001} we plot the quasi-2D band energies at the $\Gamma$ point measured from the Fermi level for odd-layer-number Sn thin films {\it vs.} the number of layers. As expected the energy separations tend to decrease with increasing film thickness, but are suitably small only occasionally. For the films with thickness of 7, 9 and 11 layers, band extrema are within tens of meV of the Fermi level. The calculated $g$-factors at the $\Gamma$ point for these thicknesses are up to around 600. (The $g$-factors of the films highlighted by arrows in Fig. \ref{EnLayerSn001} are presented in Table \ref{g_table}.) For bulk Pb the bandwidth from $ \Gamma $ to L is $\sim 10$ eV, implying larger typical energy separation values. The band separation plot for Pb $(111)$ thin films is presented in Fig.~\ref{EnLayerPb111}, which shows apparent quantum size effect oscillations due to confinement of electron wave functions along the thickness direction \cite{Wei2002}. In spite of the larger typical separations, we find that at some thicknesses band extrema at $\Gamma$ and $K$ points can be within a few tens of meV of the Fermi level. Since the $g$-factors we calculated for Pb thin films, listed in Table \ref{g_table}, are as large as $\sim 200$, topological superconductivity is still a possibility.

\ifpdf  \begin{figure}
\centering
\begin{subfigure}{0.45\linewidth}
\includegraphics[width=\linewidth,height=0.75\linewidth]{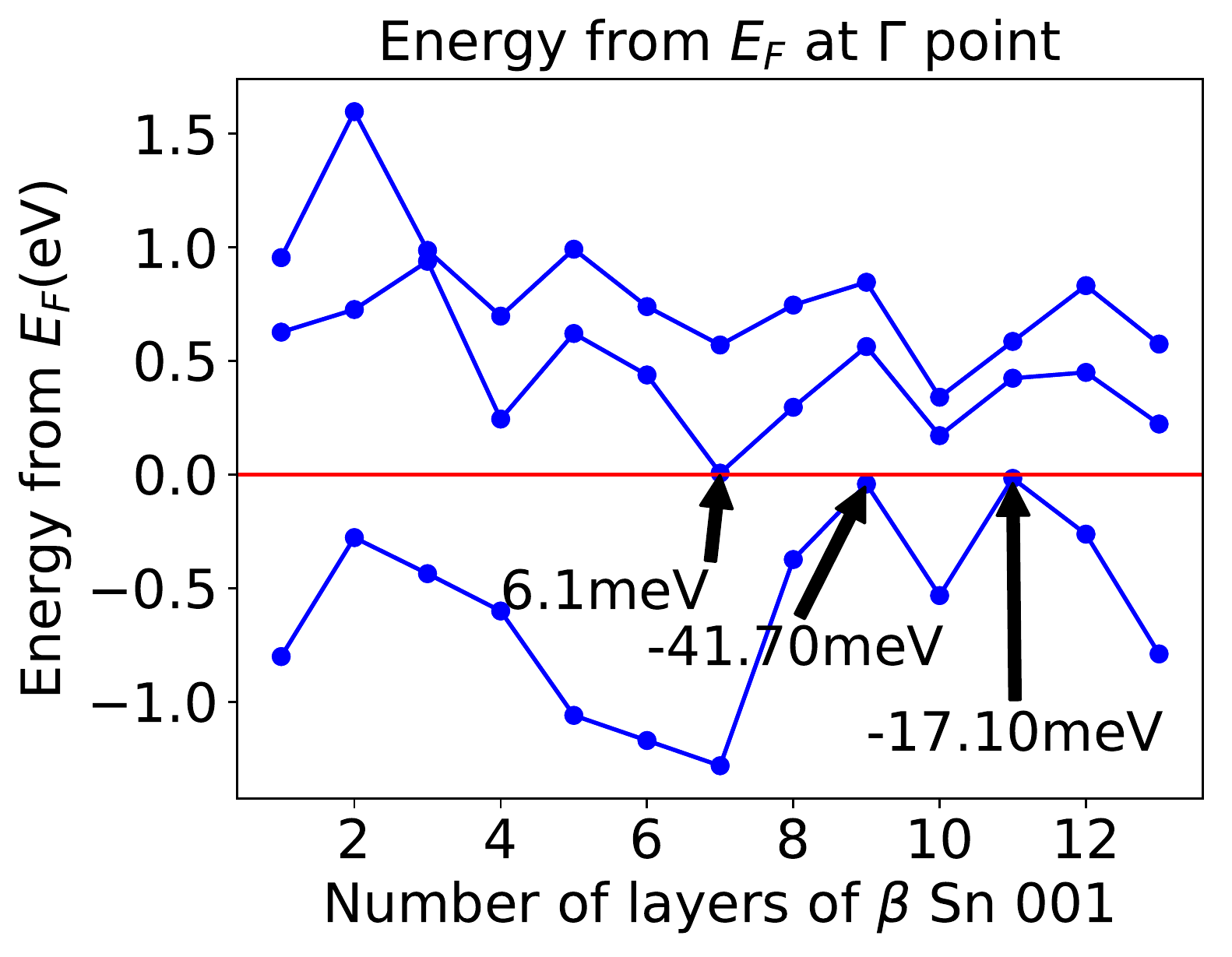}
\subcaption{Energy to $E_F$ [Sn(001)]} \label{EnLayerSn001}
\end{subfigure}
\begin{subfigure}{0.45\linewidth}
\includegraphics[width=\linewidth,height=0.75\linewidth]{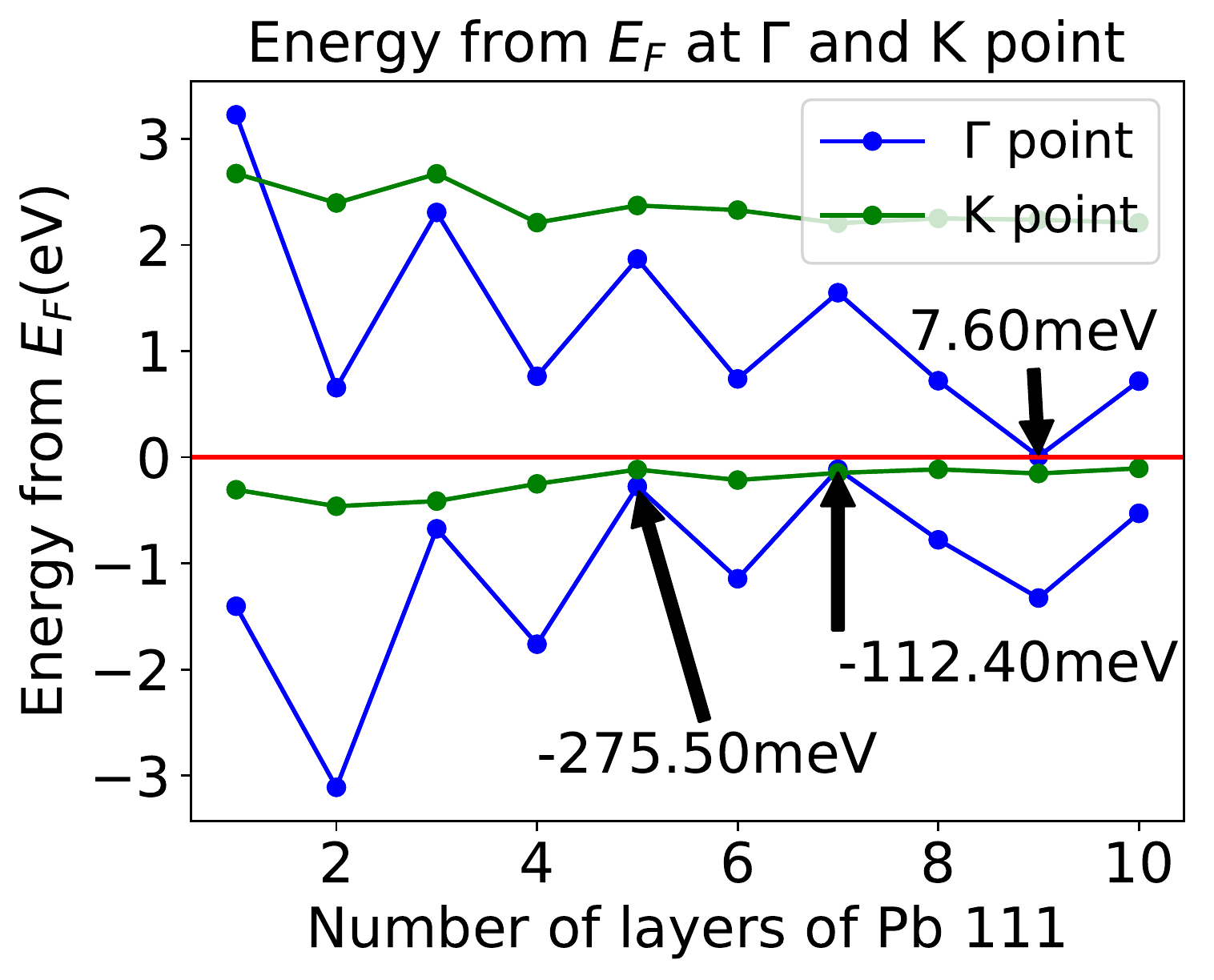}
\subcaption{Energy to $E_F$ [Pb(111)]} \label{EnLayerPb111}
\end{subfigure}
\caption{
(a) Subband energy extrema measured from $E_F$ in $ \beta $-Sn thin films grown in $(001)$ direction. Small energy separations are highlighted by numerical values attached to arrows pointing to the position at which they are plotted. The smallest separation is $6.1$ meV for a seven layer film. (b) $\Gamma$ and $K$-point band extrema relative to the Fermi level for Pb (111) thin films.
} \label{thick_depend}
\end{figure} \fi

%\textit{Electric Control and Strain Effect ---}
Because DFT is not likely to be perfectly predictive, and because energy separations are likely to be influenced by uncontrolled environmental effects, practical searches for TSC in metal thin films will be greatly assisted by {\it in situ} control. We have examined the efficacy of two possibilities. In Fig. \ref{Field_Sn001} and \ref{Field_Pb111} we show that, in spite of the strong screening effects expected in metals, external electric fields of $\sim 1$ V/nm in magnitude can still shift subband energy positions by $\sim 10$ meV for $ \beta $-Sn (001) and by $\sim 20$ meV for Pb $(111)$, which might be large enough to tune into topological states in some instances. The field scale of these calculations are however larger than what is typically practical. Assuming linear response a field of $10^{-1} $V/nm \cite{Oshiki1975} would typically change level separations by only $\sim 1 $meV. We have therefore also examined strain effects. In Figs. \ref{Strain_Sn001} and \ref{Strain_Pb111}, energy separations at the $\Gamma$ point in $ \beta $-Sn $(001)$ and Pb $(111)$ films are plotted {\it vs.} strain. The sensitivity of energy separation to a 1\% strain is typically more than $50$ meV for $ \beta $-Sn $(001)$ and around $\sim 200$ meV for Pb $(111)$ (the case with a substrate is similar\cite{supplement}), suggesting that strains in this range could successfully tune a thin film into a TSC state. Strains of this size can be induced electrically by applying an electric field across a piezoelectric substrate. If strain could be transferred from a substrate with a large piezoelectric effect ($1.6 nm/V$)\cite{Grupp1997}, an electric field of $10^{-2}$ V/nm could give a strain larger than 1\%. We conclude that strain is more promising than direct external electric fields for tuning metal thin films into TSC states.

\ifpdf  \begin{figure}
\centering
\begin{subfigure}{0.45\linewidth}
\includegraphics[width=\linewidth,height=0.75\linewidth]{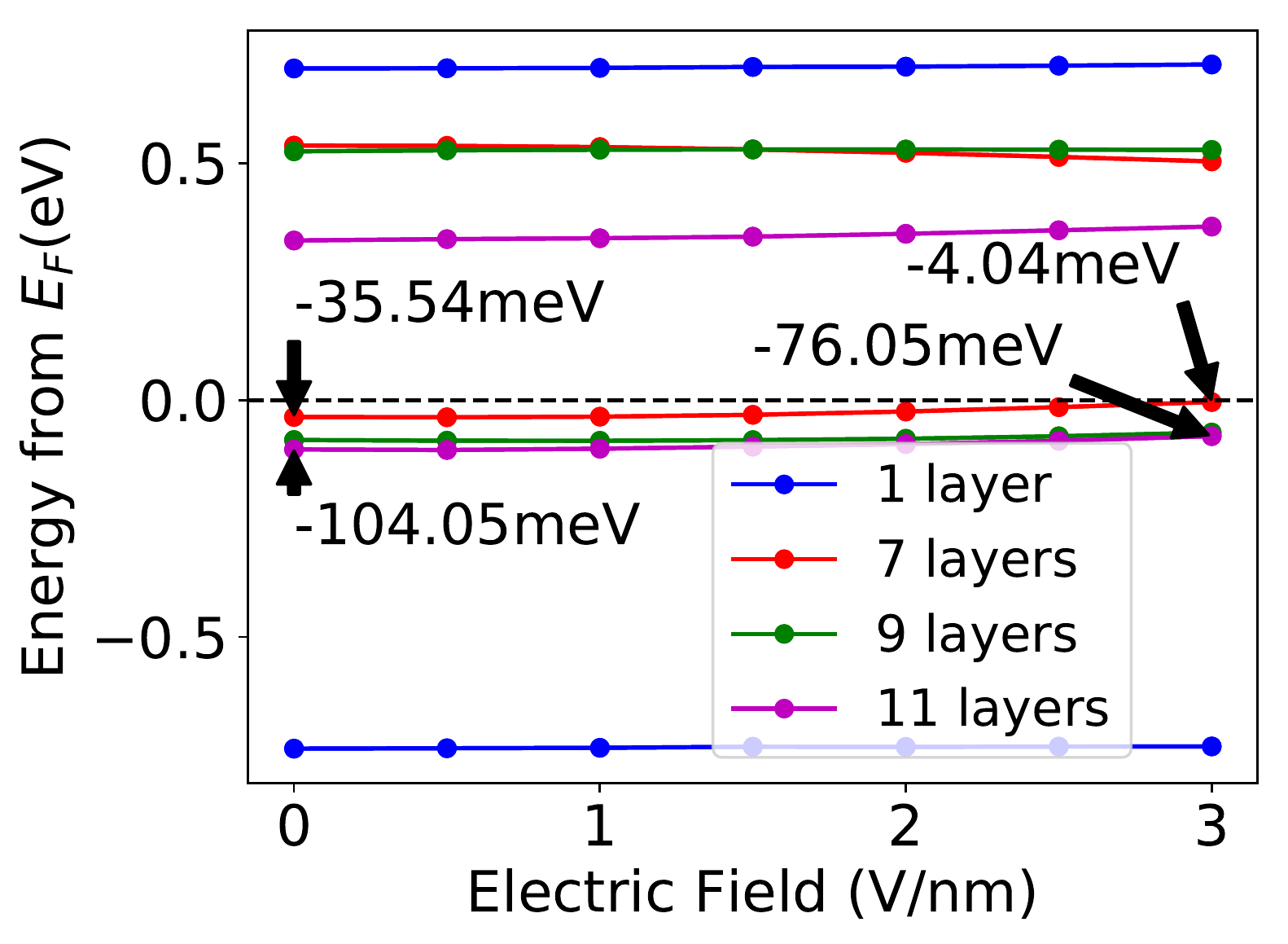}
\subcaption{Field tuned Sn (001)} \label{Field_Sn001}
\end{subfigure}
\begin{subfigure}{0.45\linewidth}
\includegraphics[width=\linewidth,height=0.75\linewidth]{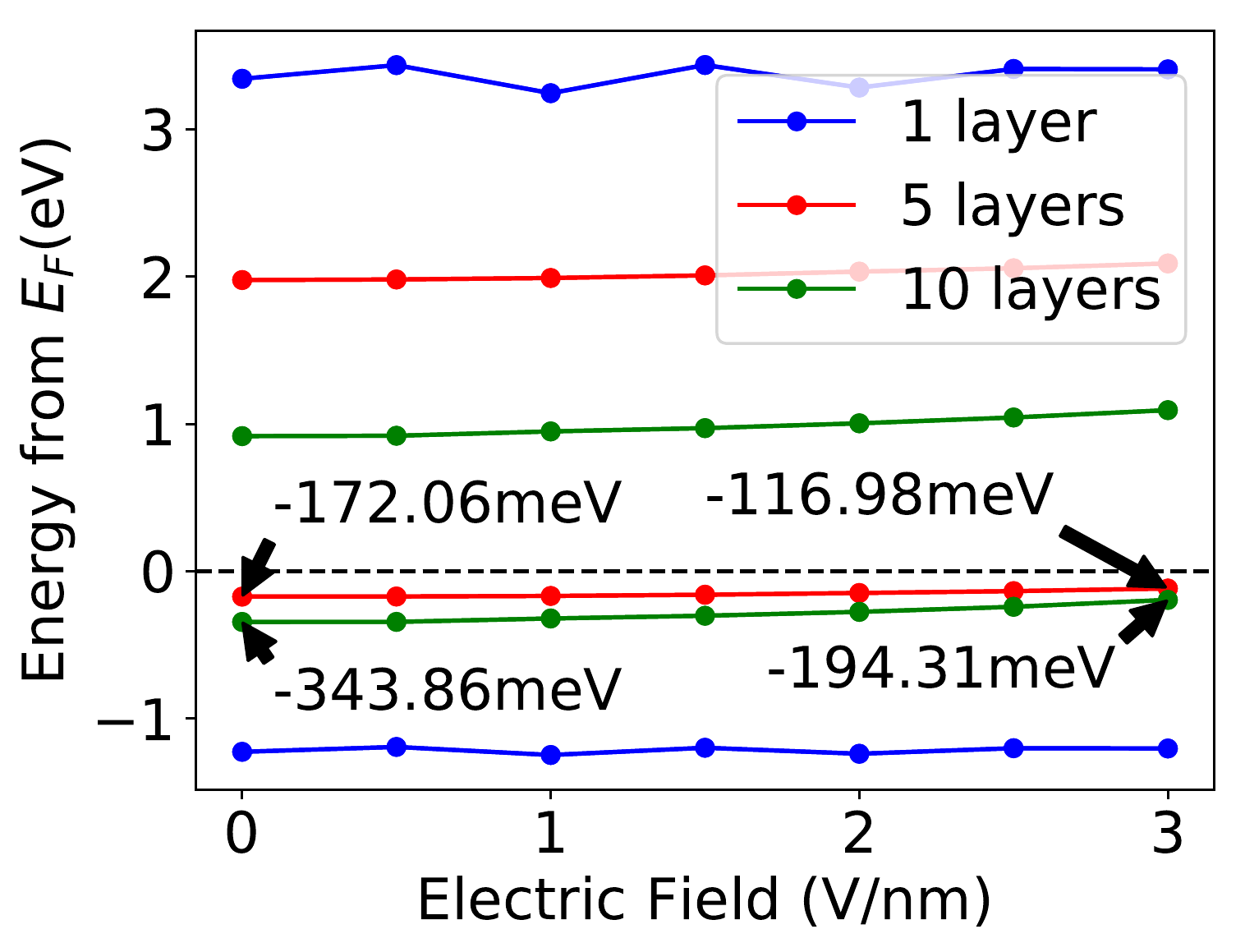}
\subcaption{Field tuned Pb (111)} \label{Field_Pb111}
\end{subfigure}
\begin{subfigure}{0.45\linewidth}
\includegraphics[width=\linewidth,height=0.75\linewidth]{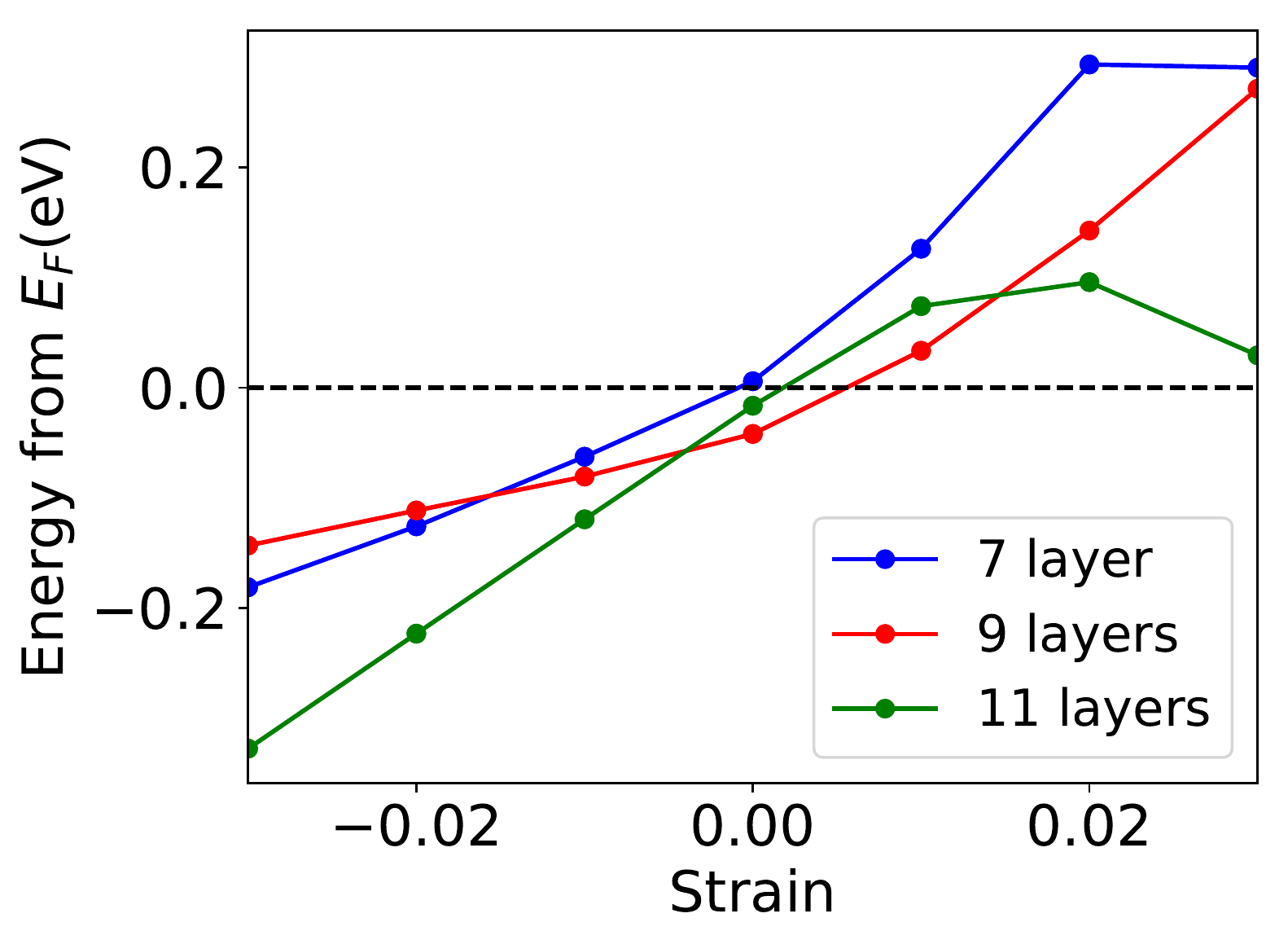}
\subcaption{Strain tuned Sn (001)} \label{Strain_Sn001}
\end{subfigure}
\begin{subfigure}{0.45\linewidth}
\includegraphics[width=\linewidth,height=0.75\linewidth]{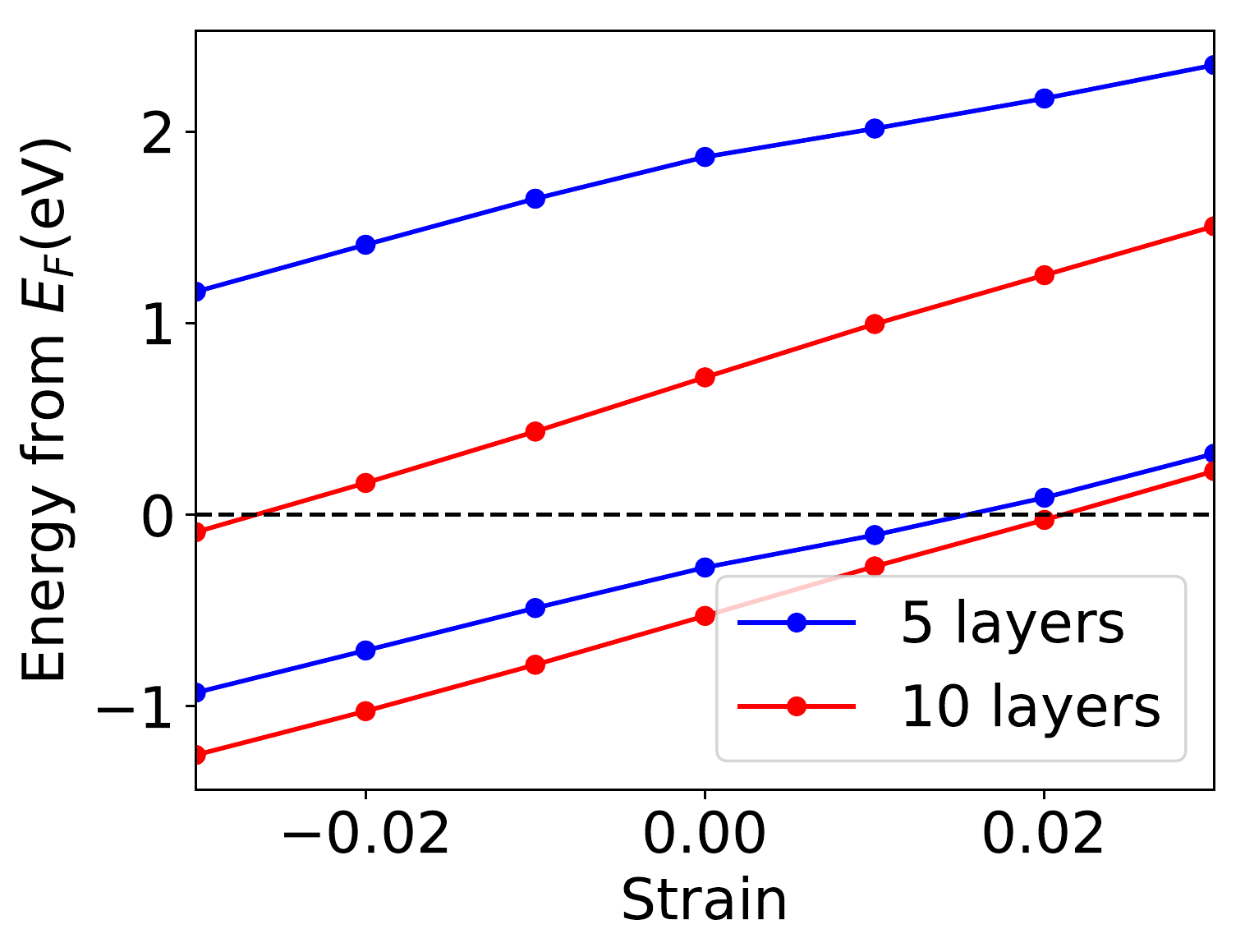}
\subcaption{Strain tuned Pb (111)} \label{Strain_Pb111}
\end{subfigure}
\caption{
Band extrema tuning via electric field and strain. (a) $\Gamma$-point band energy relative to the Fermi level as a function of electric field for a $ \beta $-Sn $(001)$ thin film. (b)$\Gamma$ point band energy relative to the Fermi level as a function of electric field for a Pb $(111)$ thin film. (c) $\Gamma$ point band energy as a function of strain in the $\pm 3\%$ range for $ \beta $-Sn $(001)$ thin films. (d) $\Gamma$ point band energy as a function of strain in the $\pm 3\%$  range for Pb $(111)$ thin films.
} \label{strain_depend}
\end{figure} \fi

\textit{Discussion ---}
Ultra-thin films of strongly spin-orbit-coupled superconducting metals have the advantage, compared to the commonly studied systems composed of semiconductors on superconducting substrates, that no interface or proximity effect is needed to achieve superconductivity in a strongly
spin-orbit coupled system. We have shown that superconducting thin films with strong spin-orbit coupling can be driven into a topological superconductor state by tuning with external electric fields or strains. We have evaluated $g$-factors\cite{gfactor} at the extrema of the quasi-2D bands of Pb and $\beta $-Sn, demonstrating that they have large values that limit the accuracy with which the band extrema energies need to be tuned to the Fermi level.

Ultra-thin film growth\cite{ThinFilmGrowth} is a key challenge that must be met to realize this proposal for topological superconductivity. Metal thin films growth is strongly influenced by quantum-size effects \cite{ThinFilmGrowth} that determine a discrete set of magic thicknesses at which smooth growth is possible. Further restrictions are imposed by the requirement that the film thickness not be too large\cite{Strain_relaxation} to allow strain tuning to be effectively employed. To our best knowledge single crystalline $ \beta $-Sn thin film growth has not yet been achieved. Recent experiments have however already demonstrated superconductivity with strong spin-orbit coupling\cite{Nam2016} in ultrathin films of Pb. Our results motivate experimental efforts to grow the $ \beta $-Sn thin film and drive $ \beta $-Sn and Pb thin film into TSC phase with a relatively weak magnetic field, or by depositing magnetic atoms or films.

\textit{Acknowledgements.---}  The authors thank Ken Shih for helpful discussions. This work was supported by the Office of Naval Research under grant ONR-N00014-14-1-0330 and the Welch Foundation under grant TBF1473. The authors acknowledge the Texas Advanced Computing Center (TACC) at The University of Texas at Austin for providing HPC resources that have contributed to the research results reported within this paper.

\bibliography{MFmetal}

%\end{document}

\clearpage
\onecolumngrid
\appendix
\begin{center}
{\bf{Supplementary Materials}}
\end{center}
\vspace{3mm}

Density functional theory calculations were performed using Quantum Espresso\cite{QE2009,QE2017} with PAW pseudopotentials\cite{PAW2014} and the Vienna Ab initio simulation package (VASP)\cite{VASP1993,VASP1994,VASP1996a,VASP1996b} with Generalized Gradient Approximation PBE\cite{PBE1996,PBE1997} pseudopotentials.   The VASP software was used only
for the calculations assessing the influence of gate electric fields.

\section{Details for $\rm \beta-Sn$ Calculations}

\subsection{Bulk $\rm \beta-Sn$: crystal structure}

Bulk $\beta$-Sn has the tetragonal structure (A5) illustrated in Fig. \ref{Sn_bulk}
with a lattice constant of $a = 5.8179 \AA$ and $c = 3.1749 \AA$.  This structure can be viewed as
two body-center cubic structures
displaced by a vector $\frac{1}{2} \bm{a} + \frac{1}{4} \bm{c}$
where $\bm{a}$ and $ \bm{c} $ are the lattice vectors of the cubic unit cell.
The lattice vectors of the primitive unit cell are
 $-\frac{1}{2} \bm{a} + \frac{1}{2} \bm{b} + \frac{1}{2} \bm{c}$, $\frac{1}{2} \bm{a} - \frac{1}{2} \bm{b} + \frac{1}{2} \bm{c}$ and
$\frac{1}{2} \bm{a} + \frac{1}{2} \bm{b} - \frac{1}{2} \bm{c}$.
There are two Sn atoms per cell with the positions as $(0,0,0)$ and $(\frac{1}{4}, \frac{3}{4}, \frac{1}{2})$
in crystal coordinates.

\ifpdf  \begin{figure}[h]
\centering
\includegraphics[width=0.9\linewidth]{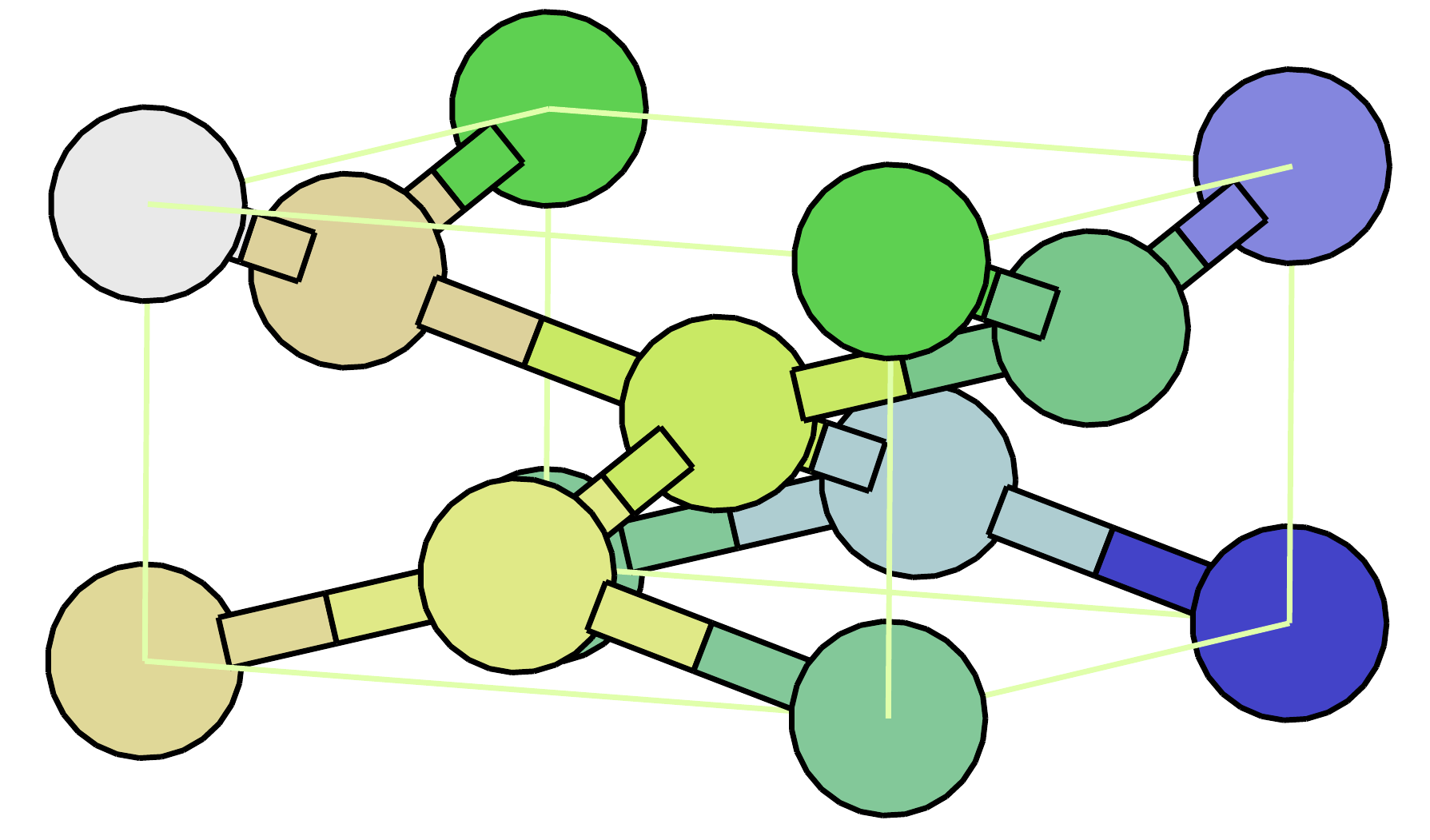}
\caption{Crystal structure of bulk $\beta$-Sn, which is a tetragonal structure (A5) with lattice constants $a = 5.8179 \AA$
and $c = 3.1749 \AA$.
} \label{Sn_bulk}
\end{figure} \fi
The k-path used for the
bulk $\beta$-Sn bandstructure plots in the main text is
specified in Table \ref{bz_sn_bulk_table} and in Fig. \ref{BZ_Sn_bulk}.
\begin{table}[h]
\caption{\label{bz_sn_bulk_table}
Symmetry points in the first Brillouin zone of bulk $\beta$-Sn with a tetragonal structure (A5),
with $\bm{k} = u \bm{a}^{\ast} + v \bm{b}^{\ast} + w \bm{c}^{\ast}$,
where $a^{\ast}$, $b^{\ast}$, $c^{\ast}$ are reciprocal lattice vectors shown as blue vectors in Fig. \ref{BZ_Sn_bulk}.
}
\begin{ruledtabular}
\begin{tabular}{c c}
  Symmetry points & k points:$(u,v,w)$ \\ [0.5ex]
 \hline
 $\Gamma$ & (0,0,0)  \\
 X & (0,0,0.5)  \\
 M & (-0.5,0.5,0.5) \\
 P & (0.25,0.25,0.25)  \\
 N & (0,0.5,0)  \\
 \end{tabular}
 \end{ruledtabular}
\end{table}

\ifpdf  \begin{figure}[h]
\centering
\includegraphics[width=0.9\linewidth]{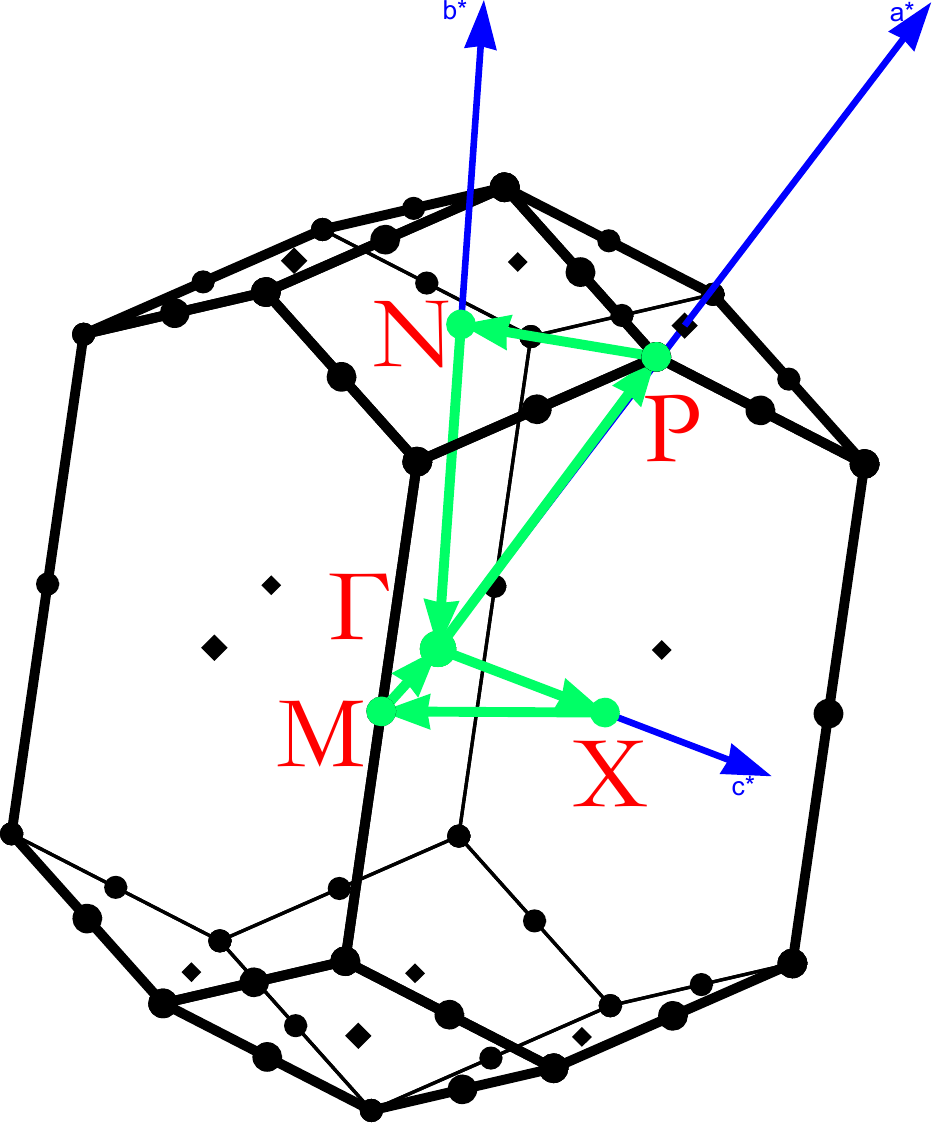}
\caption{The first Brillouin zone and the k-path selected to illustrate
 the bandstructure of bulk $\beta$-Sn.  The points labelled in red are the plotted high symmetry points
 and the green lines are the connecting paths.
} \label{BZ_Sn_bulk}
\end{figure} \fi

\subsection{$\rm \beta-Sn$ $001 $ direction thin films: crystal structure}

For thin film $\beta$-Sn, we choose the $001 $
growth direction.  The lattice constant in the $(001)$ direction is $a = 5.8197 \AA $.
To model the thin film we construct a supercell whose lattice vectors are $a \bm{x} $ and  $a \bm{y}$ and $h \bm{z}$,
where $h$ is the height along $z$ axis which is different from the lattice constant $c$ of bulk $\beta$-Sn.
In the calculations $h$ is set to be the thickness of thin film plus a 20-$\AA$-thick vacuum layer.
For the single Sn there are three Sn atoms per supercell with the positions as $(0,0,0)$, $(0.5, 0, \frac{c}{4h})$
and $(0.5, 0.5, \frac{c}{2h})$ in crystal coordinates, which form for the smallest unit cell a
single layer where every atom is bonded.
The structure of single layer is shown in Fig. \ref{Sn_film}.

\ifpdf  \begin{figure}[h]
\centering
\includegraphics[width=0.9\linewidth]{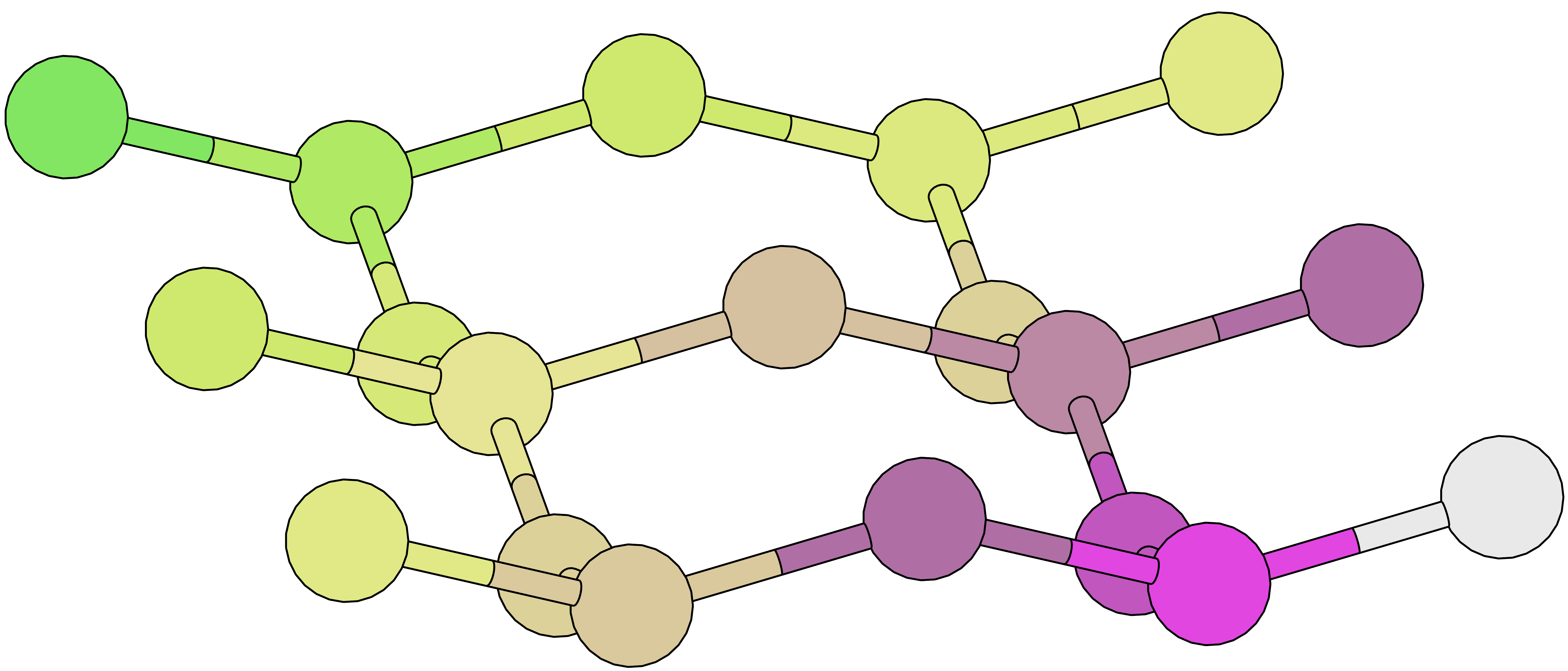}
\caption{Crystal structure of thin film $\beta$-Sn in $(001)$ direction, the lattice constant is $a = 5.8197 \AA $.} \label{Sn_film}
\end{figure} \fi

The $k$-path of the bandstructure of thin film $\beta$-Sn in $(001)$ direction in the main text is shown in table \ref{bz_sn_film_table} and Fig. \ref{BZ_Sn_film}.

\begin{table}[h]
\caption{\label{bz_sn_film_table}
Symmetry points in the first Brillouin zone of thin film $\beta$-Sn in $(001)$ direction,
with $\bm{k} = u \bm{a}^{\ast} + v \bm{b}^{\ast} + w \bm{c}^{\ast}$,
where $a^{\ast}$, $b^{\ast}$, $c^{\ast}$ are reciprocal lattice vectors shown as blue vectors in Fig. \ref{BZ_Sn_film}.
}
\begin{ruledtabular}
\begin{tabular}{c c}
  Symmetry points & k points:$(u,v,w)$ \\ [0.5ex]
 \hline
 $\Gamma$ & (0,0,0)  \\
 M & (0.5,0,0)  \\
 X & (0.5,0.5,0) \\
 \end{tabular}
 \end{ruledtabular}
\end{table}

\ifpdf  \begin{figure}[h]
\centering
\includegraphics[width=0.9\linewidth]{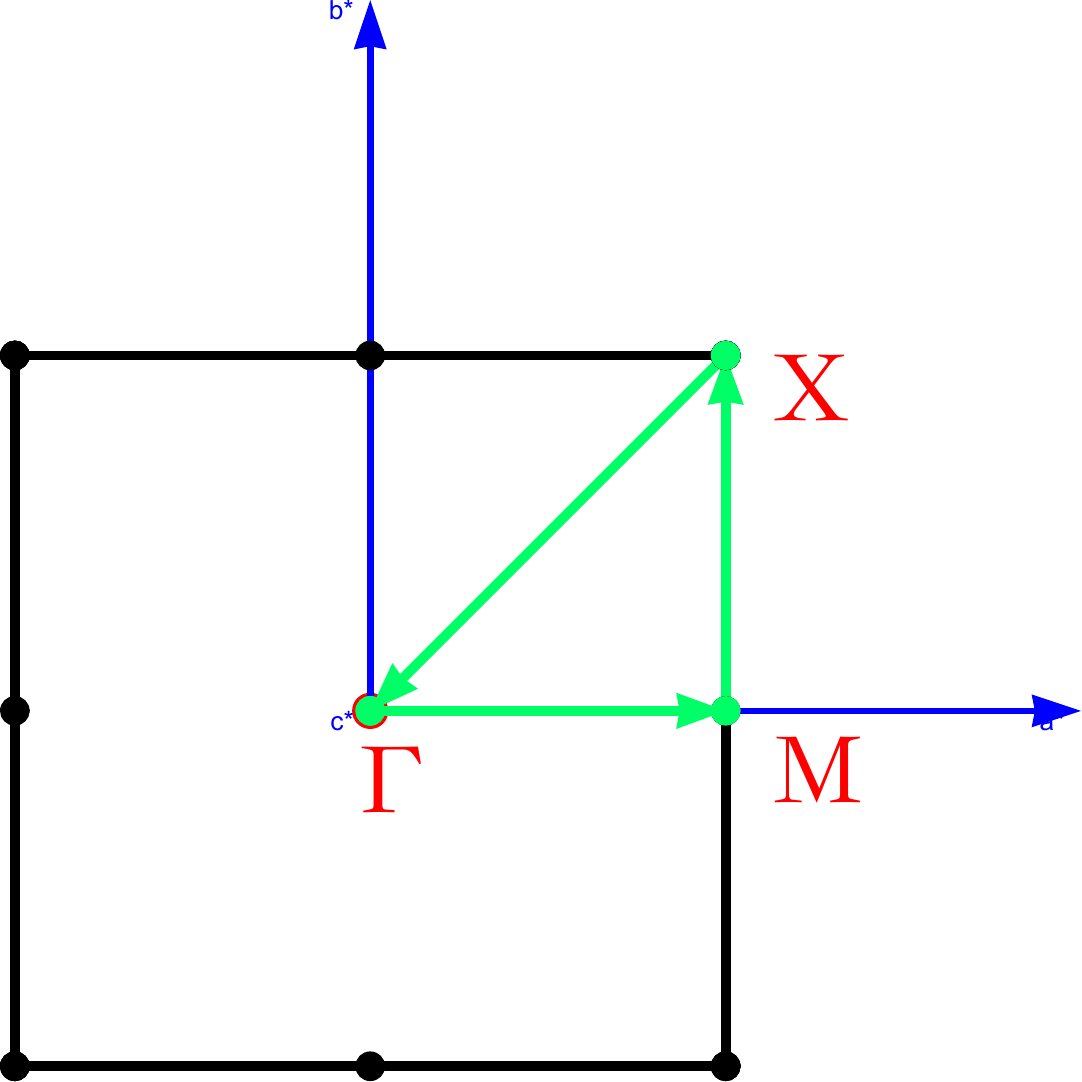}
\caption{The first Brillouin zone and the k-path used to calculate the bandstructue of the thin film $\beta$-Sn in $(001)$ direction.
} \label{BZ_Sn_film}
\end{figure} \fi

\subsection{Thin film $\rm \beta-Sn$ in $001 $ direction: electronic structure}
More bandstructures of thin film $\rm \beta-Sn$ in $001 $ direction are shown in Fig.\ref{band_sn_3}-\ref{band_sn_13},
here we choose odd number of layers thin film up to 13 layers, the bands are mainly s and p orbitals,
but s and p orbitals are highly hybridized for the layers more than 3, for the thicker thin film,
more bands are hybridized, which makes the number of subbands much larger than the s and p bands.
\ifpdf
\begin{figure*}
\centering
\begin{subfigure}{0.45\linewidth}
\includegraphics[width=\linewidth,height=0.75\linewidth]{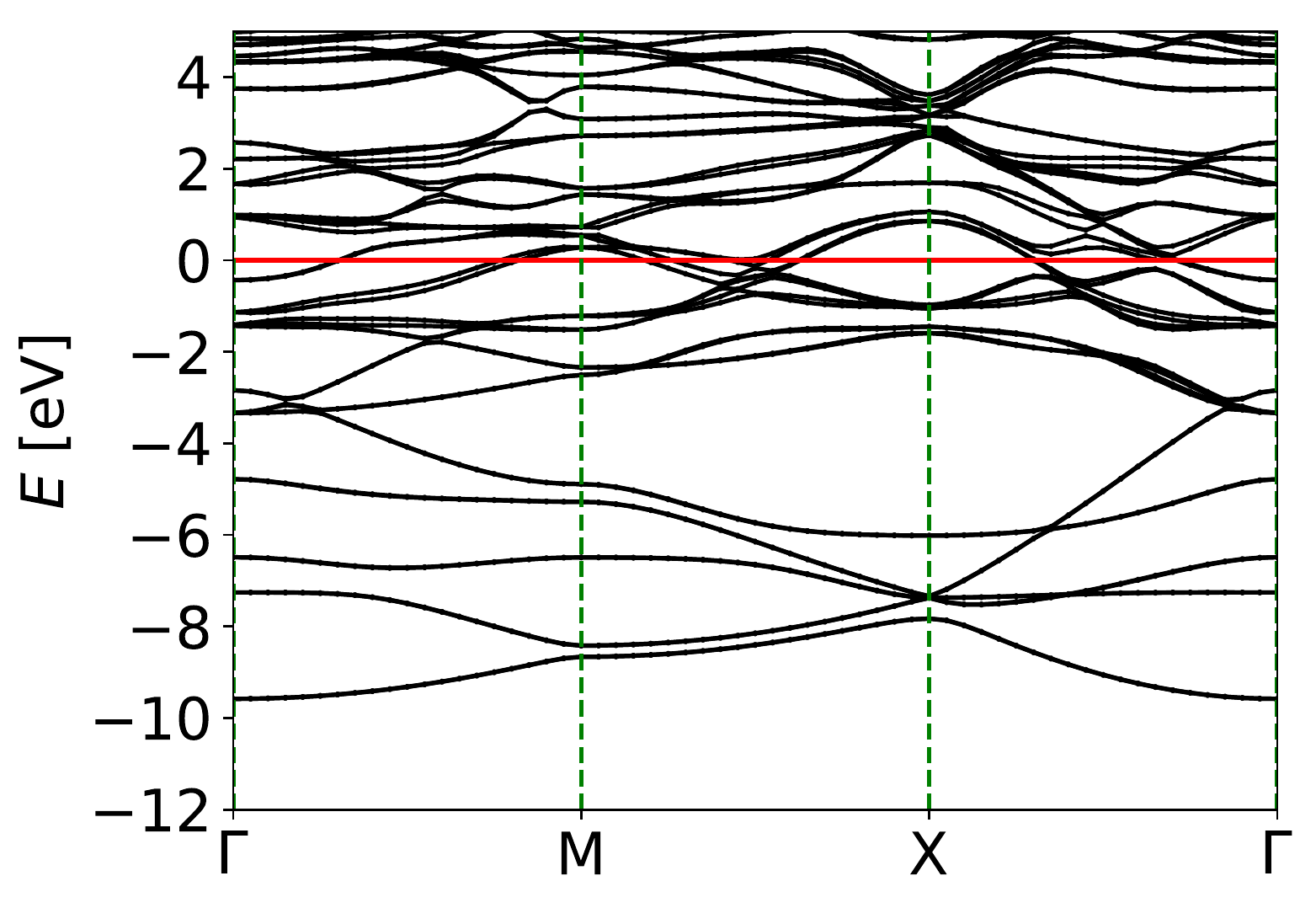}
\subcaption{3 Layer of Sn (001)} \label{band_sn_3}
\end{subfigure}
\begin{subfigure}{0.45\linewidth}
\includegraphics[width=\linewidth,height=0.75\linewidth]{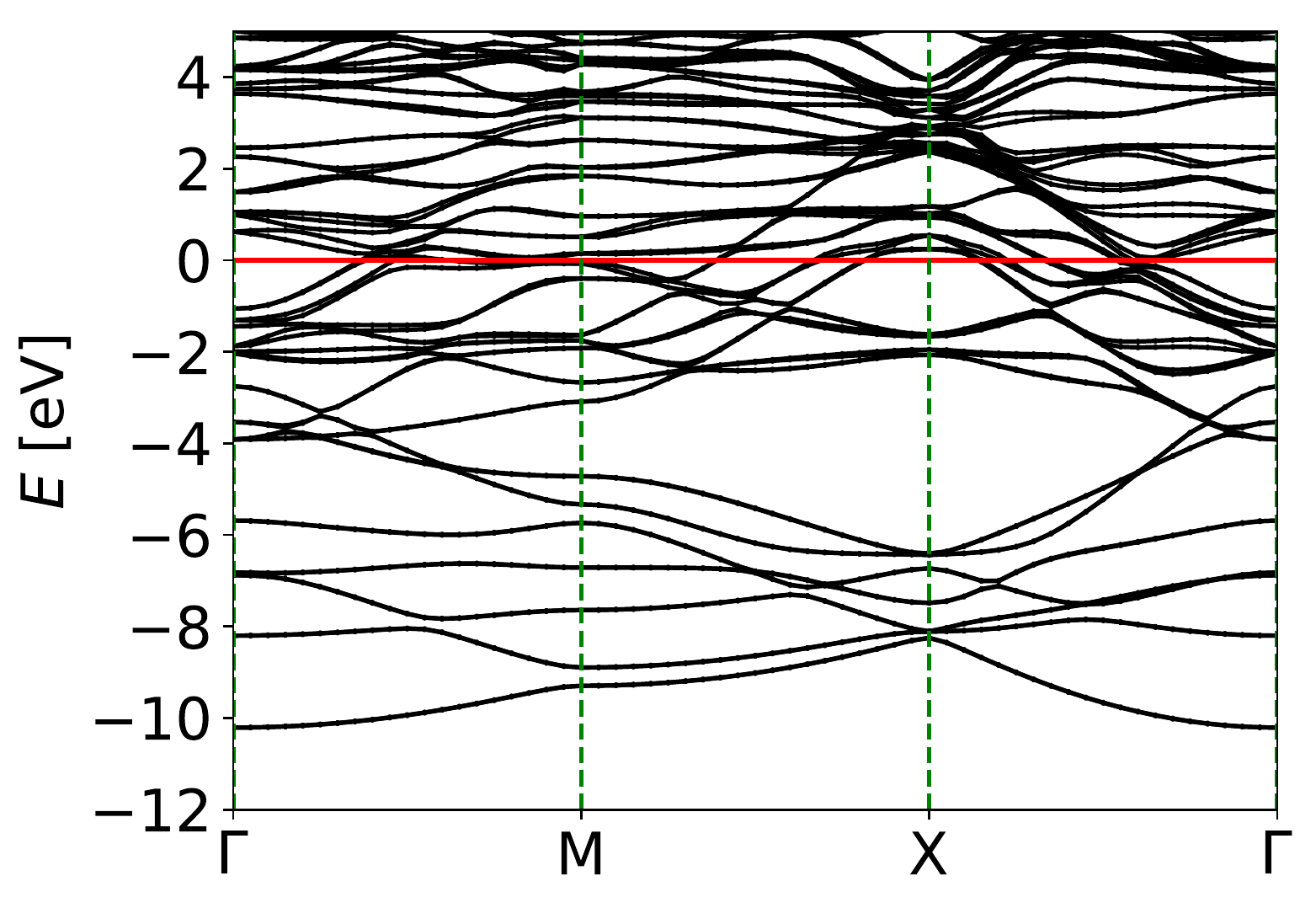}
\subcaption{5 Layer of Sn (001)} \label{band_sn_5}
\end{subfigure}
\begin{subfigure}{0.45\linewidth}
\includegraphics[width=\linewidth,height=0.75\linewidth]{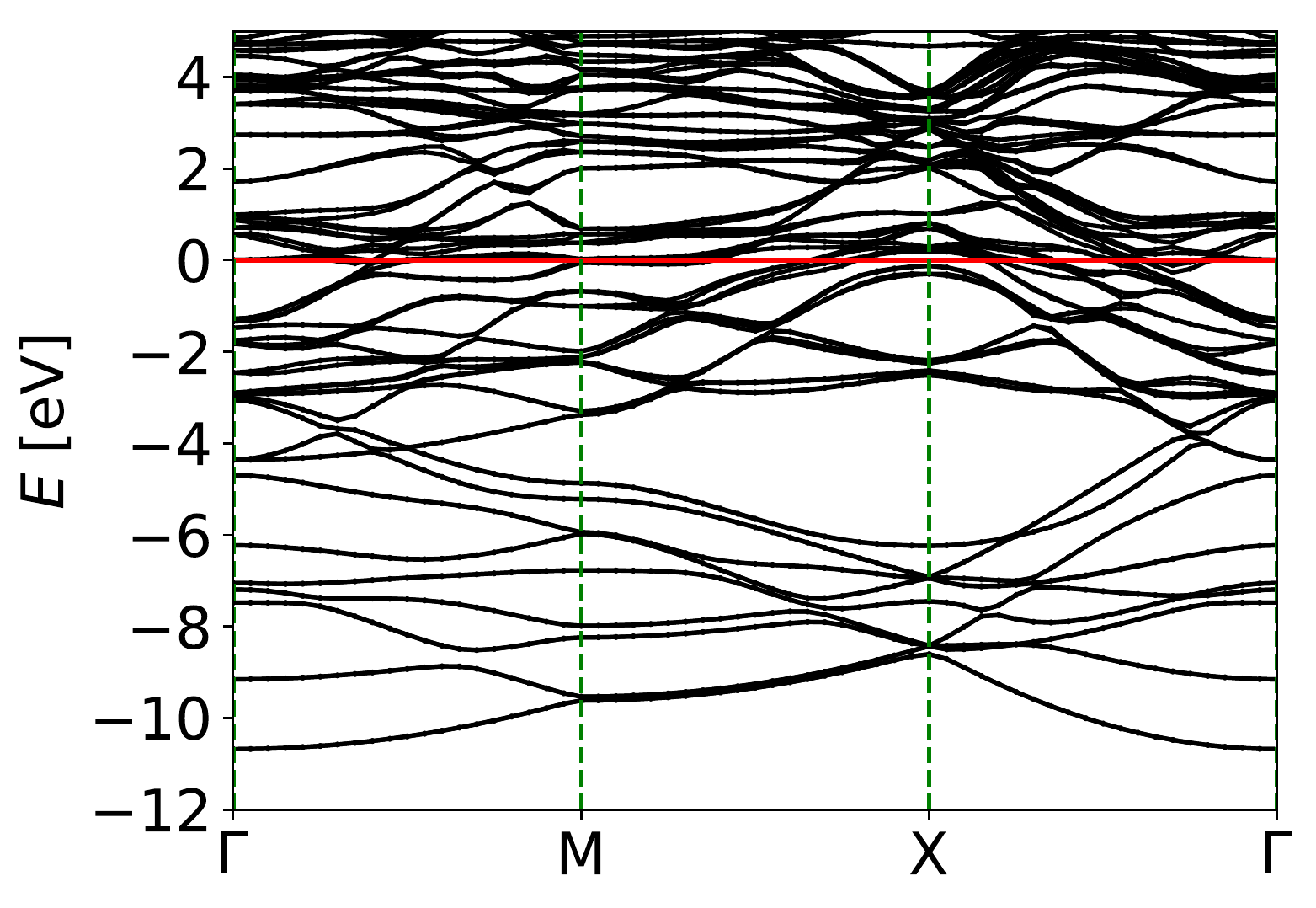}
\subcaption{7 Layer of Sn (001)} \label{band_sn_7}
\end{subfigure}
\begin{subfigure}{0.45\linewidth}
\includegraphics[width=\linewidth,height=0.75\linewidth]{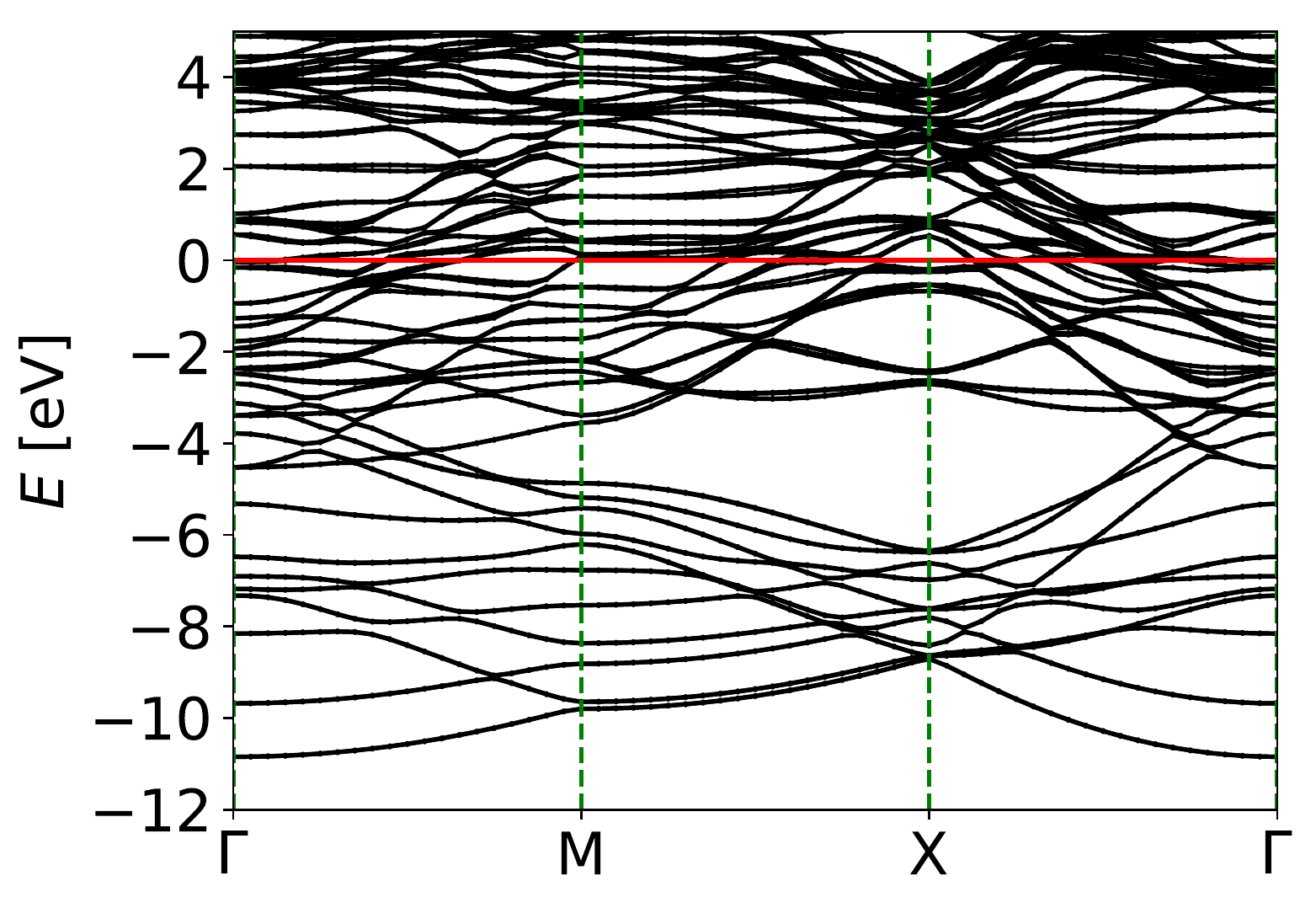}
\subcaption{9 Layer of Sn (001)} \label{band_sn_9}
\end{subfigure}
\begin{subfigure}{0.45\linewidth}
\includegraphics[width=\linewidth,height=0.75\linewidth]{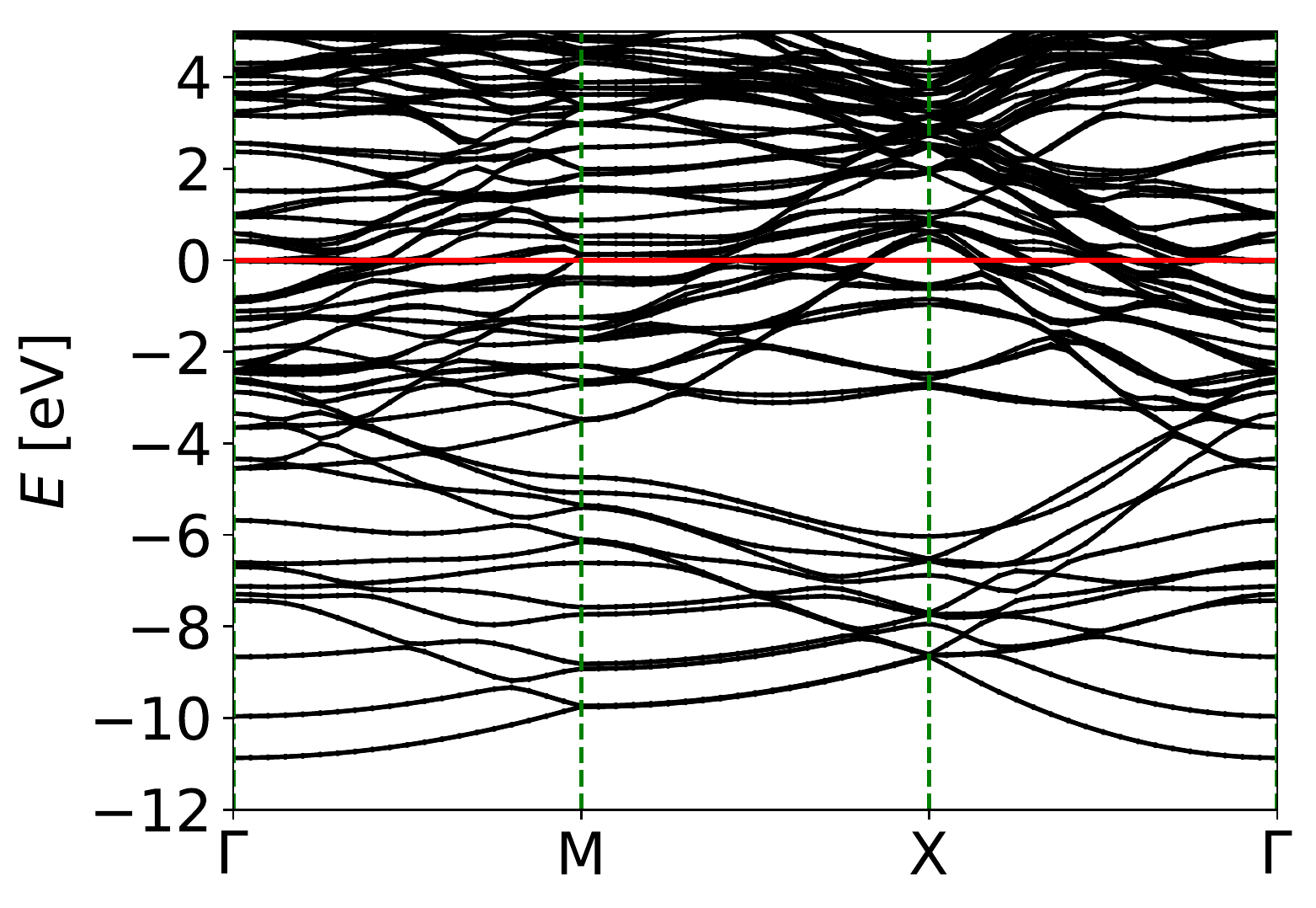}
\subcaption{11 Layer of Sn (001)} \label{band_sn_11}
\end{subfigure}
\begin{subfigure}{0.45\linewidth}
\includegraphics[width=\linewidth,height=0.75\linewidth]{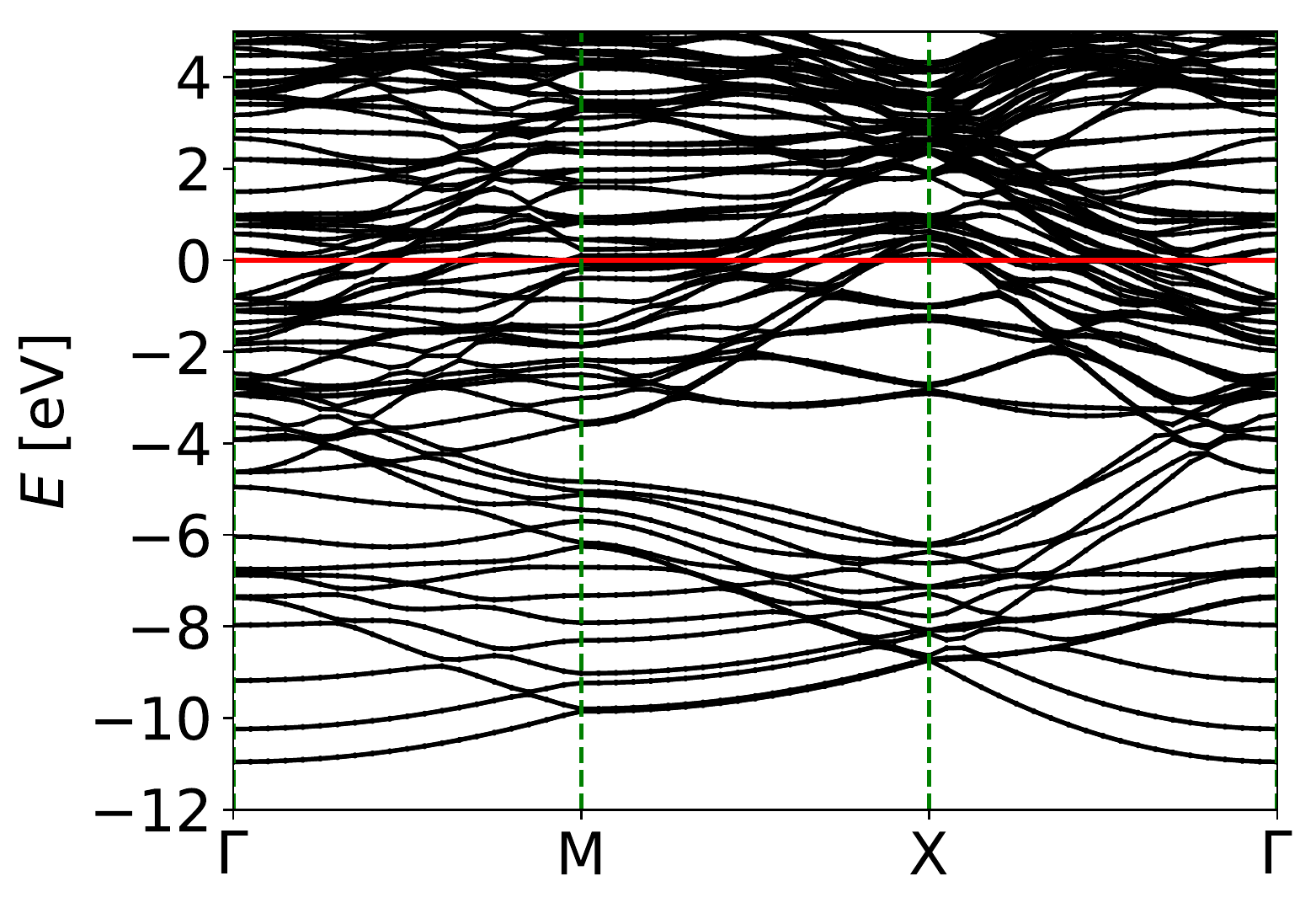}
\subcaption{13 Layer of Sn (001)} \label{band_sn_13}
\end{subfigure}
\caption{
Band structure of thin film $\beta $ Sn grown
along the $(001)$ direction, from  a 3 layers (5 atomic layer) to 13 layers}
\label{band_film_sn}
\end{figure*} \fi

\subsection{Rashba constants}
To extract the Rashba coupling constants we used the following equation:
\begin{equation}\label{rashba}
  \alpha_R \approx \frac{\Delta E}{\Delta k}
\end{equation}
where $\Delta k$ is chosen to be small enough.
The Rashba coupling is of course band-dependent.
Typical results are shown in Fig. \ref{rashba_film_sn},
where the left side plots are for the bands closest to the Fermi level
and the right sides is an average values over 12 subbands around the Fermi level: 6 above and 6 below.
We see that the Rashba coupling becomes smaller when the number of layers
increases.  Up to 15 layers, however, its value is still up to 0.85 $eV \AA$, which is several times larger than
in semiconductor quantum wires.
\ifpdf
\begin{figure*}
\centering
\begin{subfigure}{0.45\linewidth}
\includegraphics[width=\linewidth,height=0.75\linewidth]{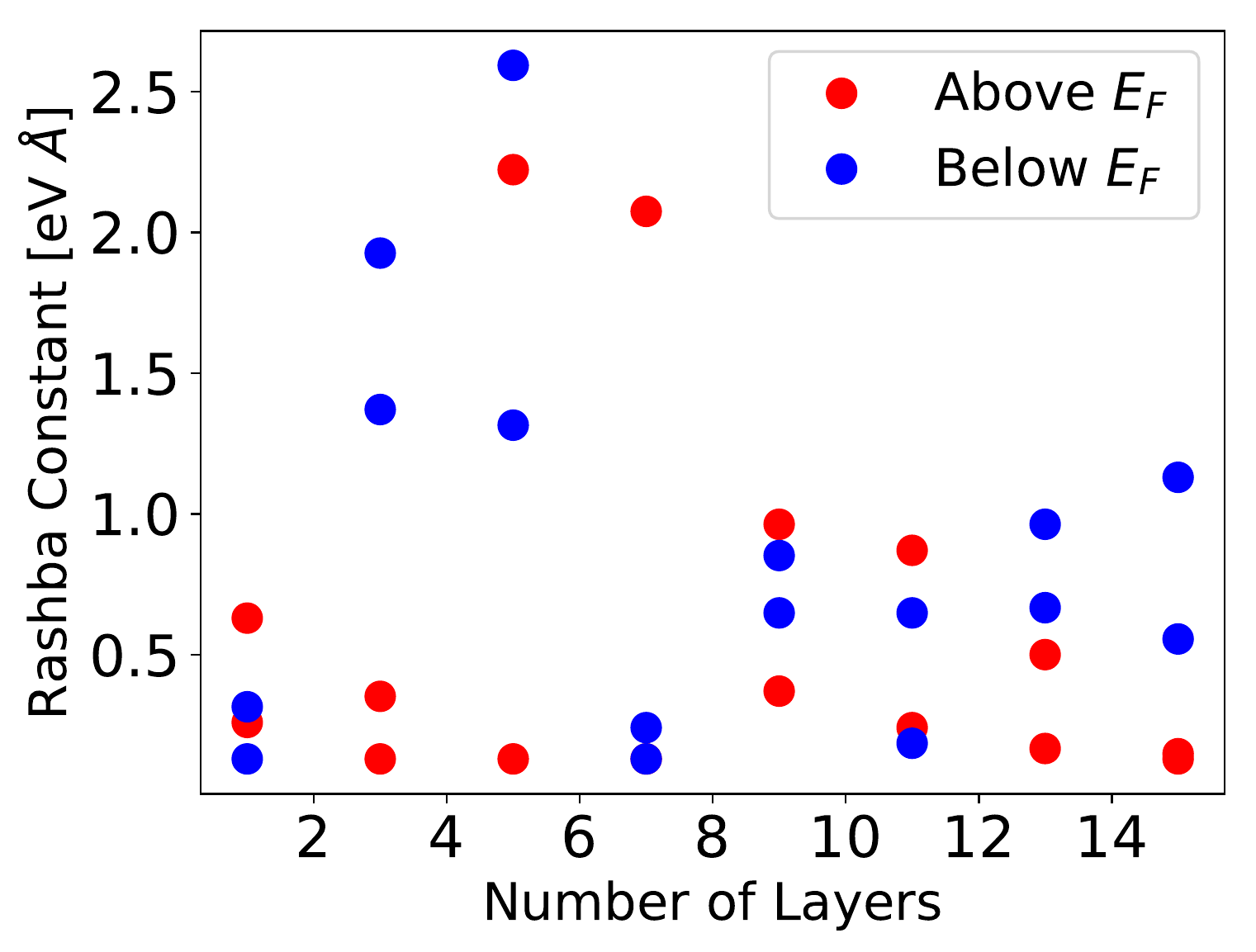}
\subcaption{Rashba coupling of Sn (001) vs layers} \label{r_sn_lay}
\end{subfigure}
\begin{subfigure}{0.45\linewidth}
\includegraphics[width=\linewidth,height=0.75\linewidth]{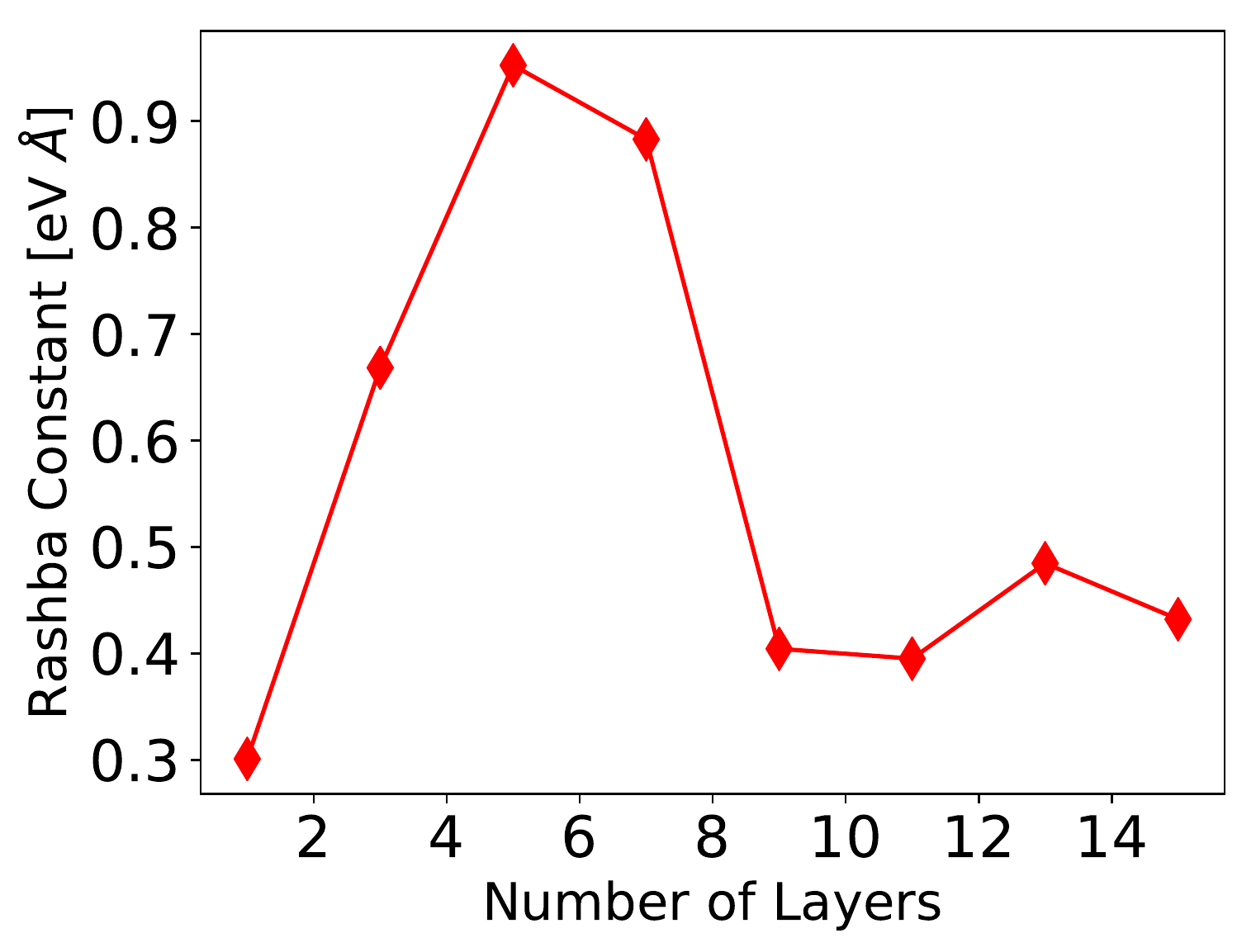}
\subcaption{Average Rashba coupling of Sn (001) vs layers} \label{r_sn_avg}
\end{subfigure}
\caption{
Rashba coupling(and the average of 10 bands around Fermi level) of Sn (001) around Fermi level vs layers}
\label{rashba_film_sn}
\end{figure*} \fi

The influence of strain on Rashba coupling for 7,9 and 11 layers of Sn thin film is shown in Fig. \ref{rashba_film_sn_strain}:
\ifpdf
\begin{figure*}
\centering
\begin{subfigure}{0.45\linewidth}
\includegraphics[width=\linewidth,height=0.75\linewidth]{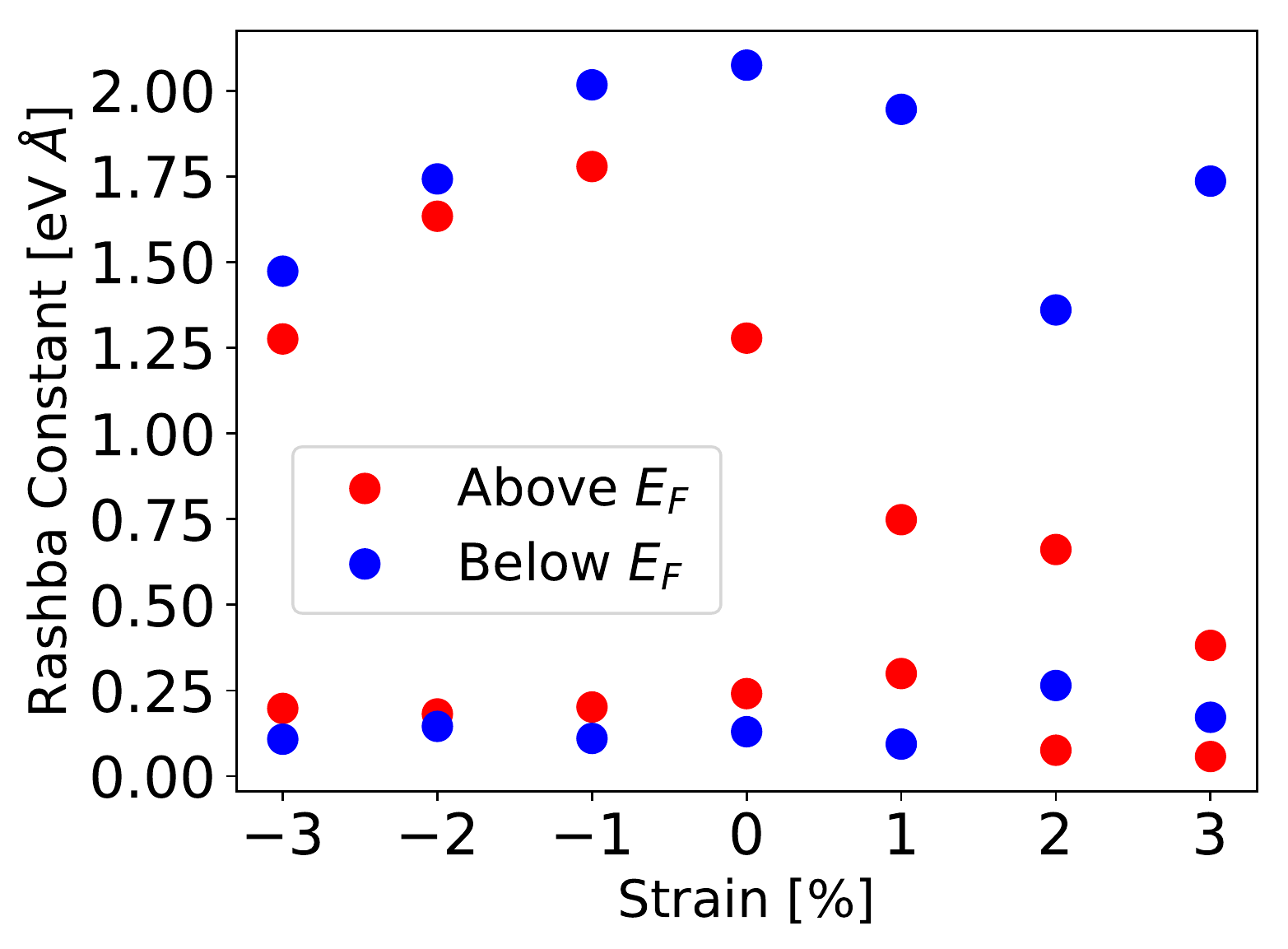}
\subcaption{Rashba coupling of Sn (001) vs strain for 7 layers case} \label{r_sn_lay_7}
\end{subfigure}
\begin{subfigure}{0.45\linewidth}
\includegraphics[width=\linewidth,height=0.75\linewidth]{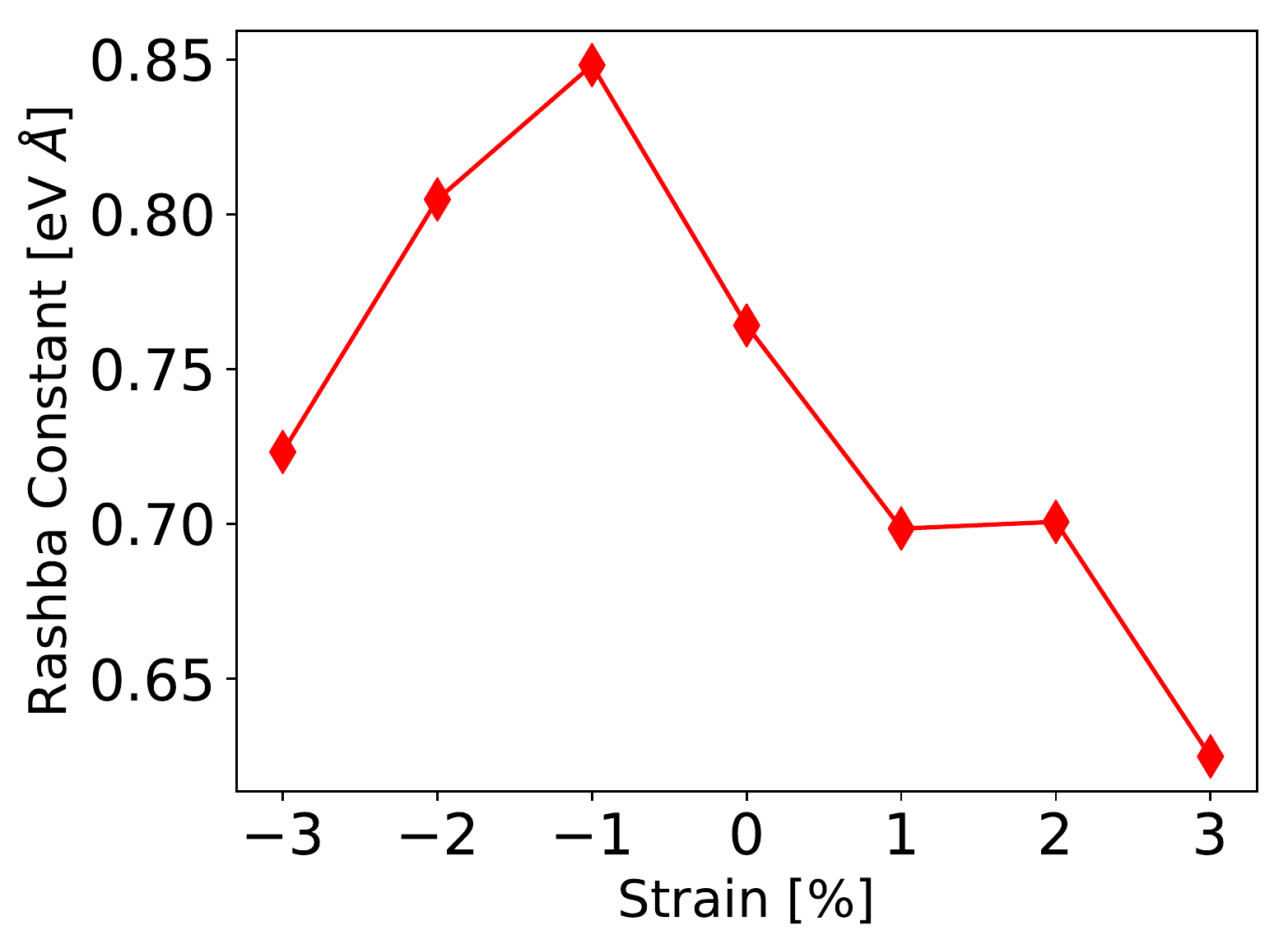}
\subcaption{Average Rashba coupling of Sn (001) vs strain for 7 layers case} \label{r_sn_avg_7}
\end{subfigure}
\begin{subfigure}{0.45\linewidth}
\includegraphics[width=\linewidth,height=0.75\linewidth]{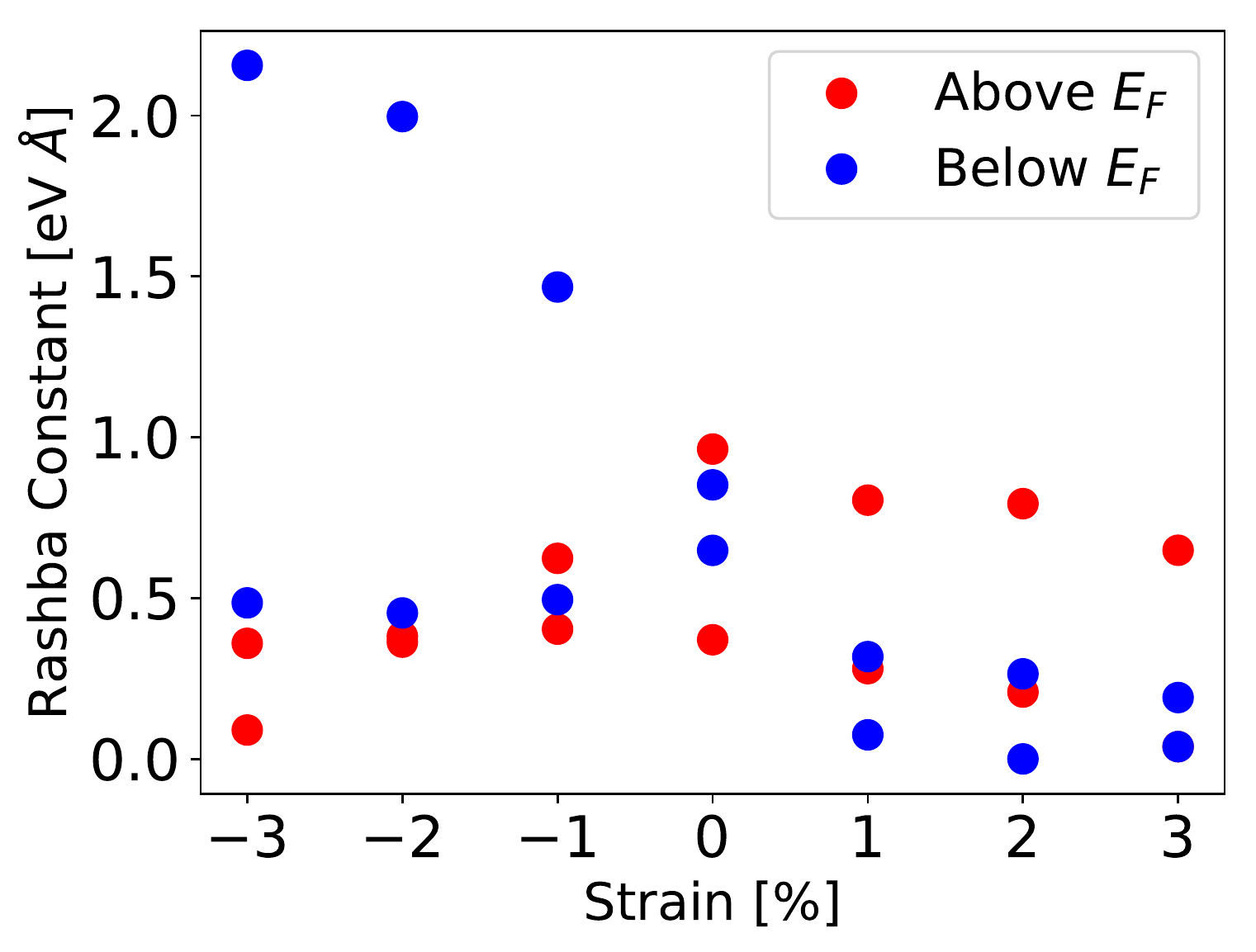}
\subcaption{Rashba coupling of Sn (001) vs strain for 9 layers case} \label{r_sn_lay_9}
\end{subfigure}
\begin{subfigure}{0.45\linewidth}
\includegraphics[width=\linewidth,height=0.75\linewidth]{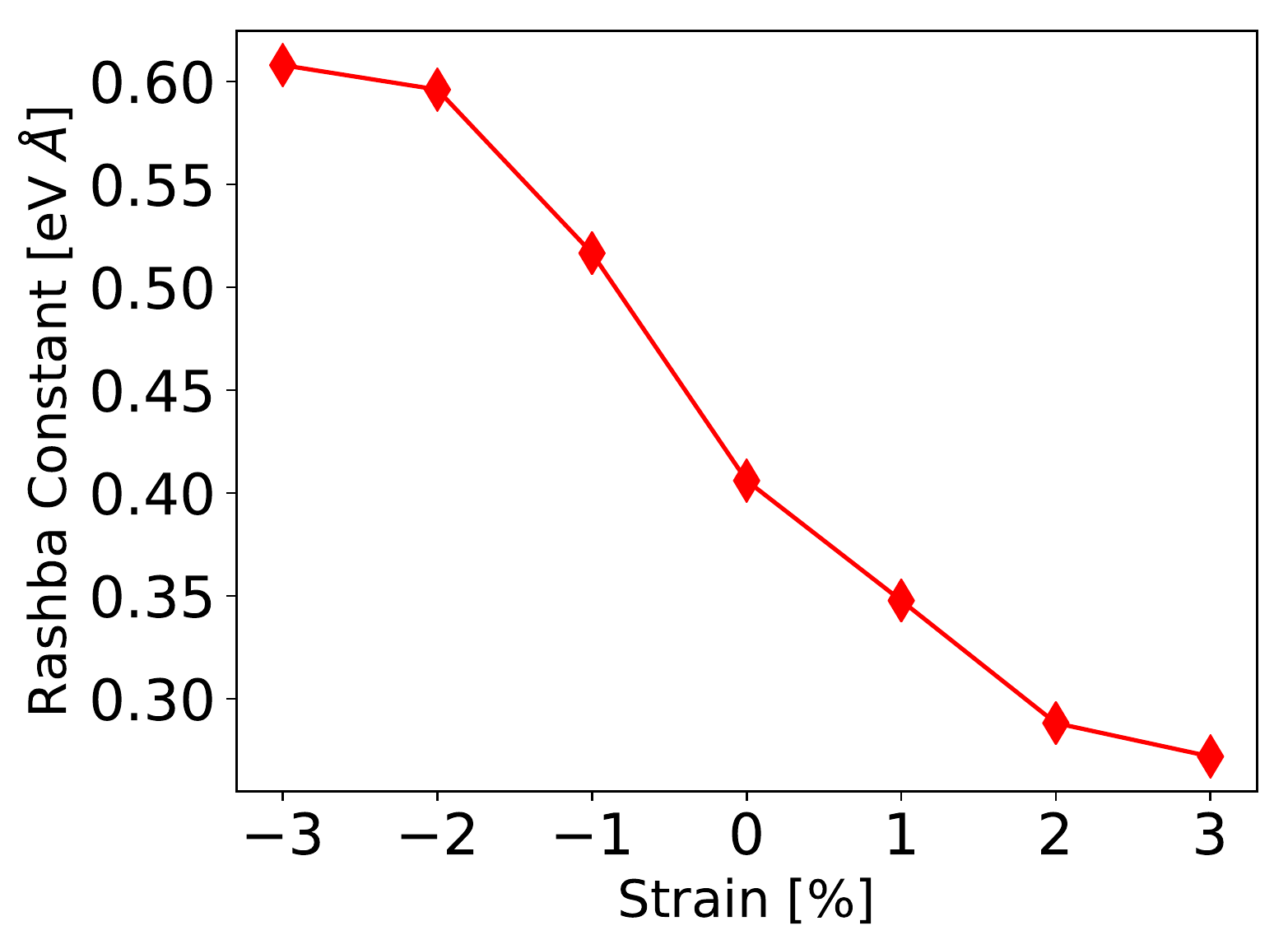}
\subcaption{Average Rashba coupling of Sn (001) vs strain for 9 layers case} \label{r_sn_avg_9}
\end{subfigure}
\begin{subfigure}{0.45\linewidth}
\includegraphics[width=\linewidth,height=0.75\linewidth]{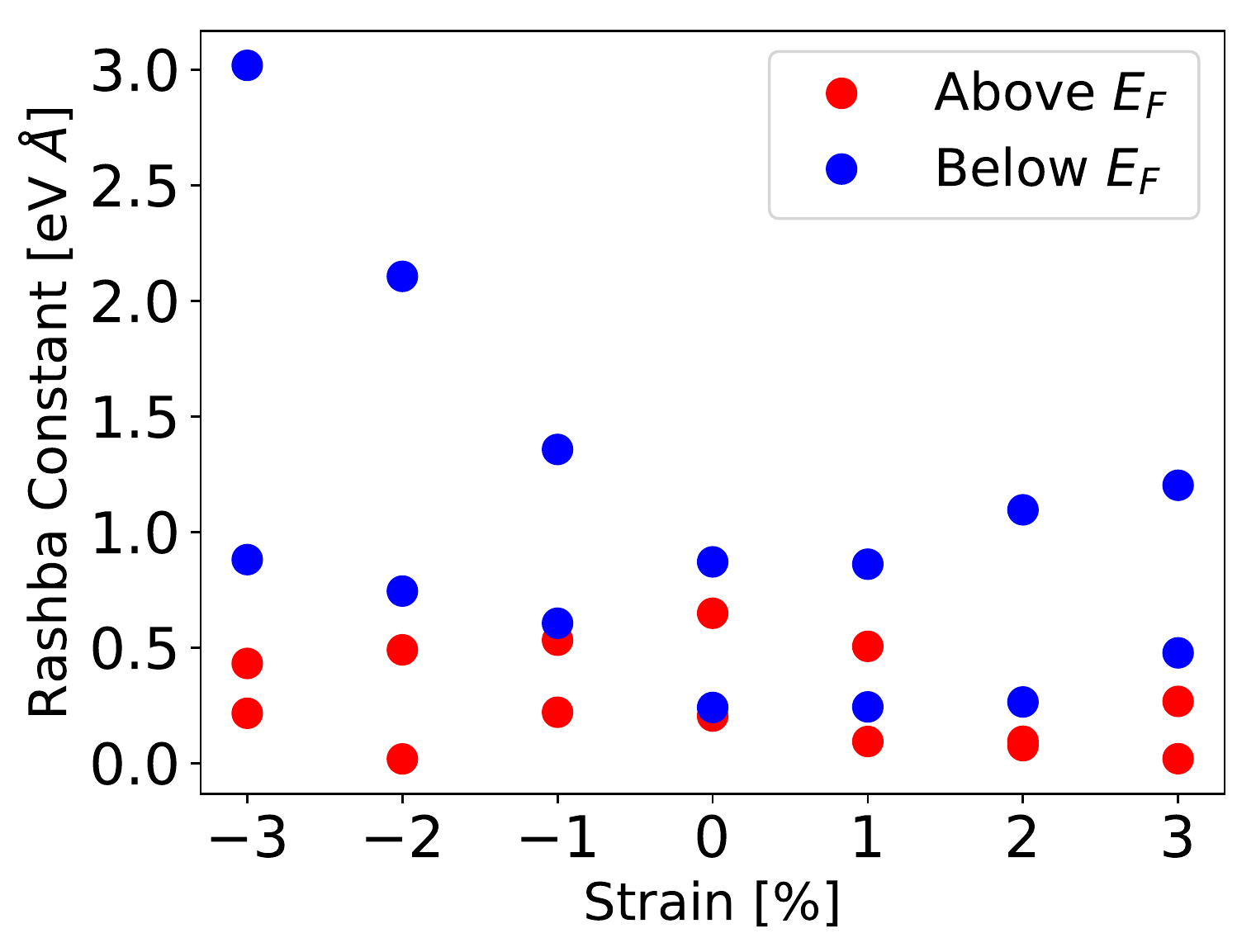}
\subcaption{Rashba coupling of Sn (001) vs strain for 11 layers case} \label{r_sn_lay_11}
\end{subfigure}
\begin{subfigure}{0.45\linewidth}
\includegraphics[width=\linewidth,height=0.75\linewidth]{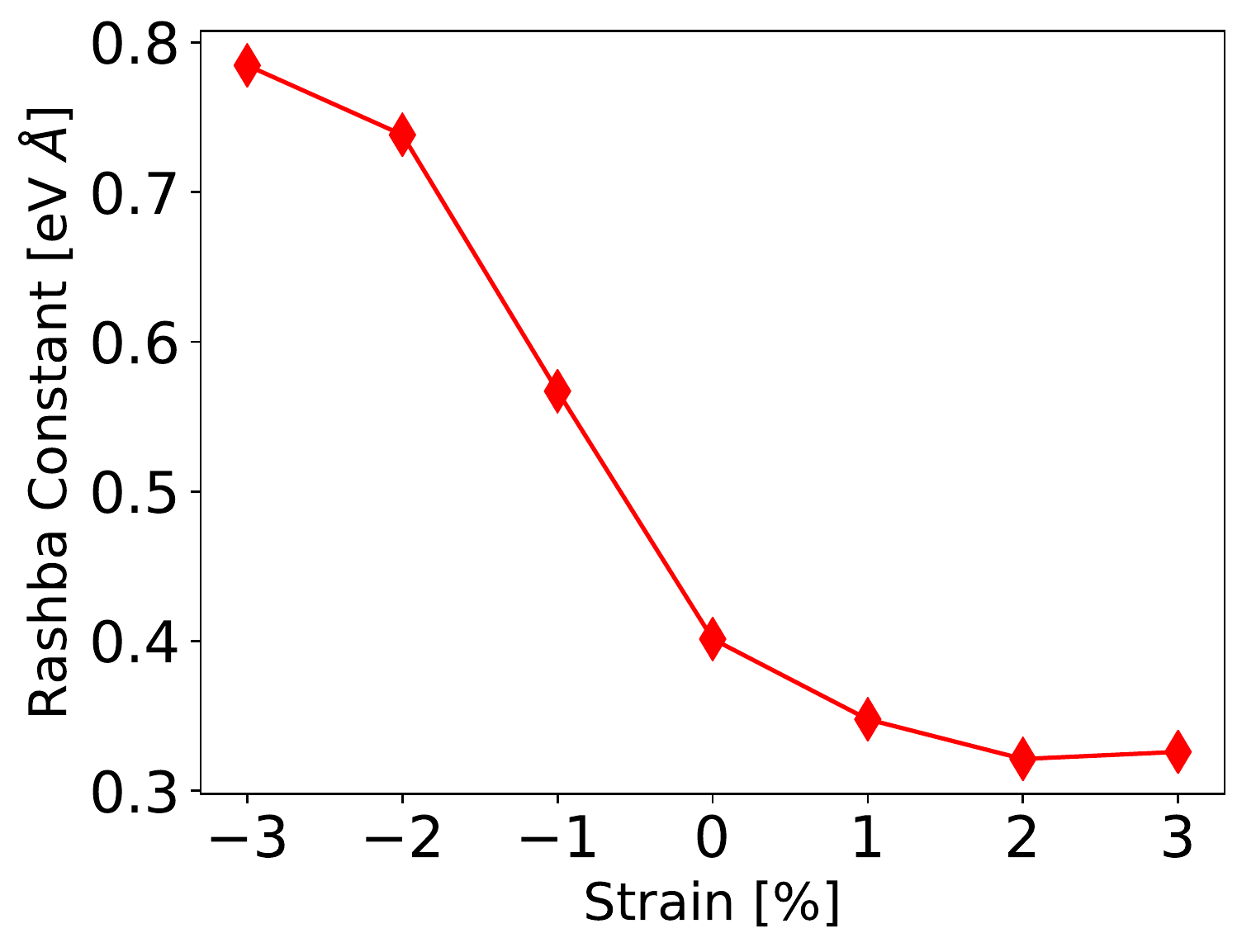}
\subcaption{Average Rashba coupling of Sn (001) vs strain for 11 layers case} \label{r_sn_avg_11}
\end{subfigure}
\caption{
Rashba coupling(and the average of 12 bands around Fermi level) of Sn (001) around Fermi level vs strain
for 7,9 and 11 layers case.}
\label{rashba_film_sn_strain}
\end{figure*} \fi
As it shows that tensile strain will weaken the Rashba effect while the compressive strain will strengthen the Rashba effect.

\section{Details for $\rm Pb$ Calculations}
\subsection{Bulk $\rm Pb$: cryatal structure}
Bulk Pb has a face centered cubic structure as illustrated in
Fig. \ref{Pb_bulk} with a lattice constant of $a = 4.8408 \AA$,
The lattice vectors of the primitive unit cell are
 $\frac{1}{2} \bm{a} + \frac{1}{2} \bm{c}$, $\frac{1}{2} \bm{a} + \frac{1}{2} \bm{b}$ and
$\frac{1}{2} \bm{b} + \frac{1}{2} \bm{c}$.
In the primitive unit cell there is one Pb atom with the positions as $(0,0,0)$ in crystal coordinates.

\ifpdf  \begin{figure}[h]
\centering
\includegraphics[width=0.9\linewidth]{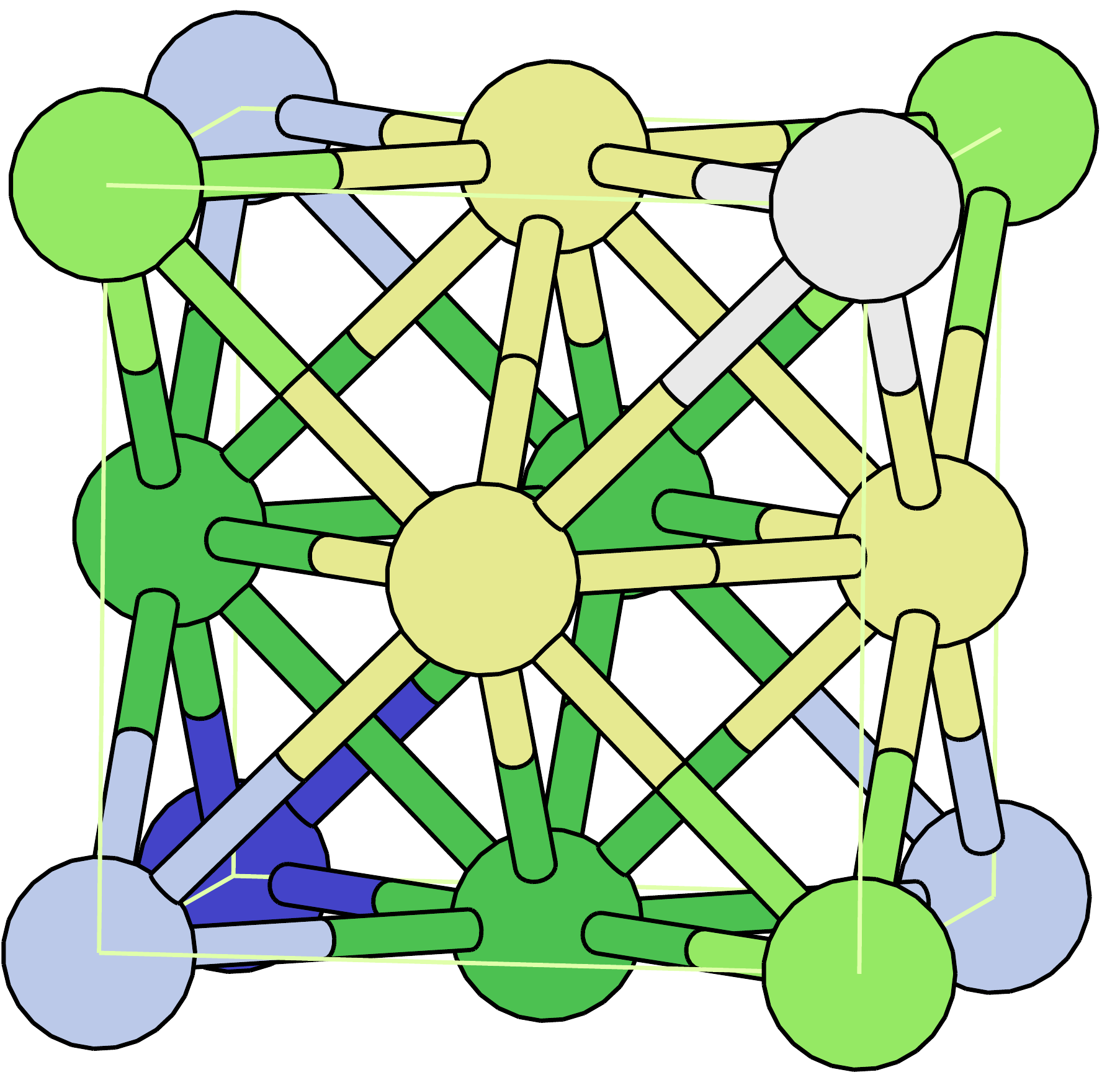}
\caption{Crystal structure of face centered cubic bulk Pb with a lattice constant of $a = 4.8408 \AA$} \label{Pb_bulk}
\end{figure} \fi

The k-path used to illustrate the bandstructure of bulk Pb in the main text is
specified in table \ref{bz_pb_bulk_table} and in Fig. \ref{BZ_Pb_bulk}.

\begin{table}[h]
\caption{\label{bz_pb_bulk_table}
Symmetry points in the first Brillouin zone of bulk Pb with a face centered cubic structure,
with $\bm{k} = u \bm{a}^{\ast} + v \bm{b}^{\ast} + w \bm{c}^{\ast}$,
where $a^{\ast}$, $b^{\ast}$, $c^{\ast}$ are reciprocal lattice vectors shown as blue vectors in Fig. \ref{BZ_Pb_bulk}.
}
\begin{ruledtabular}
\begin{tabular}{c c}
  Symmetry points & k points:$(u,v,w)$ \\ [0.5ex]
 \hline
 $\Gamma$ & (0,0,0)  \\
 X & (0.5,0,0.5)  \\
 W & (0.5,0.25,0.75) \\
 L & (0.5,0.5,0.5)  \\
 K & (0.375,0.375,0.75)  \\
 \end{tabular}
 \end{ruledtabular}
\end{table}

\ifpdf  \begin{figure}[h]
\centering
\includegraphics[width=0.9\linewidth]{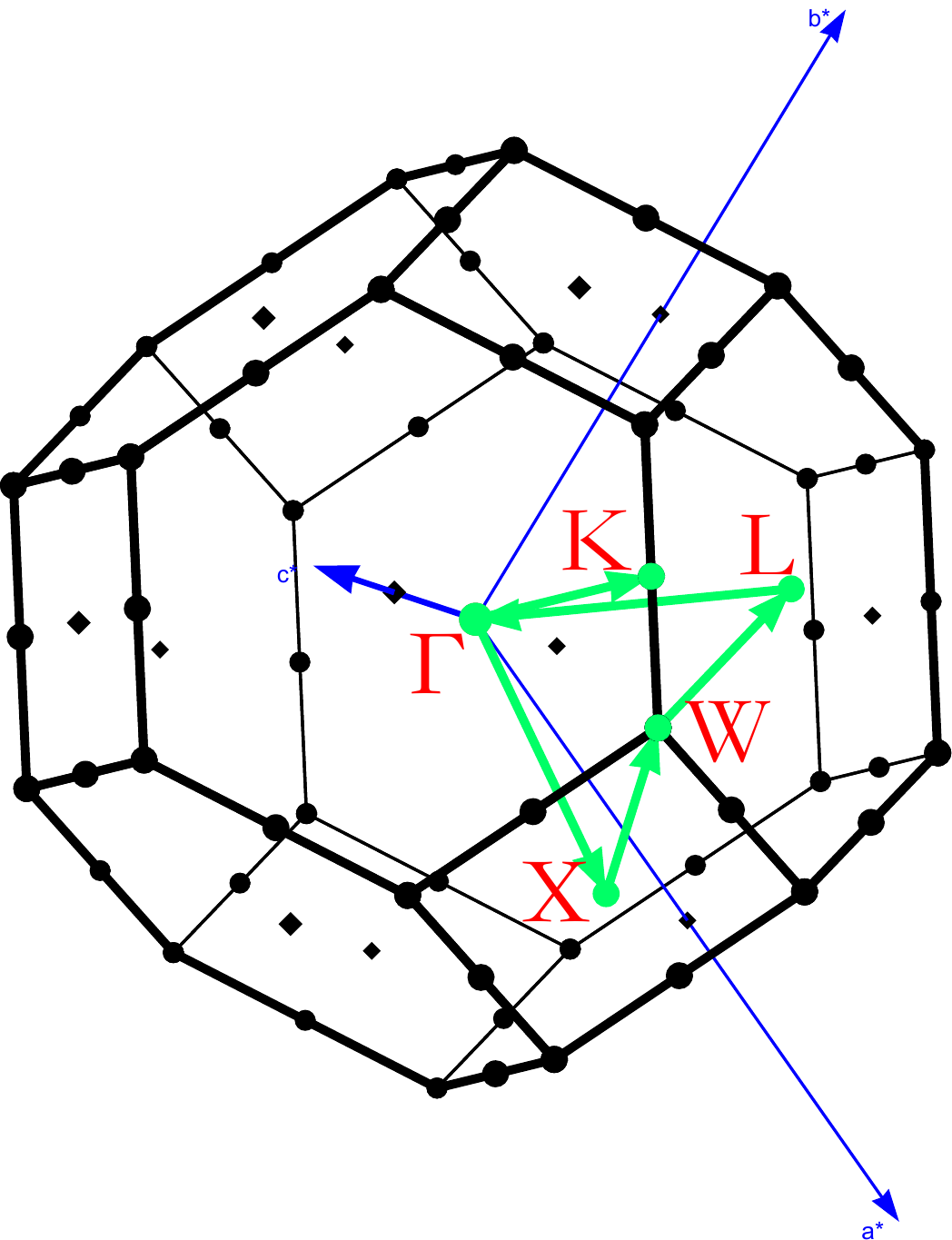}
\caption{The first Brillouin zone of bulk Pb with face centered cubic structure and the k-path used in the main text
to illustrate the bandstructure.  The green lines are the symmetry paths used to
connect high symmetry points labelled in red.
} \label{BZ_Pb_bulk}
\end{figure} \fi

\subsection{Thin film $\rm Pb$ in $111 $ direction: crystal structure}

In the calculations for Pb thin films, we chose the bulk $111 $ direction as the film normal. The lattice constant is $a = 3.42 \AA $ and the supercell constructed is similar with $\beta$-Sn film. The lattice vectors of the primitive unit cell used in the DFT calculation is $ \frac{\sqrt{3}}{2} a \bm{x} + \frac{1}{2}a \bm{y}$ and  $ \frac{\sqrt{3}}{2} a \bm{x} + \frac{1}{2}a \bm{y}$,
in which there are one Pb atom per cell located at $(0,0,0)$ in crystal coordinates for the smallest unit cell of single layer. The structure of single layer is shown in Fig. \ref{Pb_film}.

\ifpdf  \begin{figure}[h]
\centering
\includegraphics[width=0.9\linewidth]{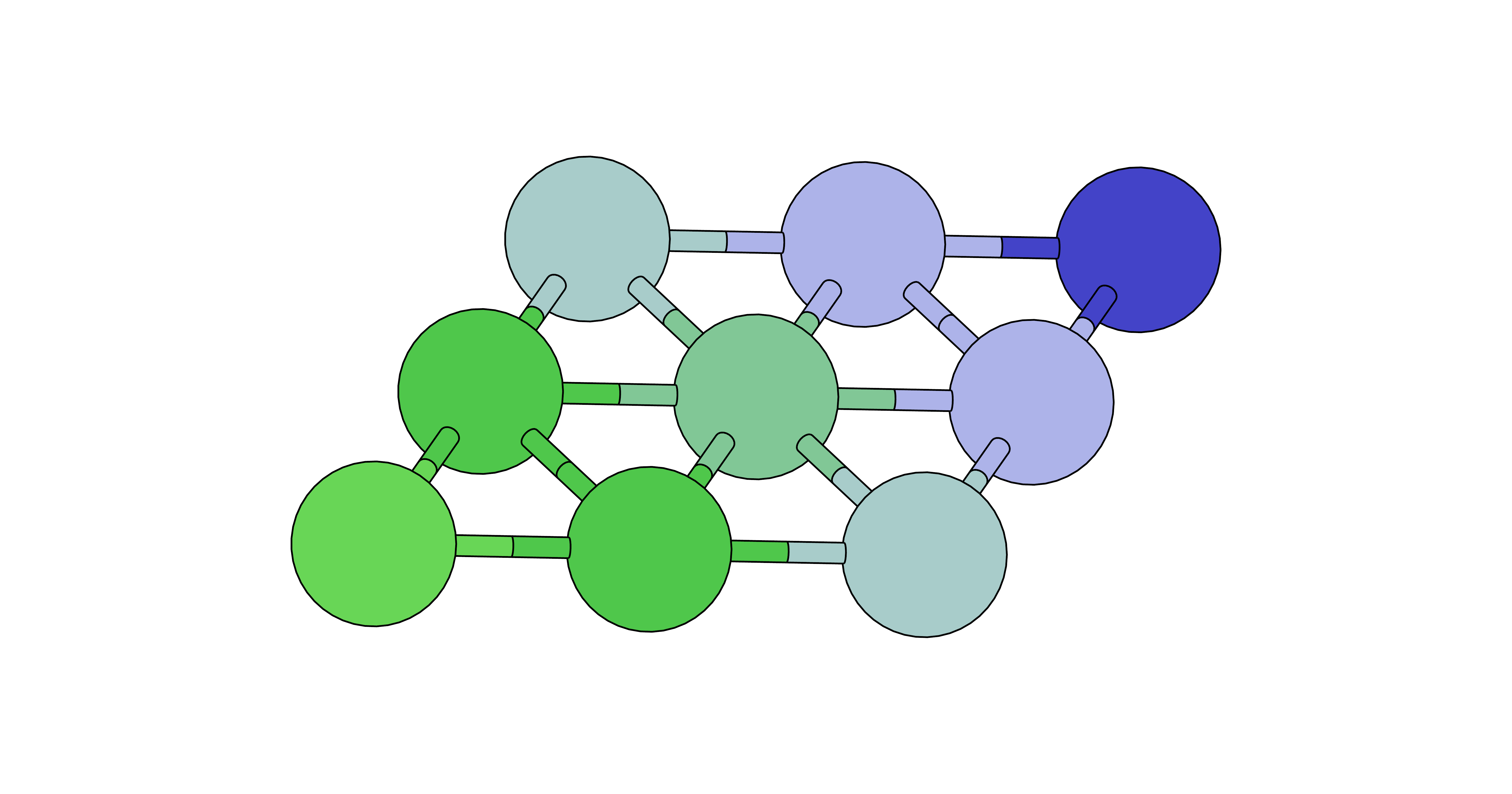}
\caption{Crystal structure of thin film Pb, whose in-plane lattice constant is the same as that of bulk Pb, $a = 3.42 \AA $.} \label{Pb_film}
\end{figure} \fi

The k-path used to illustrate the bandstructures of Pb thin films
in the main text is shown in table \ref{bz_pb111_table} and Fig. \ref{BZ_Pb111_film}.

\begin{table}[h]
\caption{\label{bz_pb111_table}
Symmetry points in the first Brillouin zone of thin film Pb in $(111)$ direction,
with $\bm{k} = u \bm{a}^{\ast} + v \bm{b}^{\ast} + w \bm{c}^{\ast}$,
where $a^{\ast}$, $b^{\ast}$, $c^{\ast}$ are reciprocal lattice vectors shown as blue vectors in Fig. \ref{BZ_Pb111_film}.
}
\begin{ruledtabular}
\begin{tabular}{c c}
  Symmetry points & k points:$(u,v,w)$ \\ [0.5ex]
 \hline
 $\Gamma$ & (0,0,0)  \\
 X & (0.5,0.5,0)  \\
 K & $(\frac{1}{3},\frac{2}{3},0)$ \\
 \end{tabular}
 \end{ruledtabular}
\end{table}

\ifpdf  \begin{figure}[h]
\centering
\includegraphics[width=0.9\linewidth]{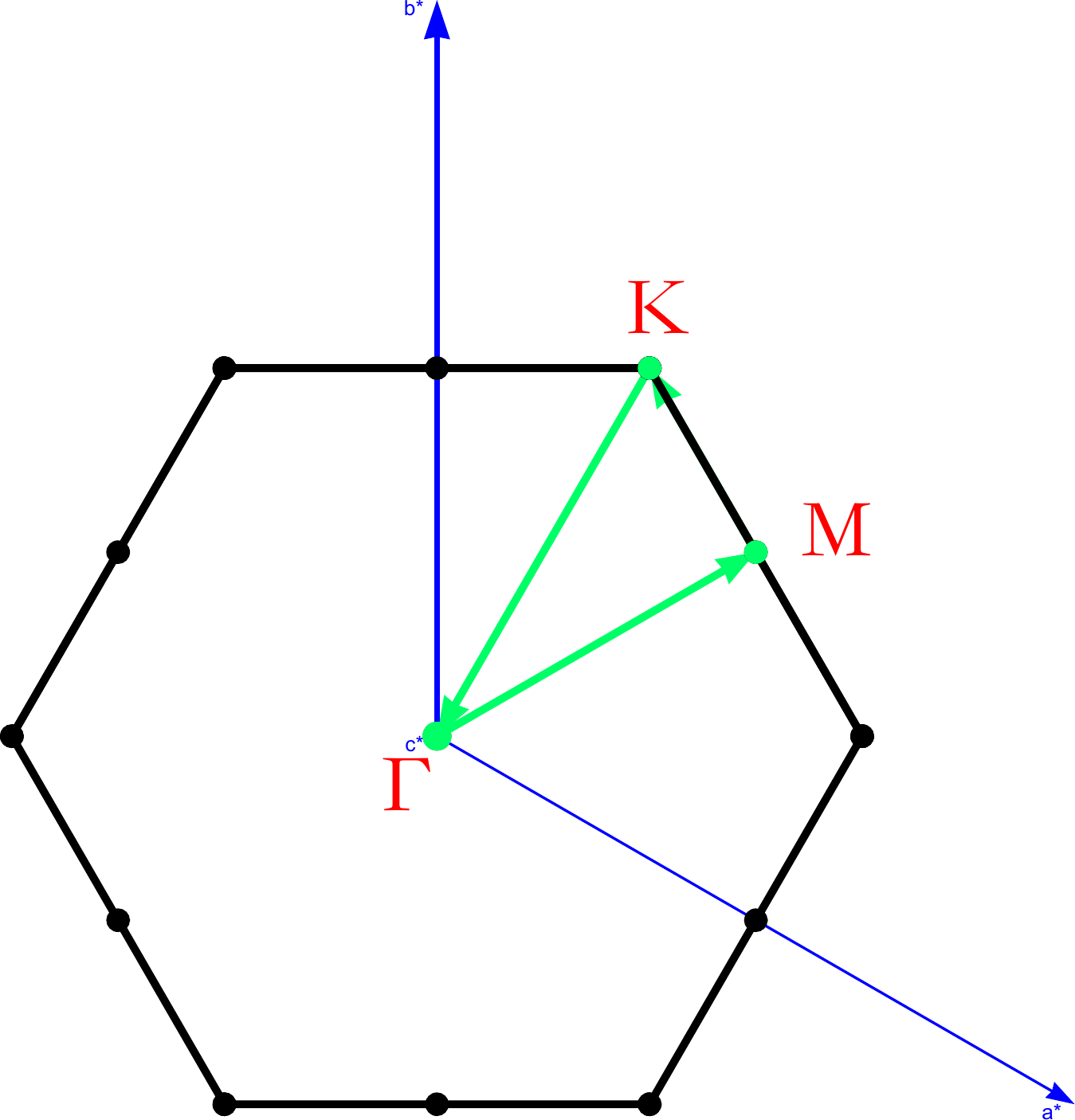}
\caption{The first Brillouin zone of $(111)$ growth direction Pb thin films
and the k-path used to calculate the bandstructure in the main text.
The k-path is labeled with green lines
and the symmetry points are indicated by red letters.
} \label{BZ_Pb111_film}
\end{figure} \fi

\subsection{Thin film $\rm Pb$ in $111 $ direction: electronic structure}
Here we selected several cases of Pb thin film to show the bandstructures, that is 3,5,7 and 9 layers thin film $\rm Pb$ in $111 $ direction,
shown in Fig.\ref{band_pb_3}-\ref{band_pb_9}:
\ifpdf
\begin{figure*}
\centering
\begin{subfigure}{0.45\linewidth}
\includegraphics[width=\linewidth,height=0.75\linewidth]{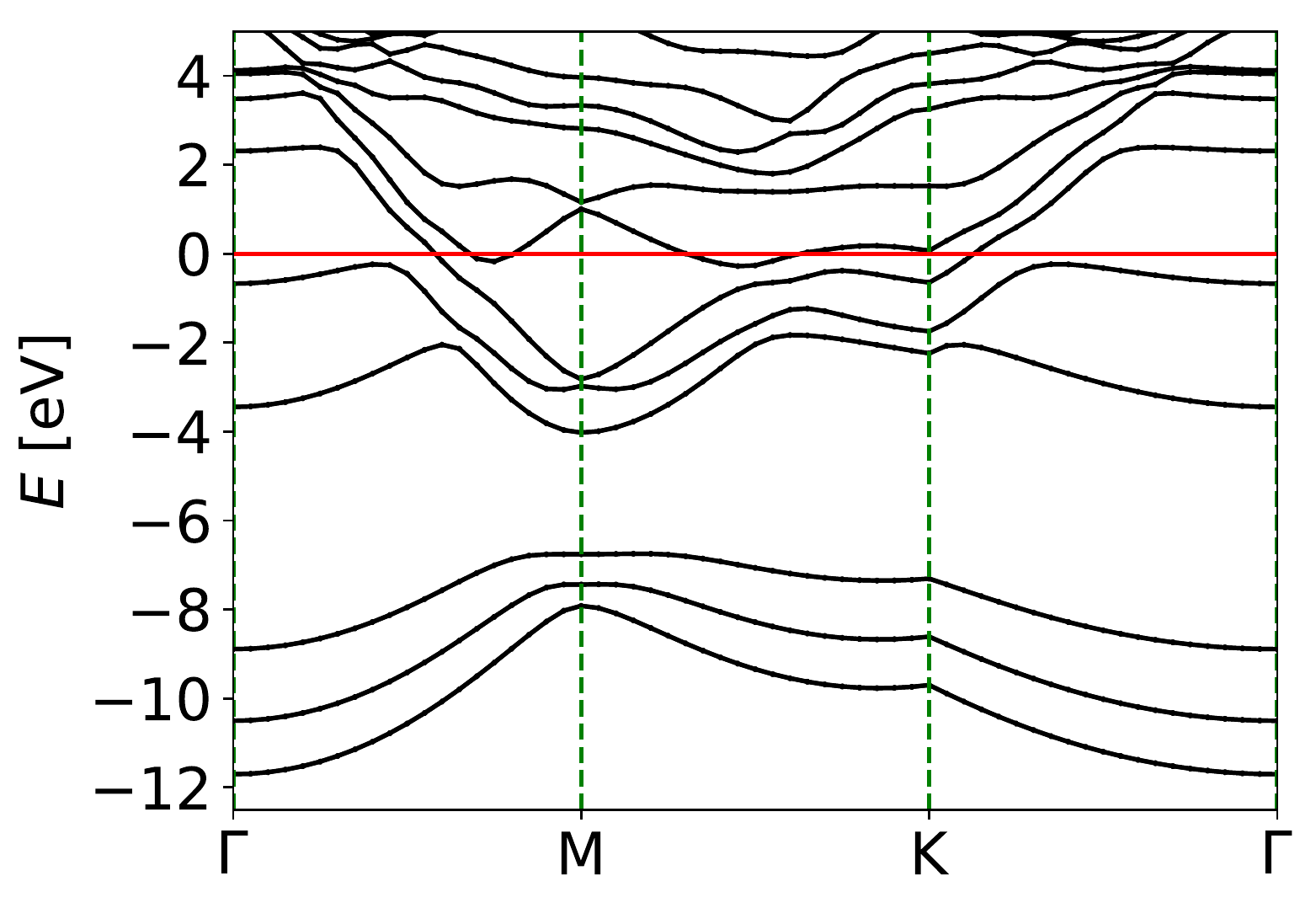}
\subcaption{3 Layers of Pb (111)} \label{band_pb_3}
\end{subfigure}
\begin{subfigure}{0.45\linewidth}
\includegraphics[width=\linewidth,height=0.75\linewidth]{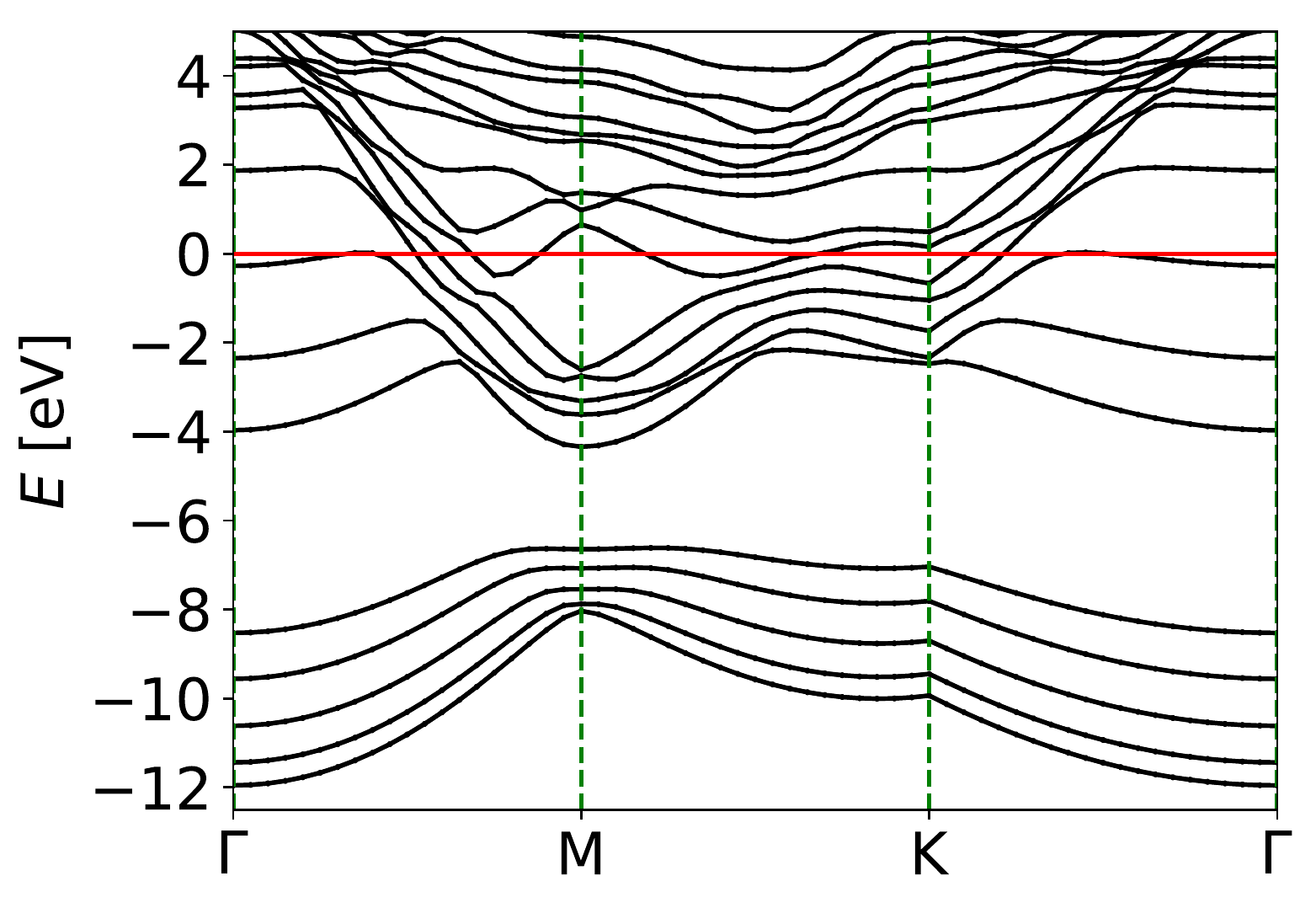}
\subcaption{5 Layers of Pb (111)} \label{band_pb_5}
\end{subfigure}
\begin{subfigure}{0.45\linewidth}
\includegraphics[width=\linewidth,height=0.75\linewidth]{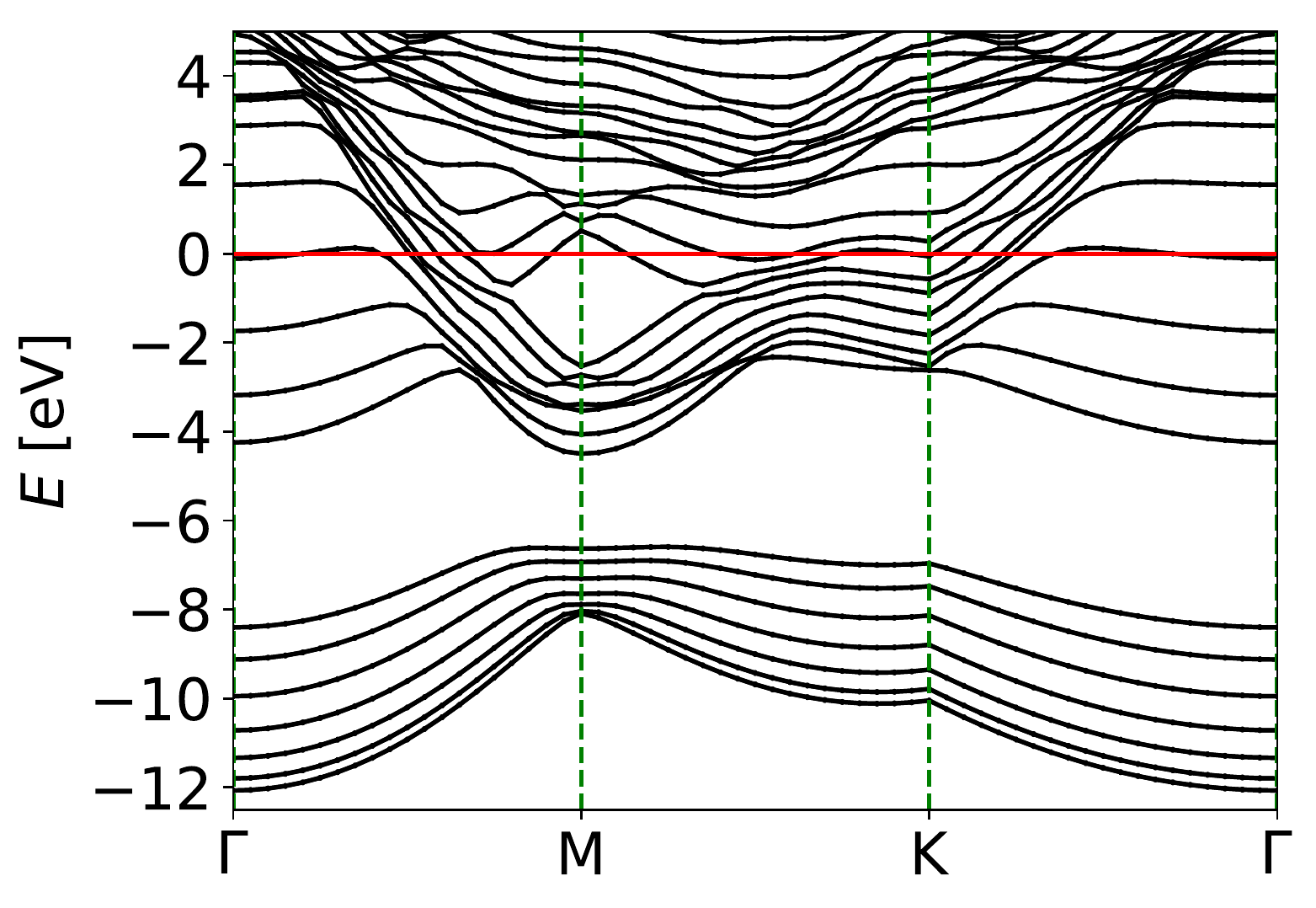}
\subcaption{7 Layers of Pb (111)} \label{band_pb_7}
\end{subfigure}
\begin{subfigure}{0.45\linewidth}
\includegraphics[width=\linewidth,height=0.75\linewidth]{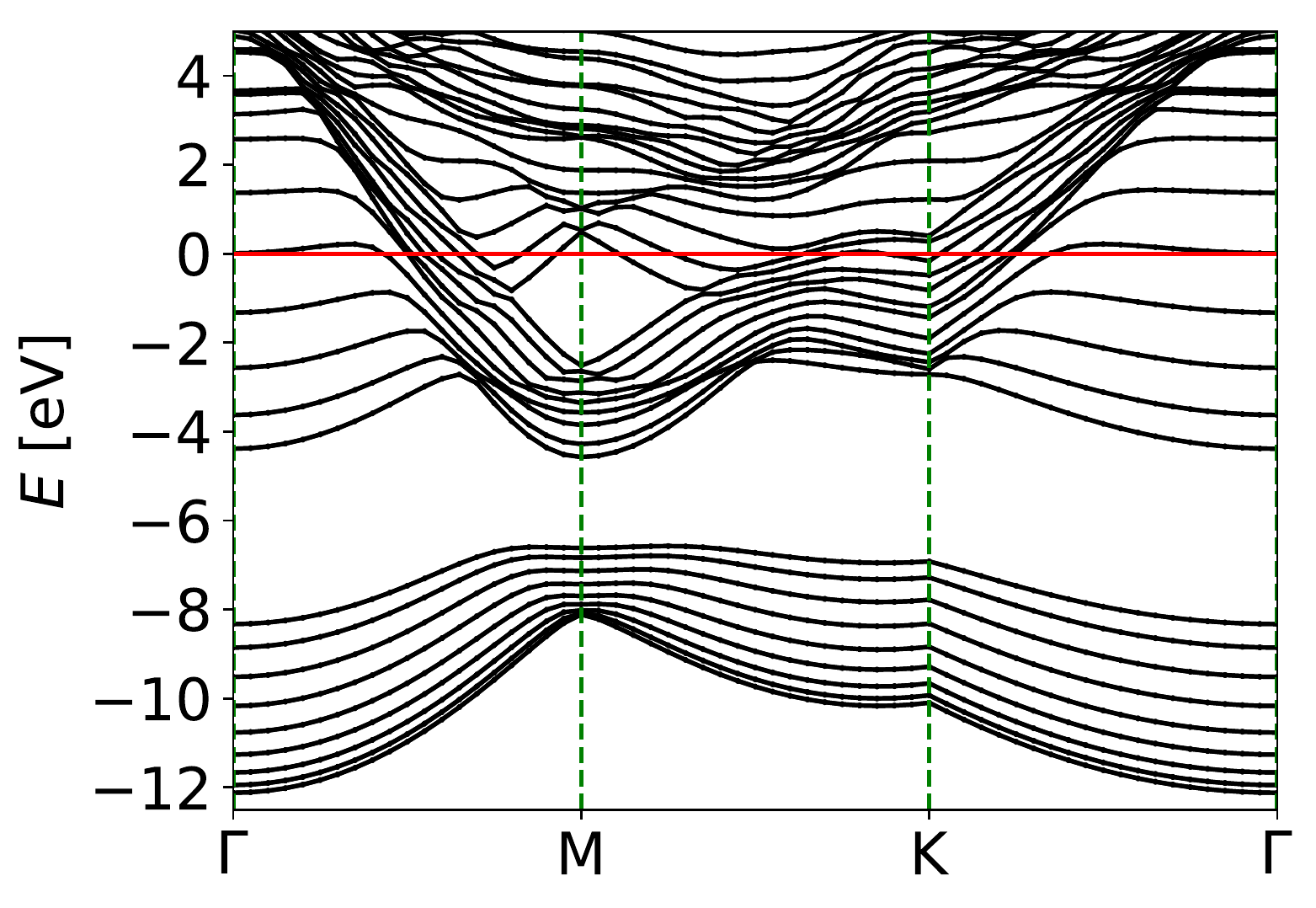}
\subcaption{9 Layers of Pb (111)} \label{band_pb_9}
\end{subfigure}
\caption{
Selected bandstructure of a thin film Pb grown
along the $(111)$ direction: 3,5,7,9 layers}
\label{band_film_pb}
\end{figure*} \fi
A main difference from $\beta$-Sn thin film is that the hybridization of s orbital with p orbital is much smaller,
and also an obvious quantum size effect has been shown, which has also been shown in the main text at $\Gamma$ points.
Another important difference is that in Pb thin film sp orbital has little hybridization with other orbitals such as d orbital,
which give a smaller g-factor as show in the follow.

\section{ $\rm Pb (111)$ thin film on  $ \rm As_2O_3 $ substrates}

To break the inversion symmetry of Pb thin film, we put it on a substrate of As$_2$O$_3$.
This choice is just a convenient theoretical model and other choices might be
preferable experimentally.  The As$_2$O$_3$ substrate has the $\rm Bi_2Te_3$ structure.
It is critical that the substrate be an insulator.
Other possible substrates and their lattice constants are listed
in table \ref{lattice_constants}.

\begin{table}[h]
\caption{\label{lattice_constants}
Lattice constants of Pb thin film in $(111)$ direction and possible substrates.
}
\begin{ruledtabular}
\begin{tabular}{c c  c}
 Materials & Lattice constant (\AA) & Reference \\ [0.5ex]
 \hline
 Pb $(111)$ & 3.42 &  \\
 Silicon $(111)$ & 3.84 &  \\
 Bi$_2$Te$_3$ & 4.3835 & \cite{BiTe_lattice} \\
 Bi$_2$Se$_3$ & 4.138 & \cite{BiTe_lattice}  \\
 Sb$_2$Te$_3$ & 4.25 & \cite{BiTe_lattice} \\
 Sb$_2$Se$_3$ & 4.076 & \cite{SbSe_lc} \\
  MnAs(NiAs structure) & 3.68 & \cite{MnAs} \\
  SrTiO$_3$ & 3.905 & \cite{sto} \\
 \end{tabular}
 \end{ruledtabular}
\end{table}

The crystal structure of a bilayer Pb $(111)$ thin film as an example on As$_2$O$_3$ substrate is shown in Fig. \ref{PbAsO_film}, the lattice constant used come from Pb $(111)$ thin film.

\ifpdf  \begin{figure}[h]
\centering
\includegraphics[width=0.9\linewidth]{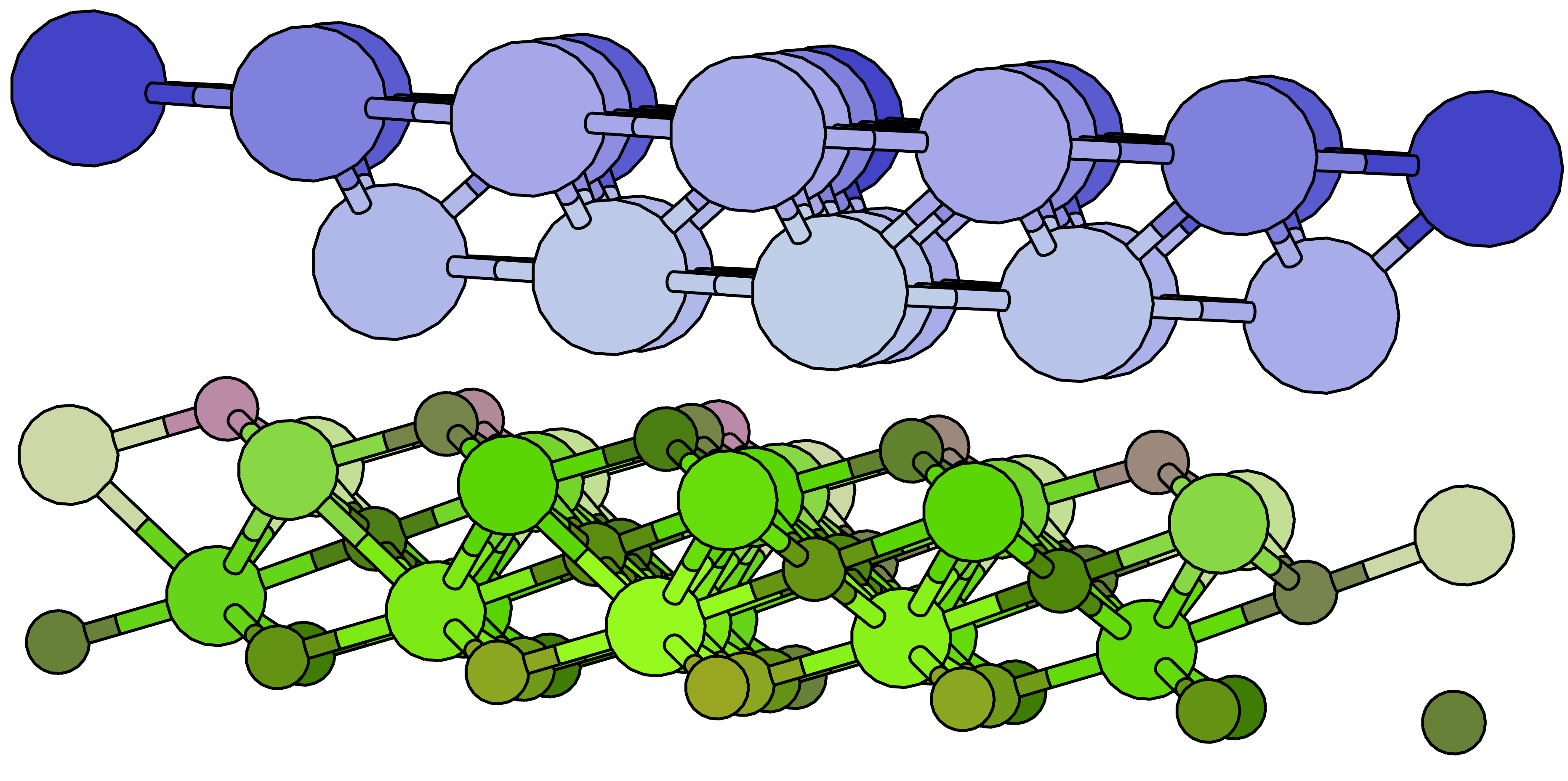}
\caption{Crystal structure of thin film Pb in $(111)$ direction on As$_2$O$_3$ substrate with $\rm Bi_2Te_3$ structure,
where blue slab on the top is the Pb thin film and the green slab on the bottom is the As$_2$O$_3$ substrate.
The lattice constant used come from Pb.} \label{PbAsO_film}
\end{figure} \fi

The k-path and symmetry points of the bandstructure of thin film Pb in $(111)$ direction on a As$_2$O$_3$ substrate in the main text
is shown in table \ref{bz_pbaso_table} and Fig. \ref{BZ_PbAsO}.

\begin{table}[h]
\caption{\label{bz_pbaso_table}
Symmetry points in the first Brillouin zone of thin film Pb in $(111)$ direction on a As$_2$O$_3$ substrate,
with $\bm{k} = u \bm{a}^{\ast} + v \bm{b}^{\ast} + w \bm{c}^{\ast}$,
where $a^{\ast}$, $b^{\ast}$, $c^{\ast}$ are reciprocal lattice vectors shown as blue vectors in Fig. \ref{BZ_PbAsO}.
}
\begin{ruledtabular}
\begin{tabular}{c c}
  Symmetry points & k points:$(u,v,w)$ \\ [0.5ex]
 \hline
 $\Gamma$ & (0,0,0)  \\
 X & (0.5,0,0)  \\
 K & $(\frac{1}{3},\frac{1}{3},0)$ \\
 \end{tabular}
 \end{ruledtabular}
\end{table}

\ifpdf  \begin{figure}[h]
\centering
\includegraphics[width=0.9\linewidth]{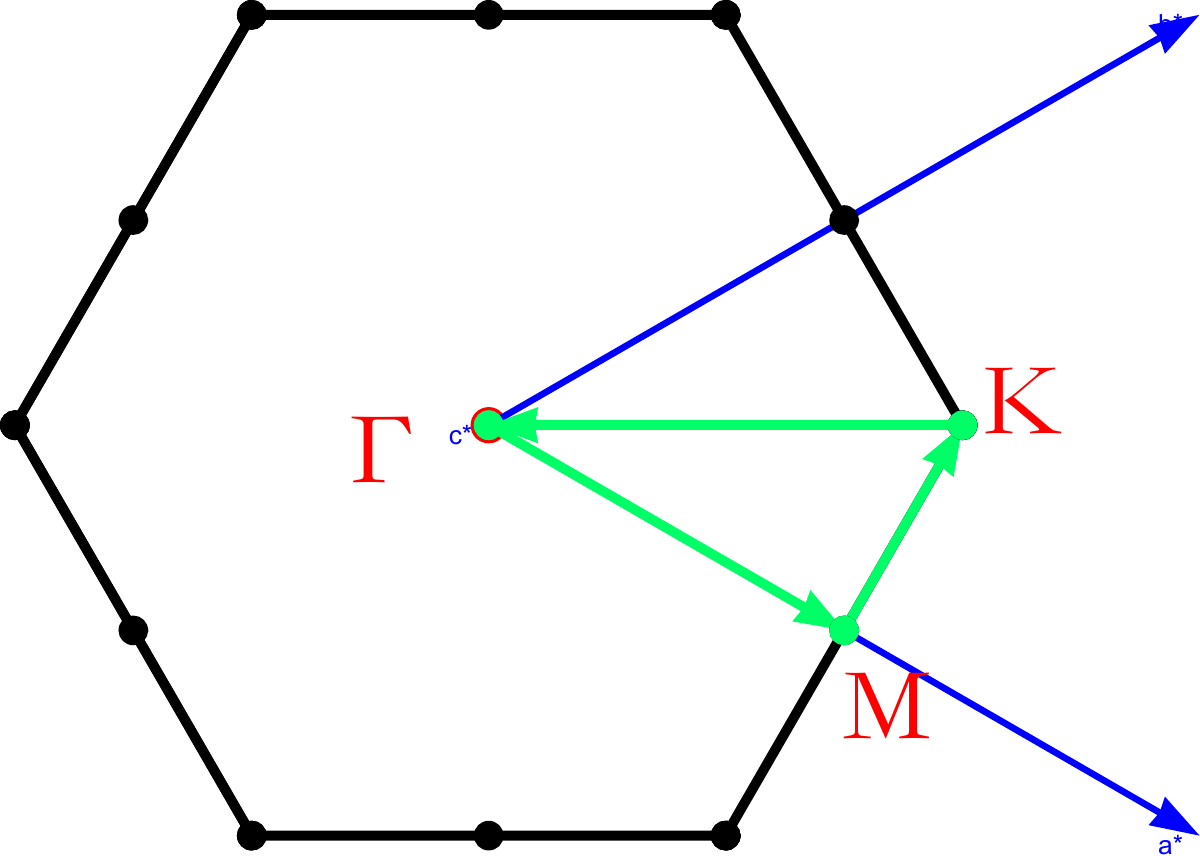}
\caption{The first Brillouin zone of thin film Pb in $(111)$ direction on a As$_2$O$_3$ substrate.
The k-path is labeled with green lines
and the symmetry points are indicated by red letters.
} \label{BZ_PbAsO}
\end{figure} \fi

\subsection{ $\rm Pb (111)$ thin film on $\rm As_2 O_3$ substrate: Electronic structure}
Electronic structure of 1-10 layers of $\rm Pb (111)$ thin film on $\rm As_2 O_3$ substrate are shown in Fig. \ref{band_film_pb_s}:
\ifpdf \begin{figure*}
\centering
\begin{subfigure}{0.30\linewidth}
\includegraphics[width=\linewidth,height=0.65\linewidth]{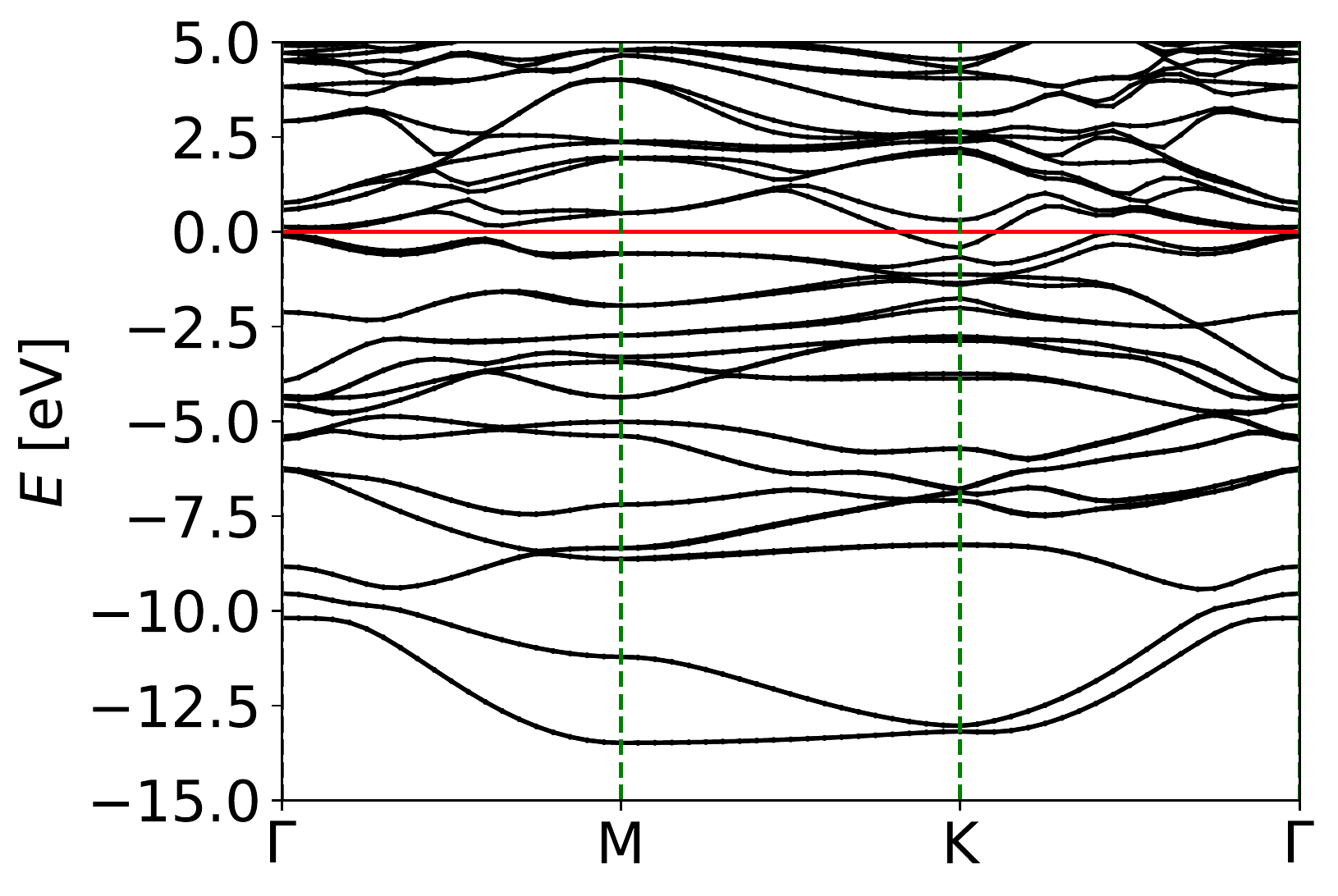}
\subcaption{1 Layer of Pb (111) on 1 quintuple layer of $As_2O_3$ substrate} \label{band_Pb111_s1}
\end{subfigure}
\begin{subfigure}{0.30\linewidth}
\includegraphics[width=\linewidth,height=0.65\linewidth]{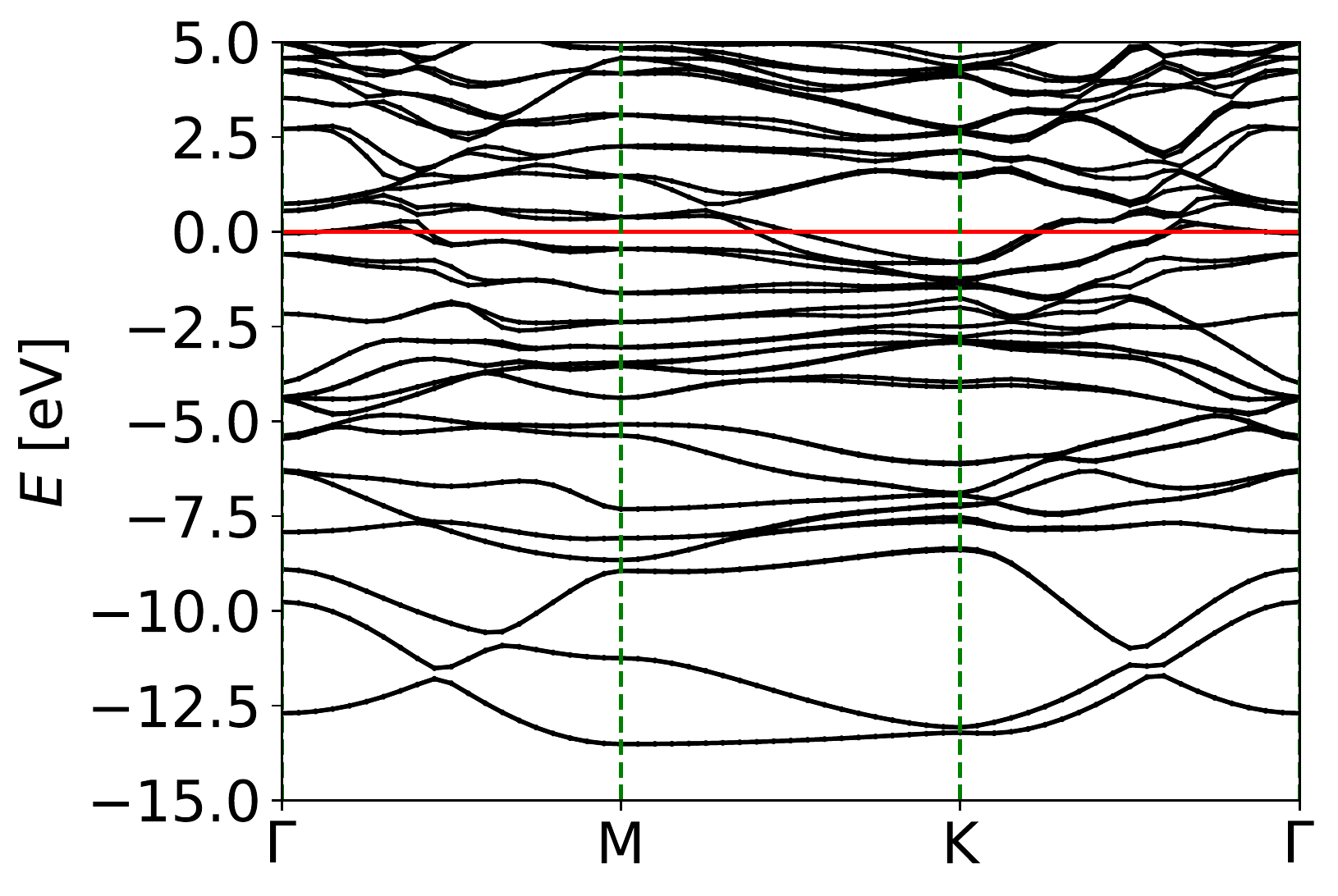}
\subcaption{2 Layer of Pb (111) on 1 quintuple layer of $As_2O_3$ substrate} \label{band_Pb111_s2}
\end{subfigure}
\begin{subfigure}{0.30\linewidth}
\includegraphics[width=\linewidth,height=0.65\linewidth]{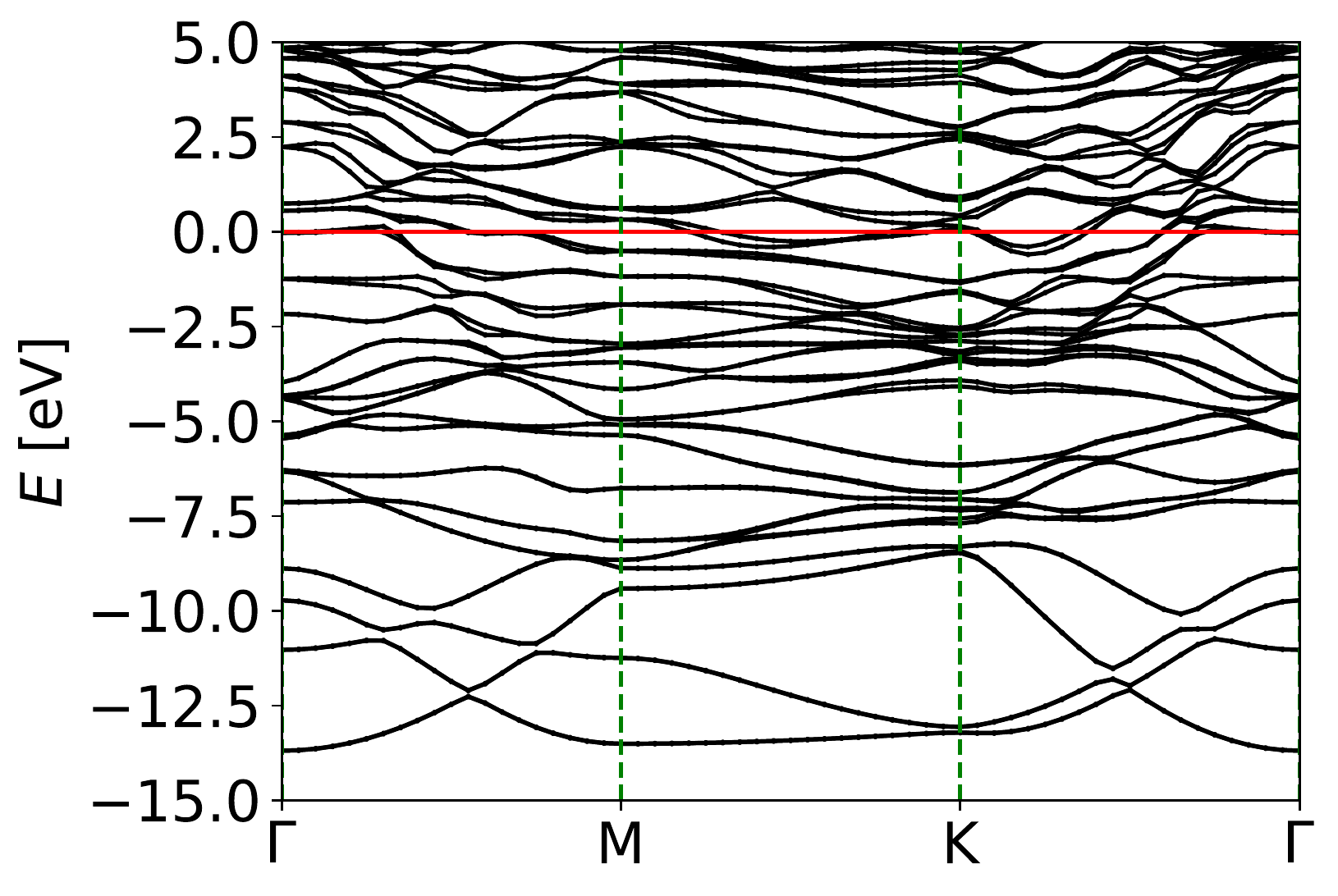}
\subcaption{3 Layer of Pb (111) on 1 quintuple layer of $As_2O_3$ substrate} \label{band_Pb111_s3}
\end{subfigure}
\begin{subfigure}{0.30\linewidth}
\includegraphics[width=\linewidth,height=0.65\linewidth]{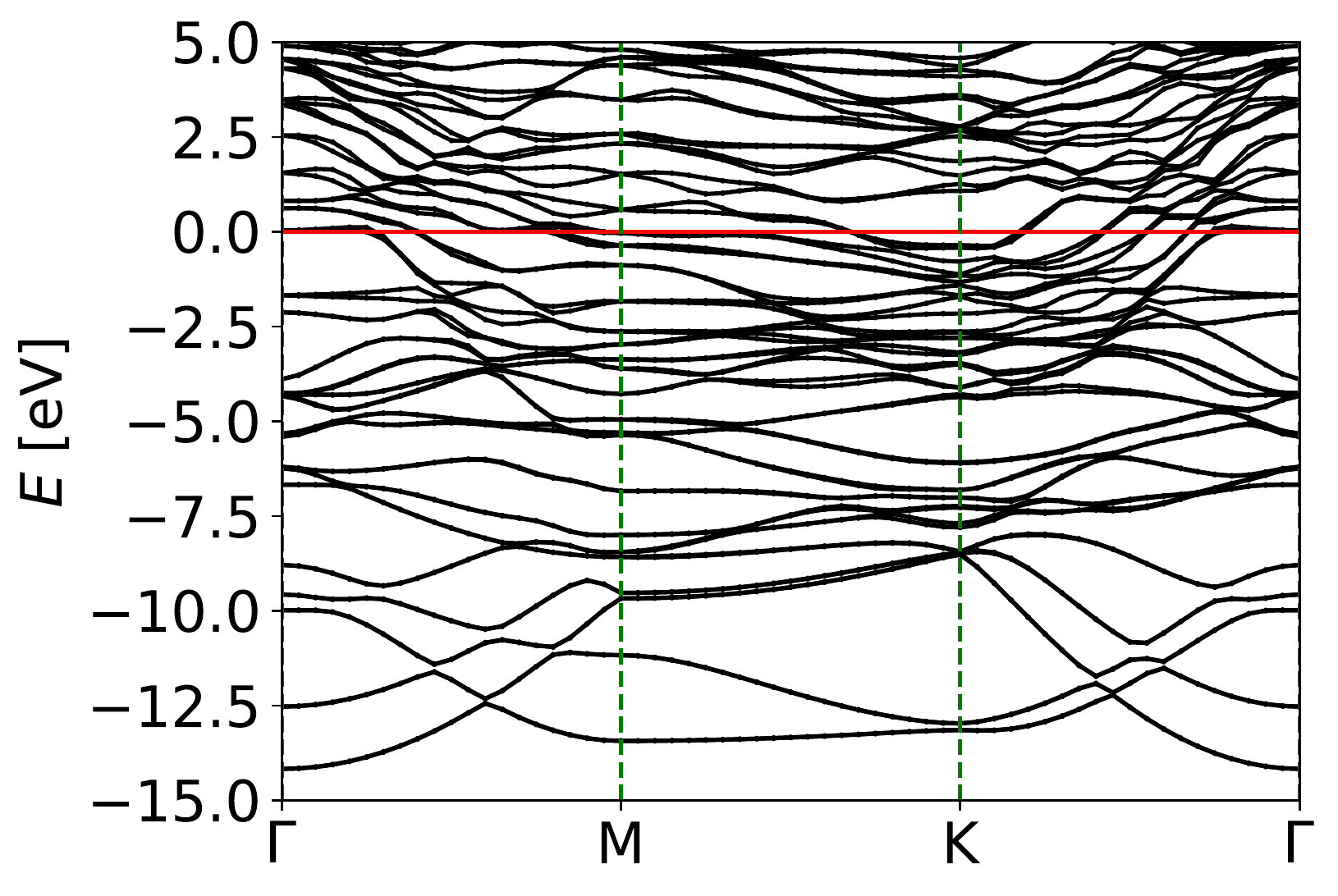}
\subcaption{4 Layer of Pb (111) on 1 quintuple layer of $As_2O_3$ substrate} \label{band_Pb111_s4}
\end{subfigure}
\begin{subfigure}{0.30\linewidth}
\includegraphics[width=\linewidth,height=0.65\linewidth]{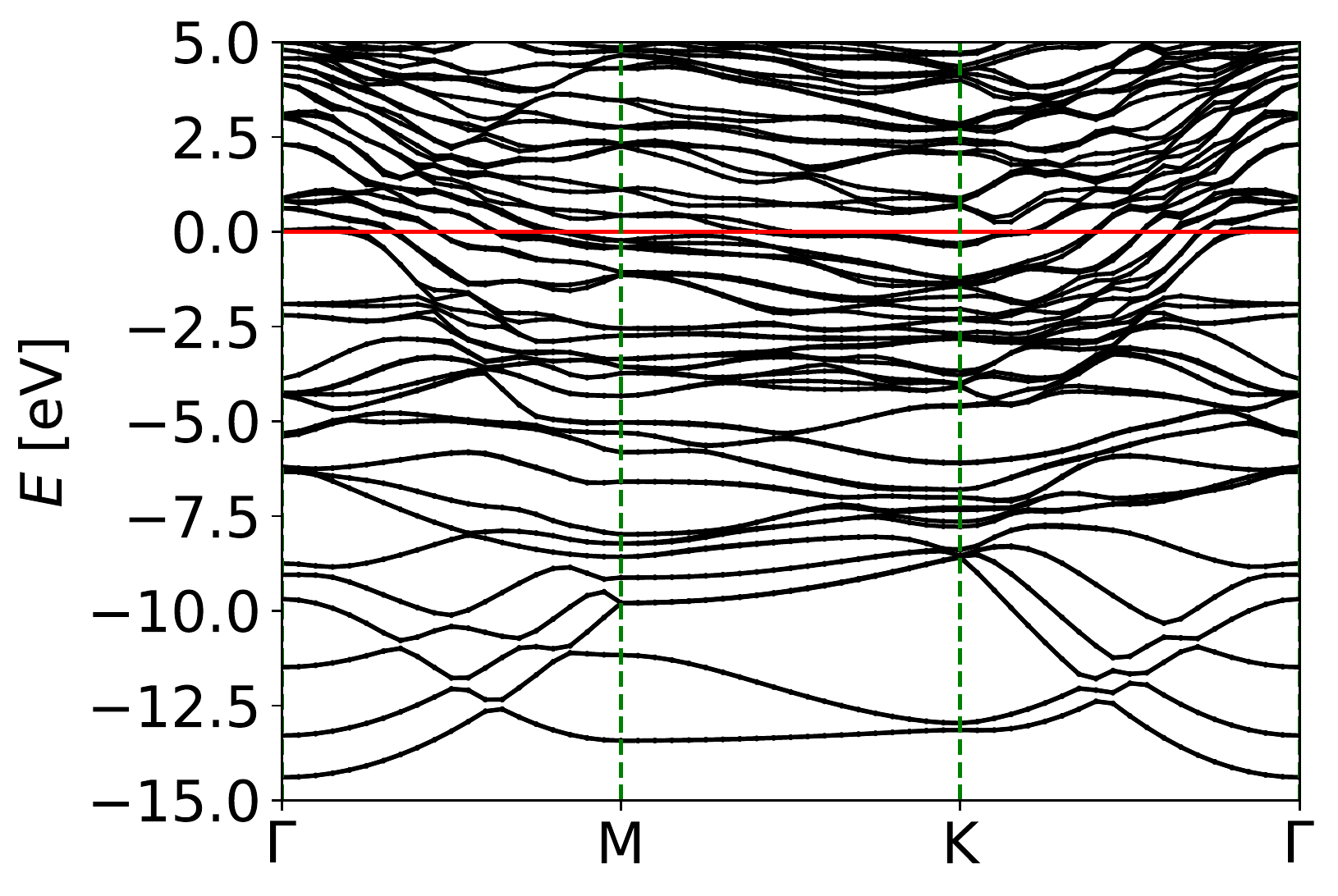}
\subcaption{5 Layer of Pb (111) on 1 quintuple layer of $As_2O_3$ substrate} \label{band_Pb111_s5}
\end{subfigure}
\begin{subfigure}{0.30\linewidth}
\includegraphics[width=\linewidth,height=0.65\linewidth]{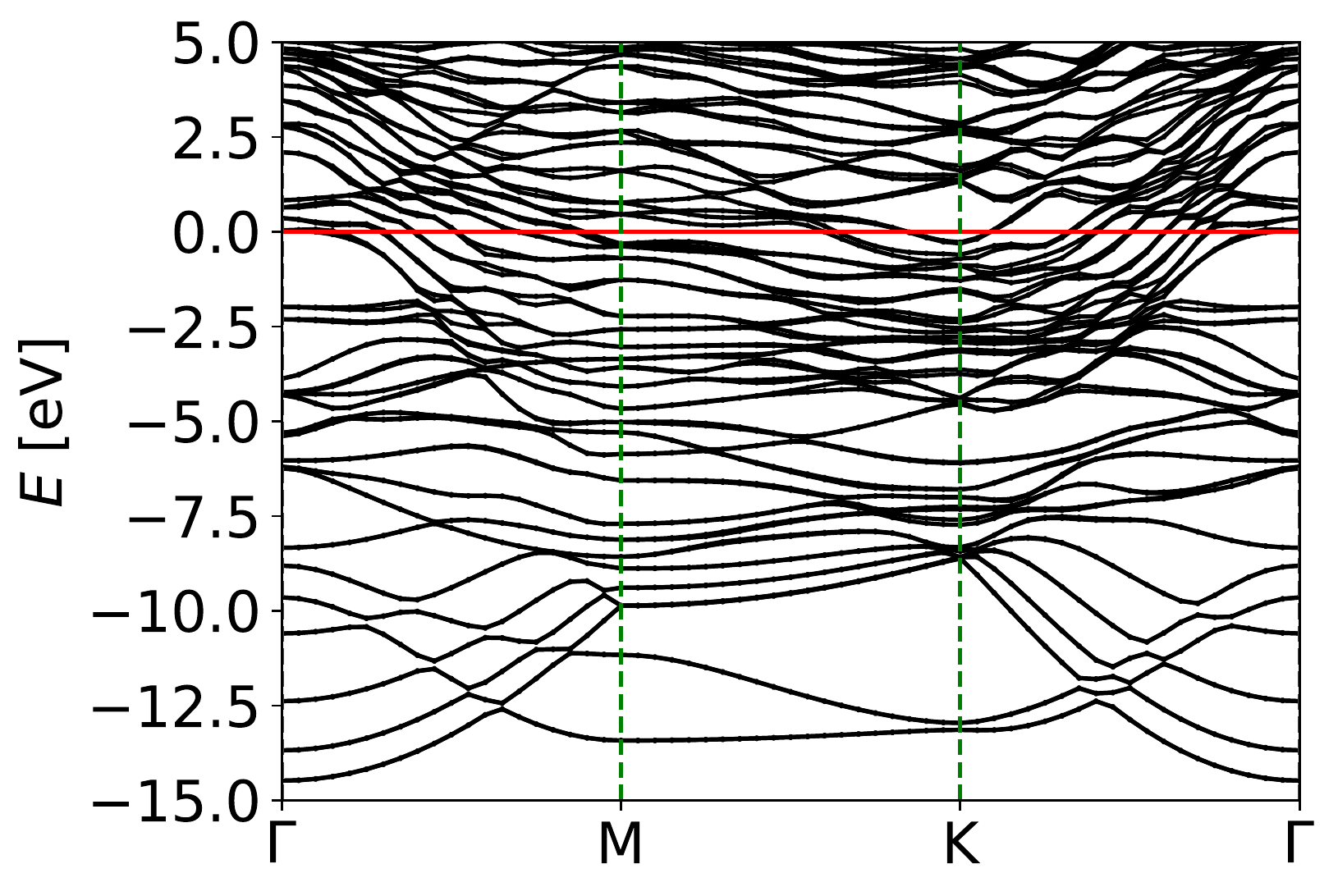}
\subcaption{6 Layer of Pb (111) on 1 quintuple layer of $As_2O_3$ substrate} \label{band_Pb111_s6}
\end{subfigure}
\begin{subfigure}{0.30\linewidth}
\includegraphics[width=\linewidth,height=0.65\linewidth]{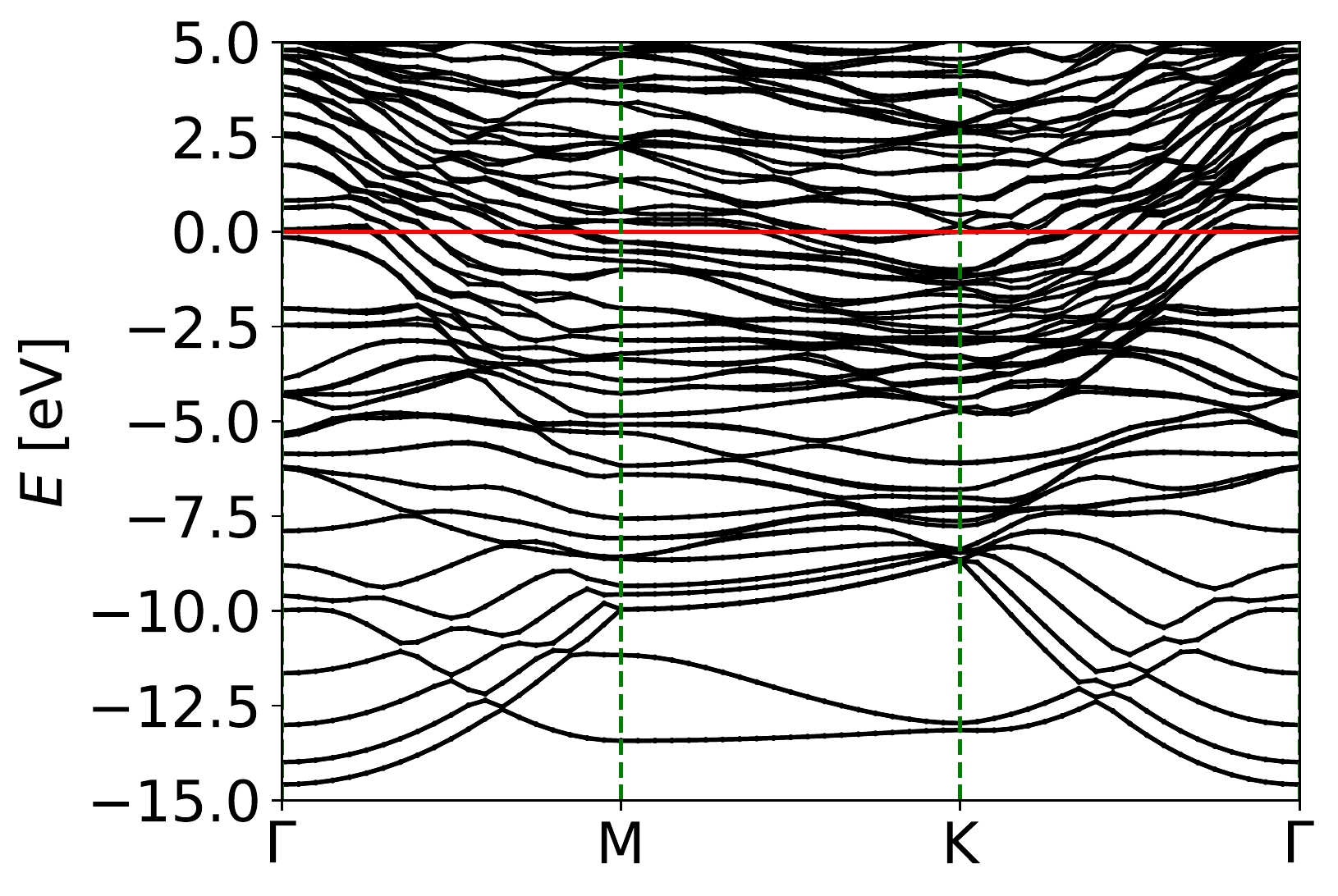}
\subcaption{7 Layer of Pb (111) on 1 quintuple layer of $As_2O_3$ substrate} \label{band_Pb111_s7}
\end{subfigure}
\begin{subfigure}{0.30\linewidth}
\includegraphics[width=\linewidth,height=0.65\linewidth]{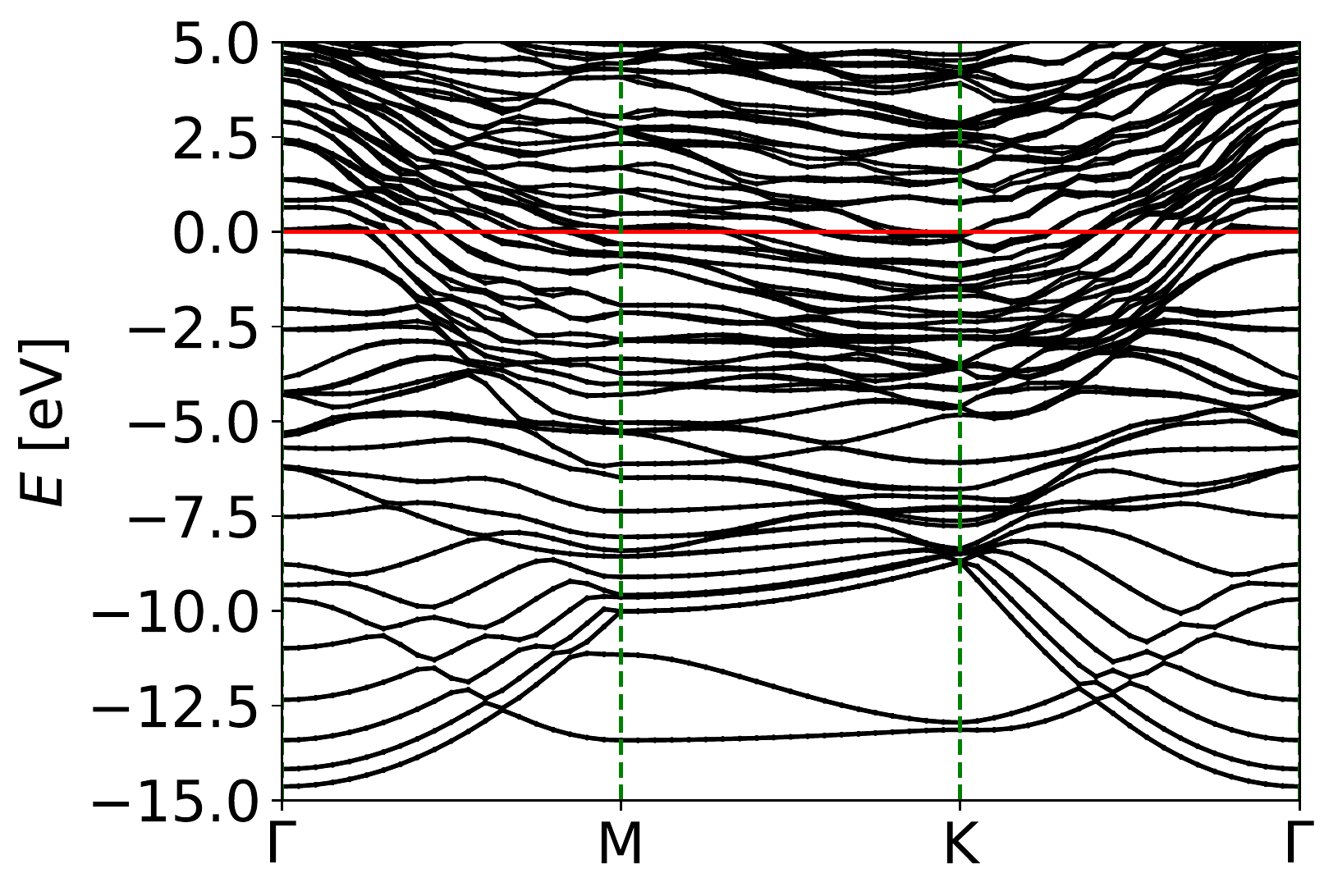}
\subcaption{8 Layer of Pb (111) on 1 quintuple layer of $As_2O_3$ substrate} \label{band_Pb111_s8}
\end{subfigure}
\begin{subfigure}{0.30\linewidth}
\includegraphics[width=\linewidth,height=0.65\linewidth]{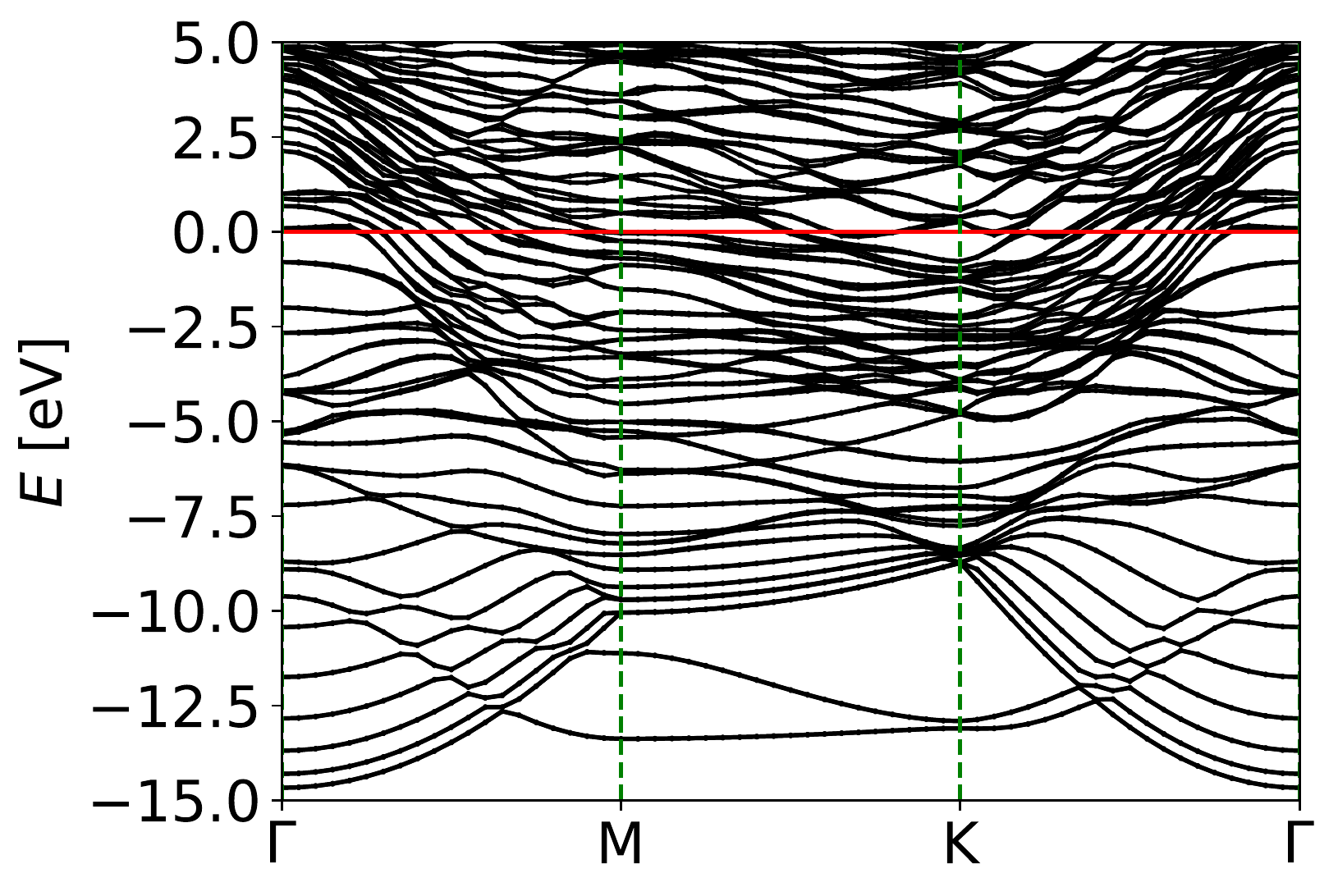}
\subcaption{9 Layer of Pb (111) on 1 quintuple layer of $As_2O_3$ substrate} \label{band_Pb111_s9}
\end{subfigure}
\begin{subfigure}{0.30\linewidth}
\includegraphics[width=\linewidth,height=0.65\linewidth]{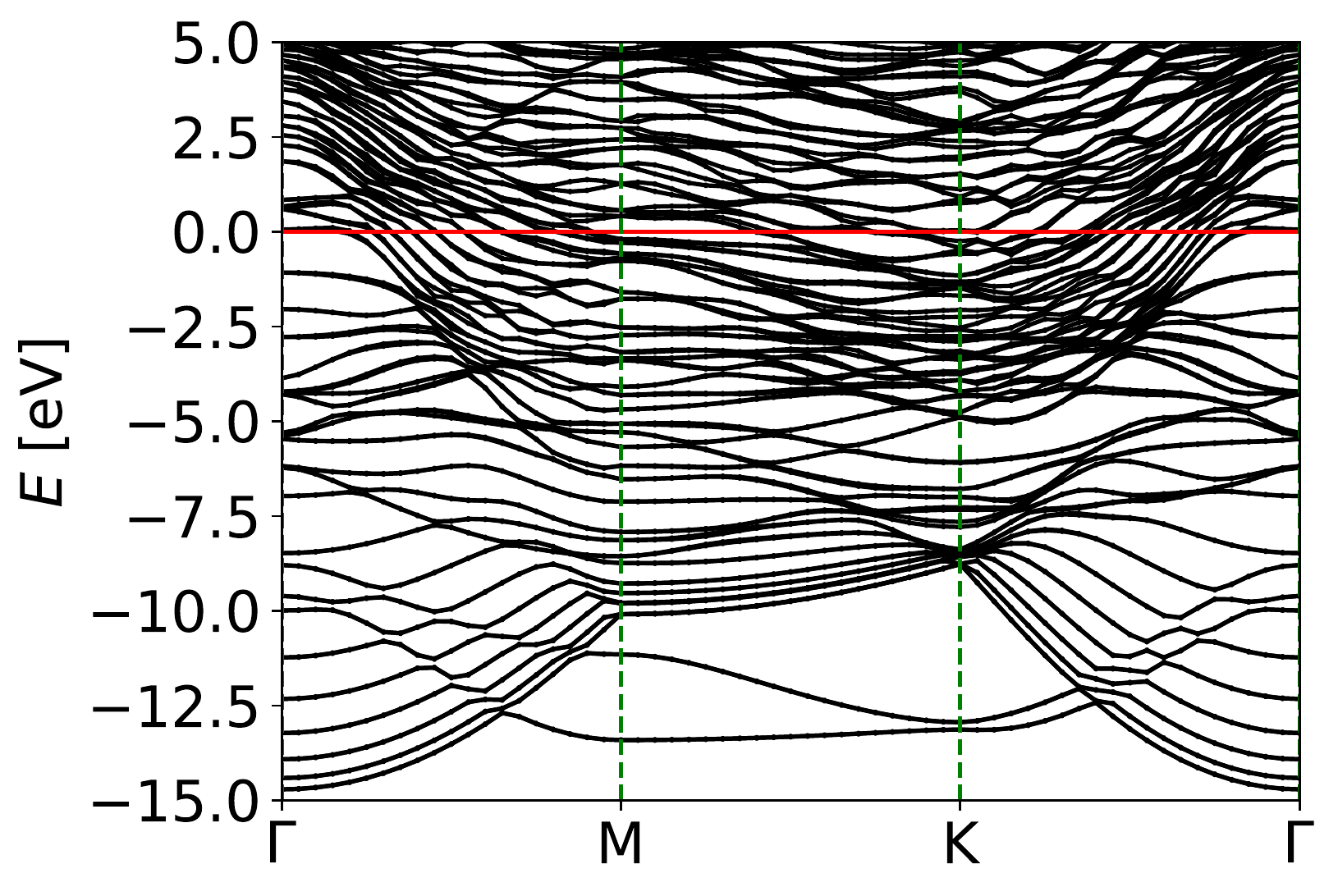}
\subcaption{10 Layer of Pb (111) on 1 quintuple layer of $As_2O_3$ substrate} \label{band_Pb111_s10}
\end{subfigure}
\caption{
Band structure of Pb thin film in (111) direction grown on 1 quintuple layer of $As_2O_3$ substrate:
1-10 layers of Pb thin film}
\label{band_film_pb_s}
\end{figure*} \fi
It can be seen that the bands of substrates are hybridized with subbands of Pb thin film,
according to the main text, for the thin film with a strain around $2\%$ the subband close to Fermi level
has a small chemical potential, here we thus show the thin film with a strain around $2\%$, that the lattice constant
is 4.9508 $\AA$.

To study the strain effect of substrates on the chemical potential, we calculated the band extremum from Fermi level at $\Gamma$ point on  $\rm As_2O_3$ substrate,
shown as in Fig. \ref{strain_substrate} for 5 and 10 layers of Pb thin film:
\ifpdf  \begin{figure}
 \centering
 \includegraphics[width=0.85\linewidth,height=0.65\linewidth]{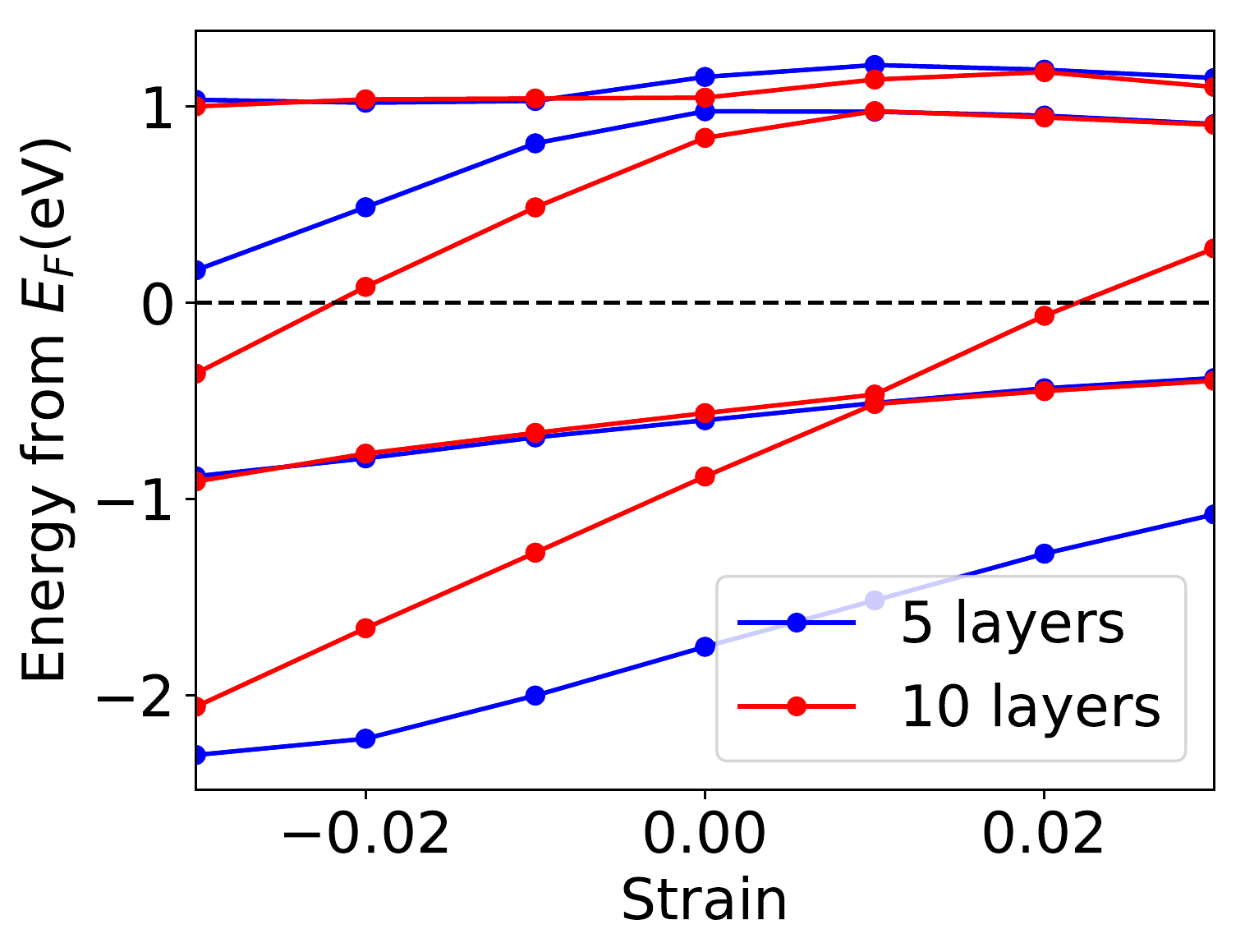}
\caption{Strain dependence of energy separations between quasi-2D band extrema at $\Gamma$ point and
Fermi energies for Pb (111) thin film on $As_2O_3$ substrate}\label{strain_substrate}
\end{figure} \fi

\subsection{Rashba constants}
The Rashba coupling constants vs the number of layers are shown as in Fig. \ref{rashba_film_pb},
as it shows that the Rashba coupling becomes smaller when increasing the number of layers,
and the average values of Rashba coupling around Fermi level is about $0.15-0.4 eV \AA$.

\ifpdf \begin{figure*}
\centering
\begin{subfigure}{0.45\linewidth}
\includegraphics[width=\linewidth,height=0.75\linewidth]{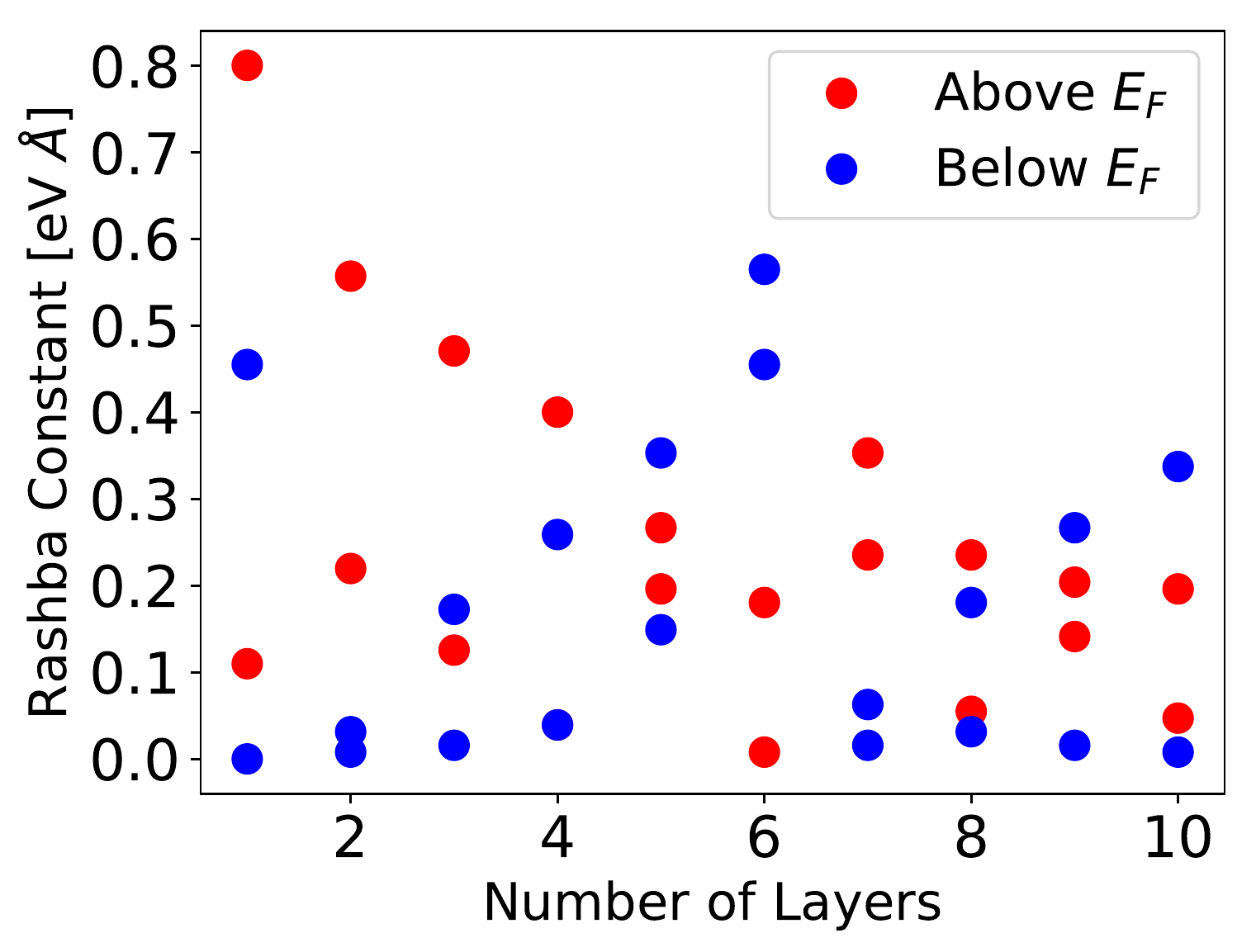}
\subcaption{Rashba coupling of Pb (111) vs layers} \label{r_pb_lay}
\end{subfigure}
\begin{subfigure}{0.45\linewidth}
\includegraphics[width=\linewidth,height=0.75\linewidth]{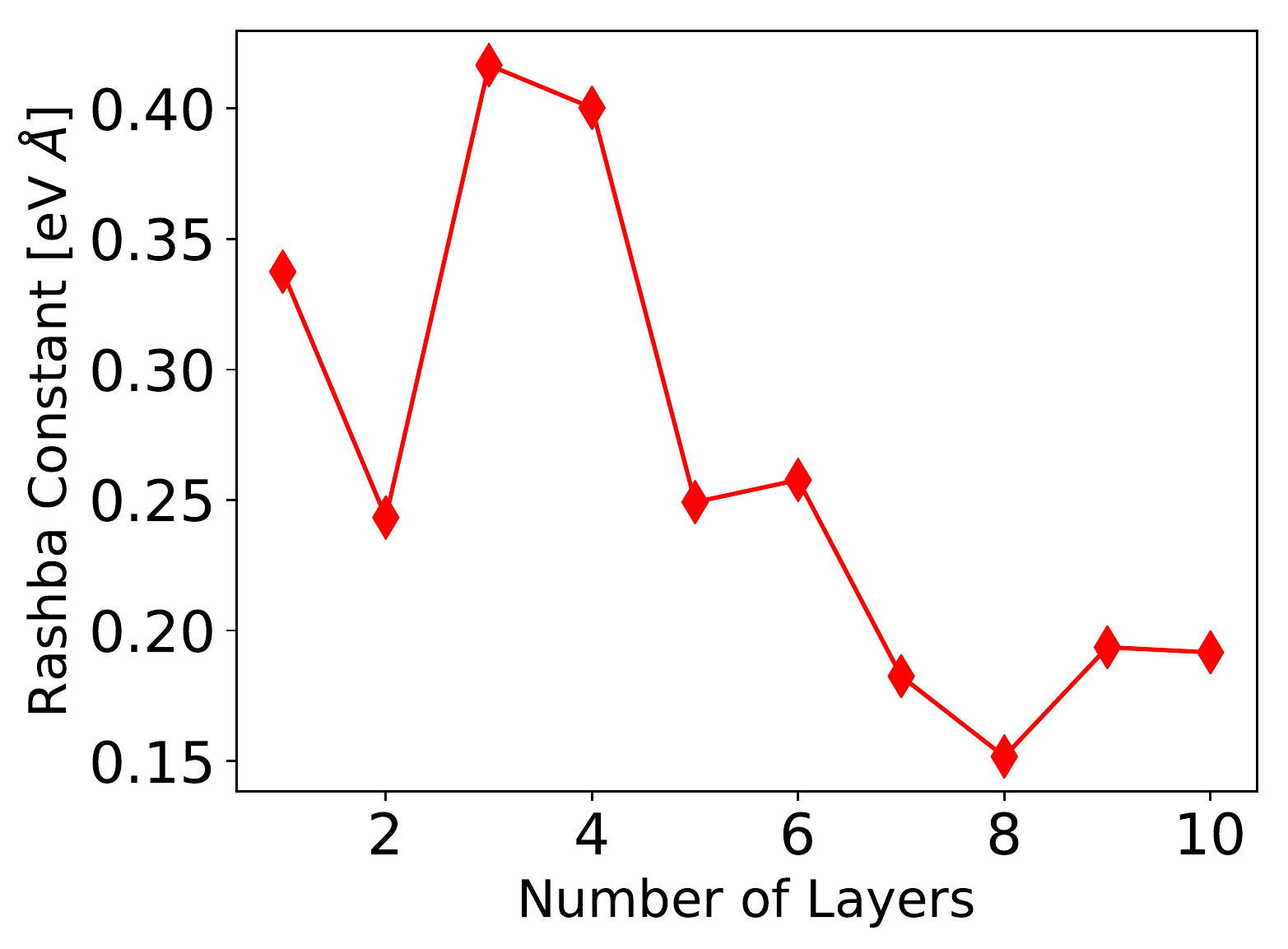}
\subcaption{Average Rashba coupling of Pb (111) vs layers} \label{r_pb_avg}
\end{subfigure}
\caption{
Rashba coupling(and the average of 10 bands around Fermi level) of Pb (111) around Fermi level vs layers}
\label{rashba_film_pb}
\end{figure*} \fi

We also studied the strain effect on the Rashba effect, as an example the Rashba coupling for 5 and 10 layers of Pb thin film
are shown in Fig. \ref{rashba_film_pb_strain}:

\ifpdf \begin{figure*}
\centering
\begin{subfigure}{0.45\linewidth}
\includegraphics[width=\linewidth,height=0.75\linewidth]{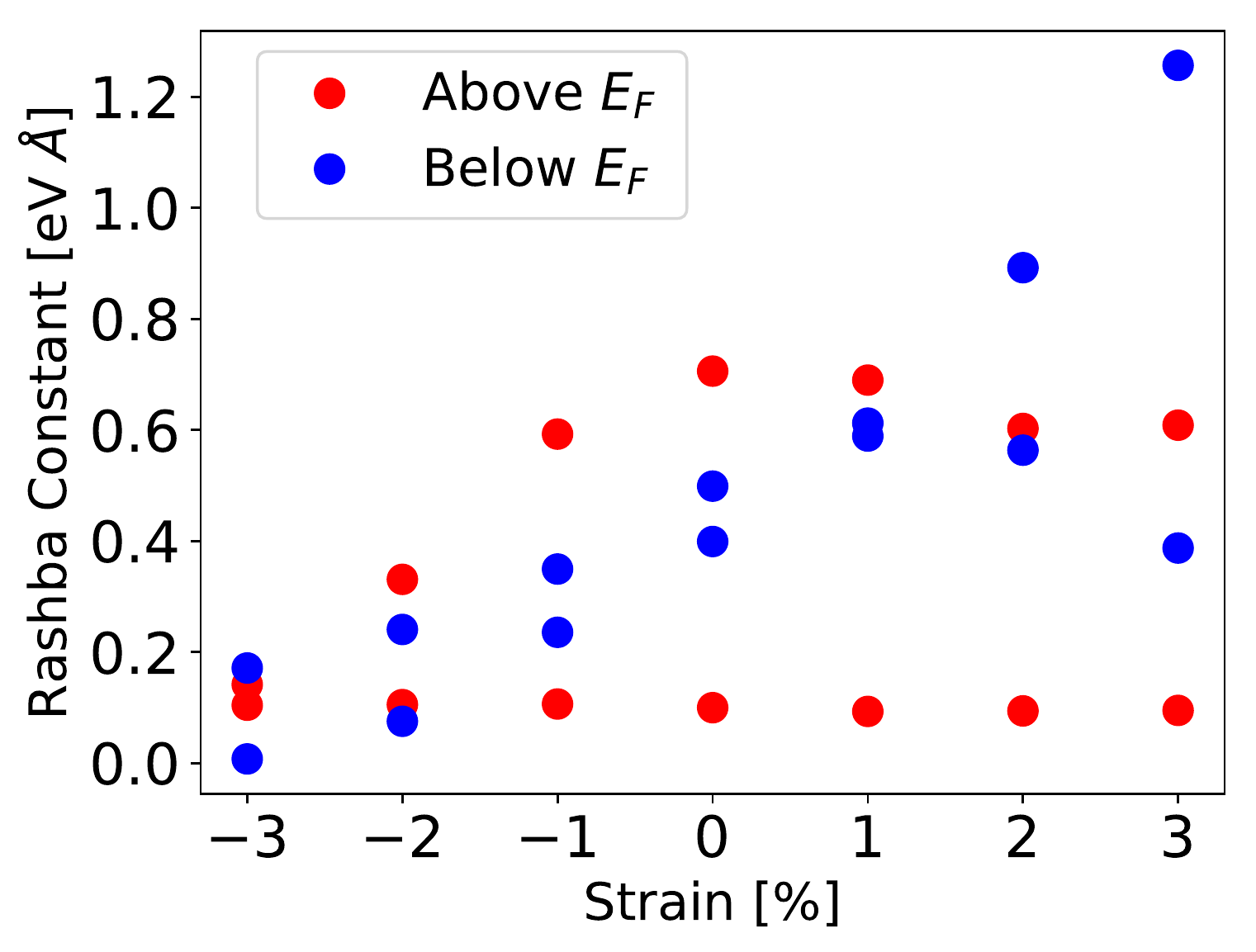}
\subcaption{Rashba coupling of Pb (111) vs strain for 5 layers case} \label{r_pb_lay_5}
\end{subfigure}
\begin{subfigure}{0.45\linewidth}
\includegraphics[width=\linewidth,height=0.75\linewidth]{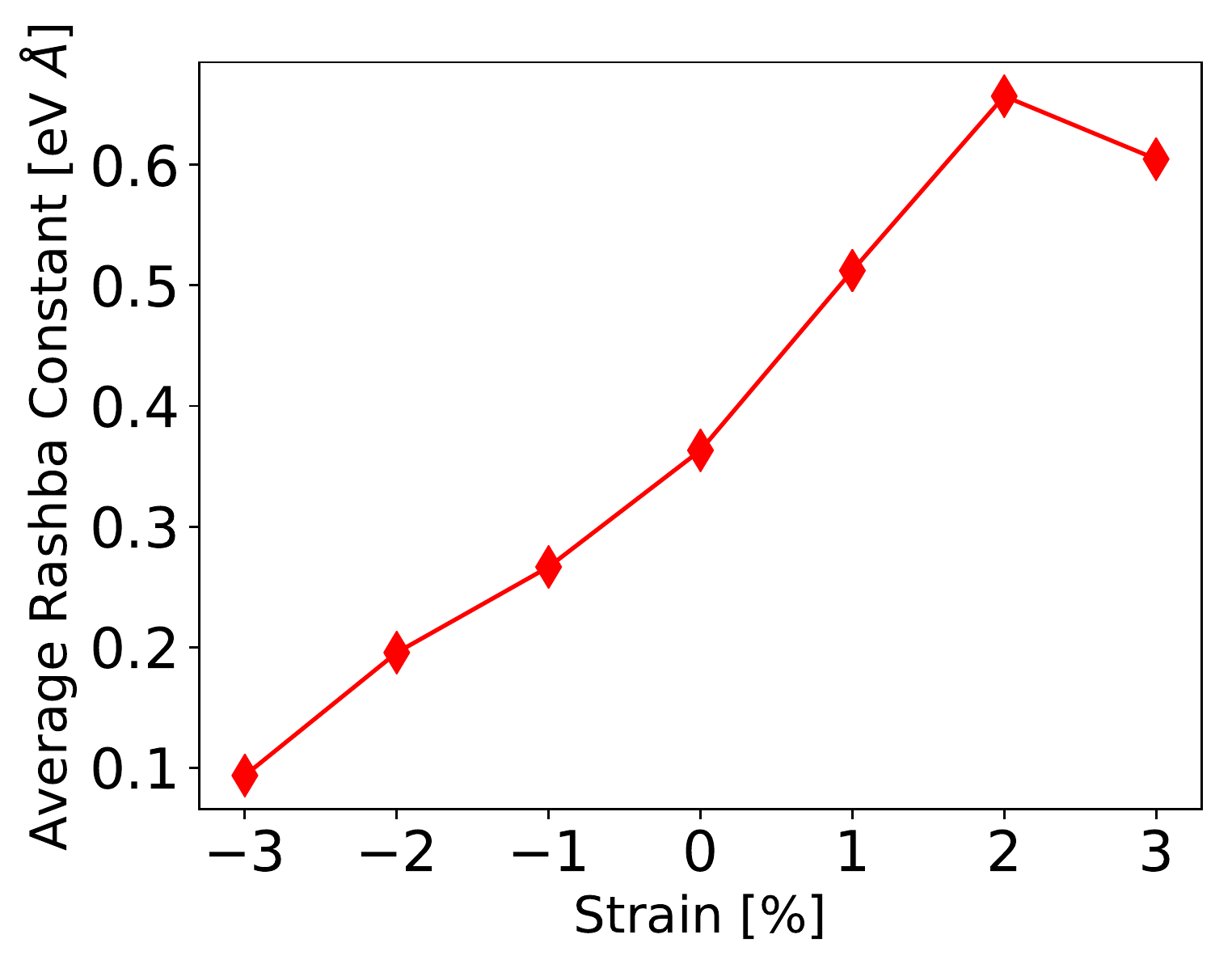}
\subcaption{Average Rashba coupling of Pb (111) vs strain for 5 layers case} \label{r_pb_avg_5}
\end{subfigure}
\begin{subfigure}{0.45\linewidth}
\includegraphics[width=\linewidth,height=0.75\linewidth]{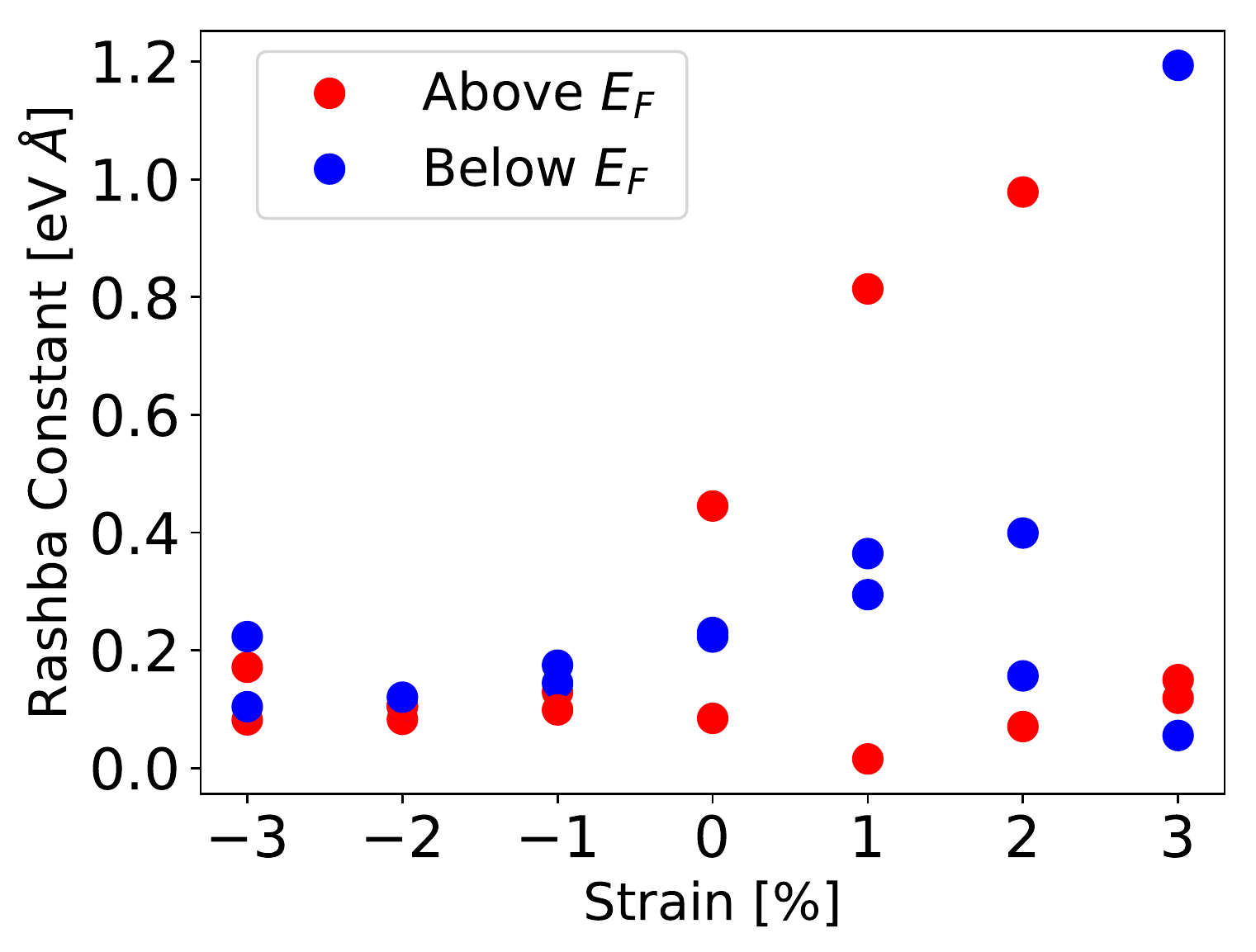}
\subcaption{Rashba coupling of Pb (111) vs strain for 10 layers case} \label{r_pb_lay_10}
\end{subfigure}
\begin{subfigure}{0.45\linewidth}
\includegraphics[width=\linewidth,height=0.75\linewidth]{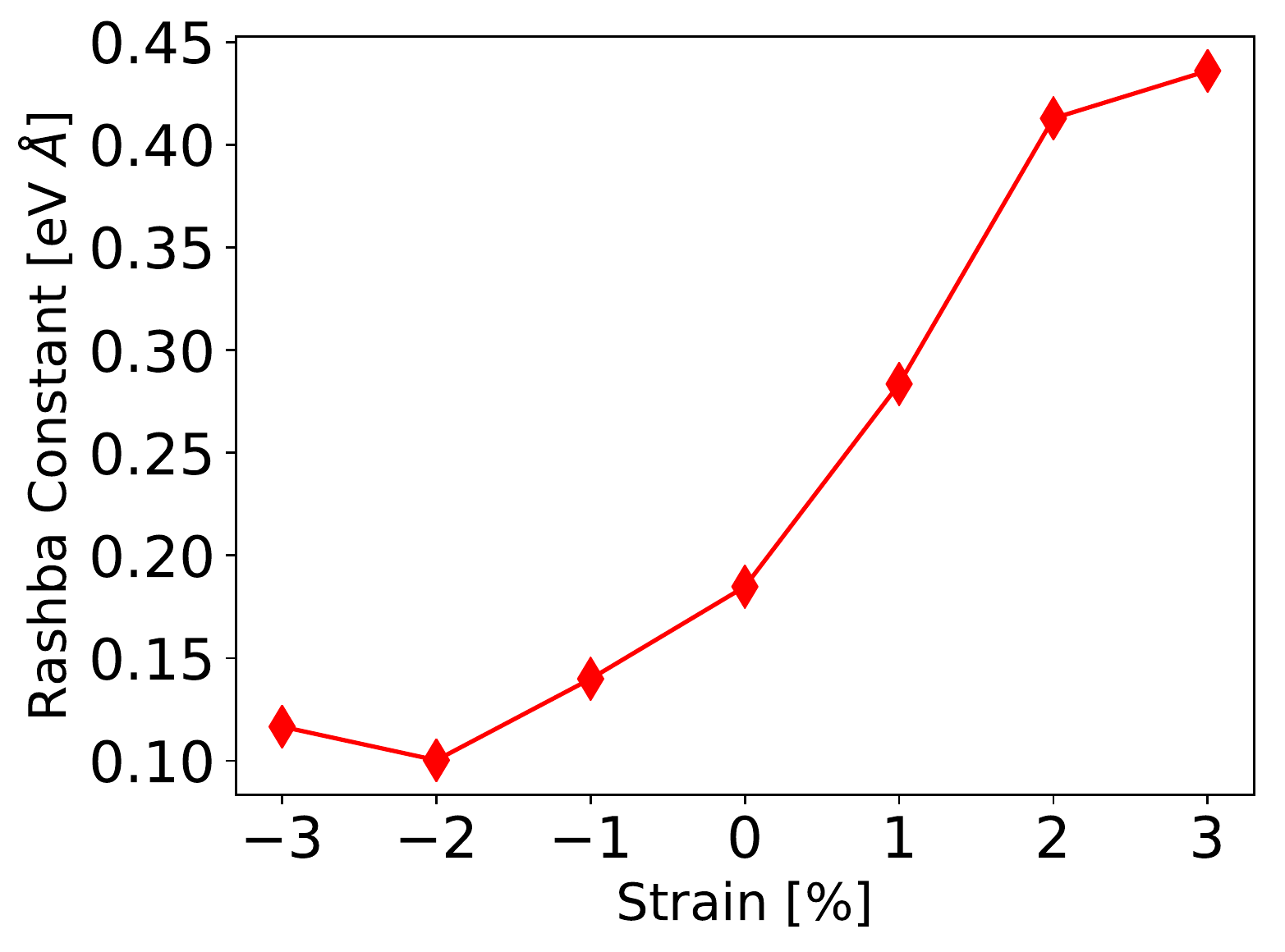}
\subcaption{Average Rashba coupling of Pb (111) vs strain for 10 layers case} \label{r_pb_avg_10}
\end{subfigure}
\caption{
Rashba coupling(and the average of 12 bands around Fermi level) of Pb (111) around Fermi level vs strain
for 5 and 10 layers case on $\rm As_2O_3$ substrates.}
\label{rashba_film_pb_strain}
\end{figure*} \fi
As it shows that tensile strain will strengthen the Rashba effect while the compressive strain will weaken the Rashba effect,
which has an opposite effect compared with Sn film.

\section{$\rm Pb (111)$ thin film on different substates}
\subsection{Electronic Structure}
To study the effect of substrates, we calculated the electronic structure of Pnictogen Chalcogenides substrates such as
$\rm As_2O_3$, $\rm Sb_2S_3$, $\rm Sb_2Se_3$, $\rm Bi_2Se_3$, $\rm Bi_2Te_3$, the results are shown in Fig.\ref{band_film1_substrate}
 for 1 layer of Pb thin film.

\ifpdf \begin{figure*}
\centering
\begin{subfigure}{0.45\linewidth}
\includegraphics[width=\linewidth,height=0.75\linewidth]{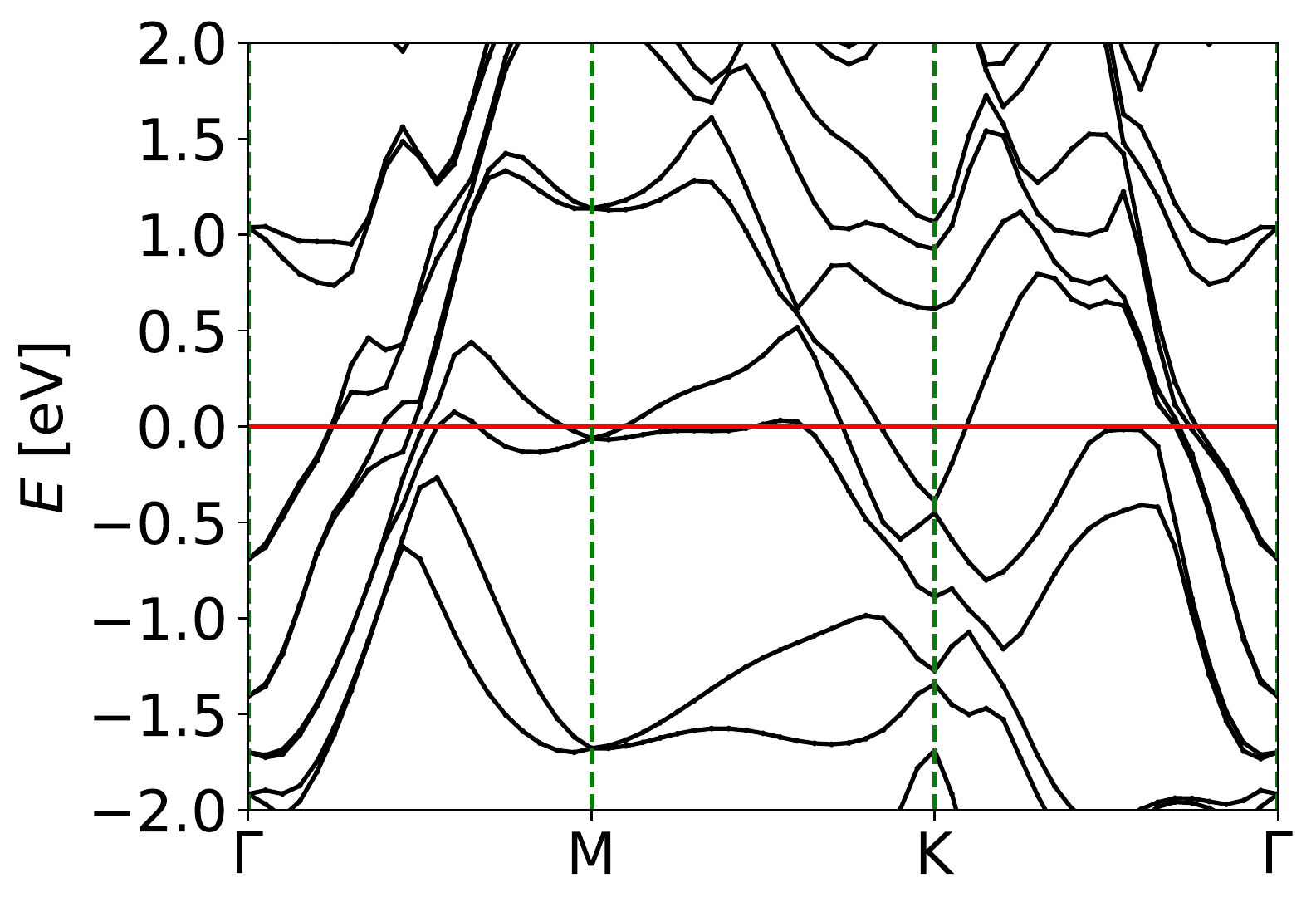}
\subcaption{1 Layer of Pb (111) grown on 1 quintuple layer of $\rm Sb_2S_3$ substrate} \label{band_PbSb2S3}
\end{subfigure}
\begin{subfigure}{0.45\linewidth}
\includegraphics[width=\linewidth,height=0.75\linewidth]{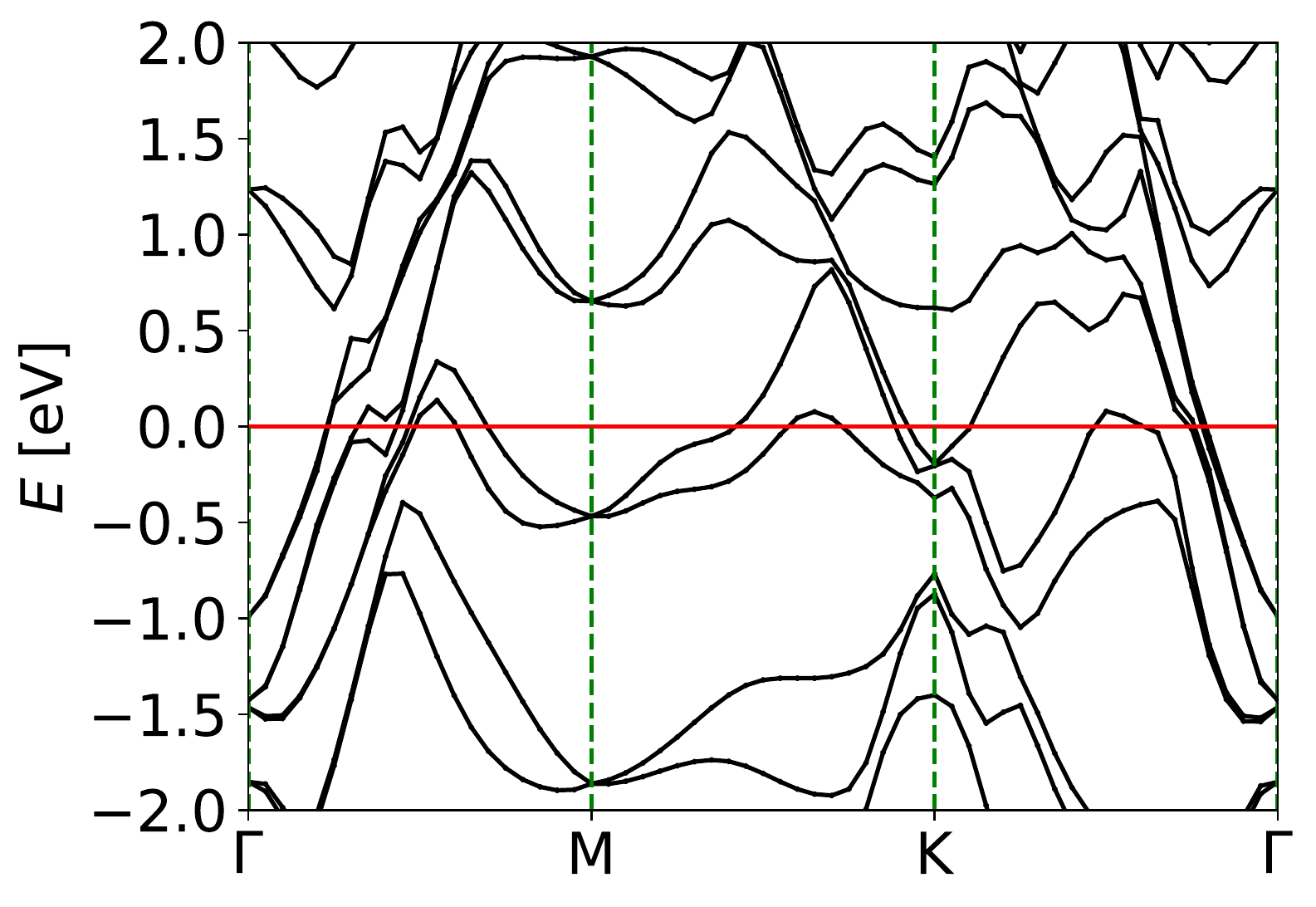}
\subcaption{1 Layer of Pb (111) grown on 1 quintuple layer of $\rm Sb_2Se_3$ substrate} \label{band_PbSb2Se3}
\end{subfigure}
\begin{subfigure}{0.45\linewidth}
\includegraphics[width=\linewidth,height=0.75\linewidth]{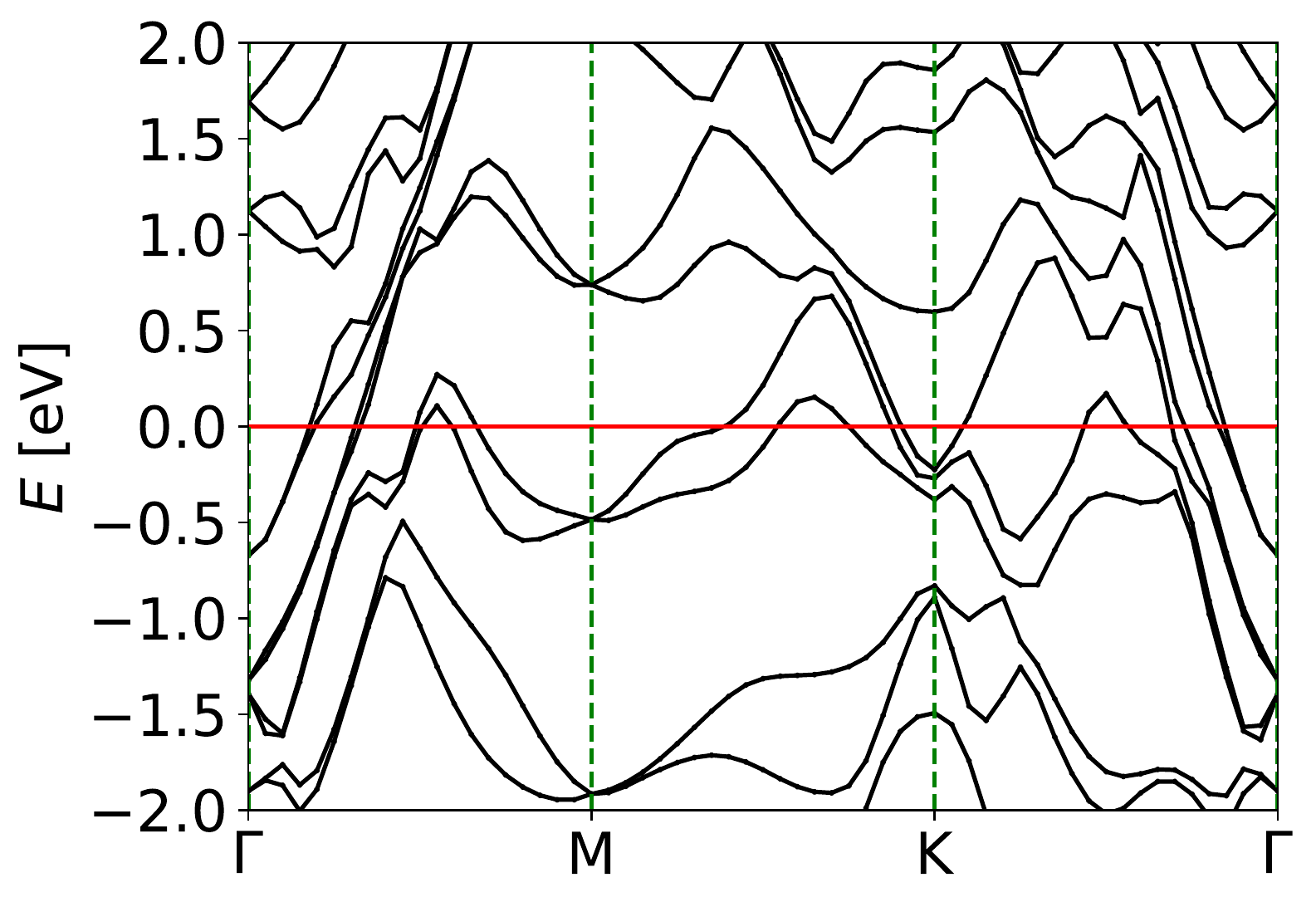}
\subcaption{1 Layer of Pb (111) grown on 1 quintuple layer of $\rm Bi_2Se_3$ substrate} \label{band_PbBi2Se3}
\end{subfigure}
\begin{subfigure}{0.45\linewidth}
\includegraphics[width=\linewidth,height=0.75\linewidth]{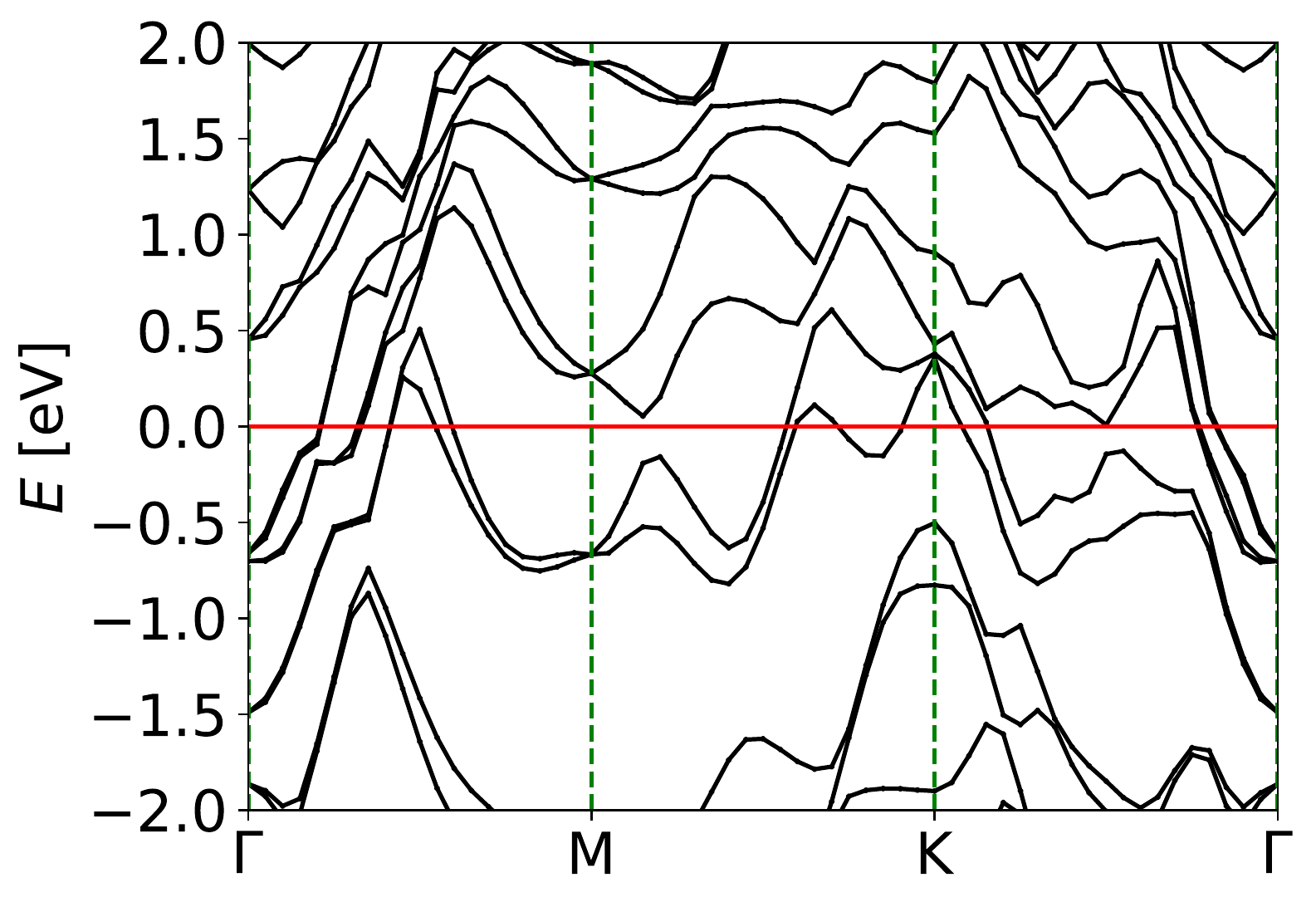}
\subcaption{1 Layer of Pb (111) grown on 1 quintuple layer of $\rm Bi_2Te_3$ substrate} \label{band_PbBi2Te3}
\end{subfigure}
\caption{
Band structure of a single layer Pb (111) grown on different substates}
\label{band_film1_substrate}
\end{figure*} \fi
For 2 layers of Pb thin film, the bandstructures are shown in Fig. \ref{band_film2_substrate}:
\ifpdf \begin{figure*}
\centering
\begin{subfigure}{0.45\linewidth}
\includegraphics[width=\linewidth,height=0.75\linewidth]{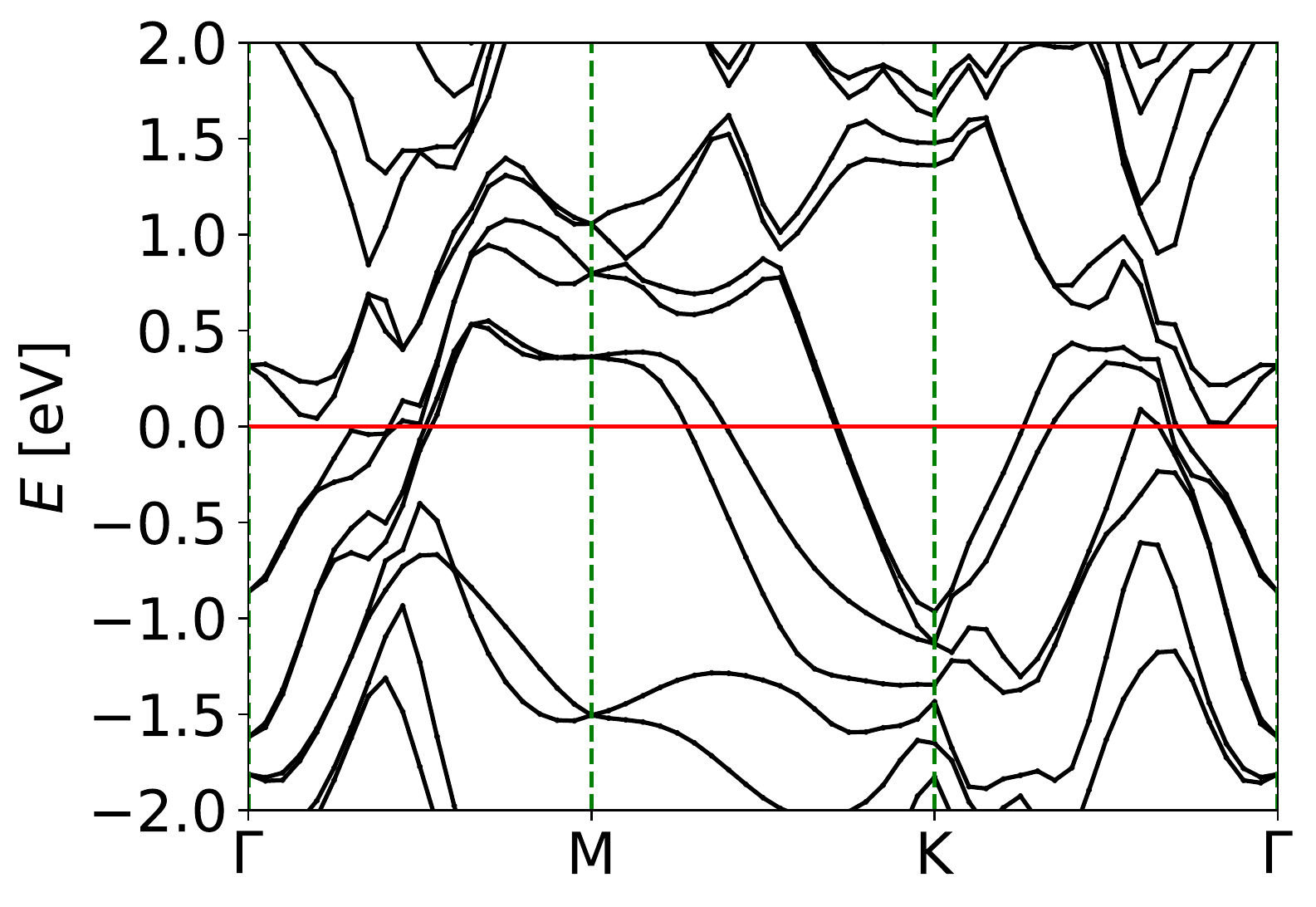}
\subcaption{2 Layer of Pb (111) grown on 1 quintuple layer of $\rm Sb_2S_3$ substrate} \label{band_PbSb2S3}
\end{subfigure}
\begin{subfigure}{0.45\linewidth}
\includegraphics[width=\linewidth,height=0.75\linewidth]{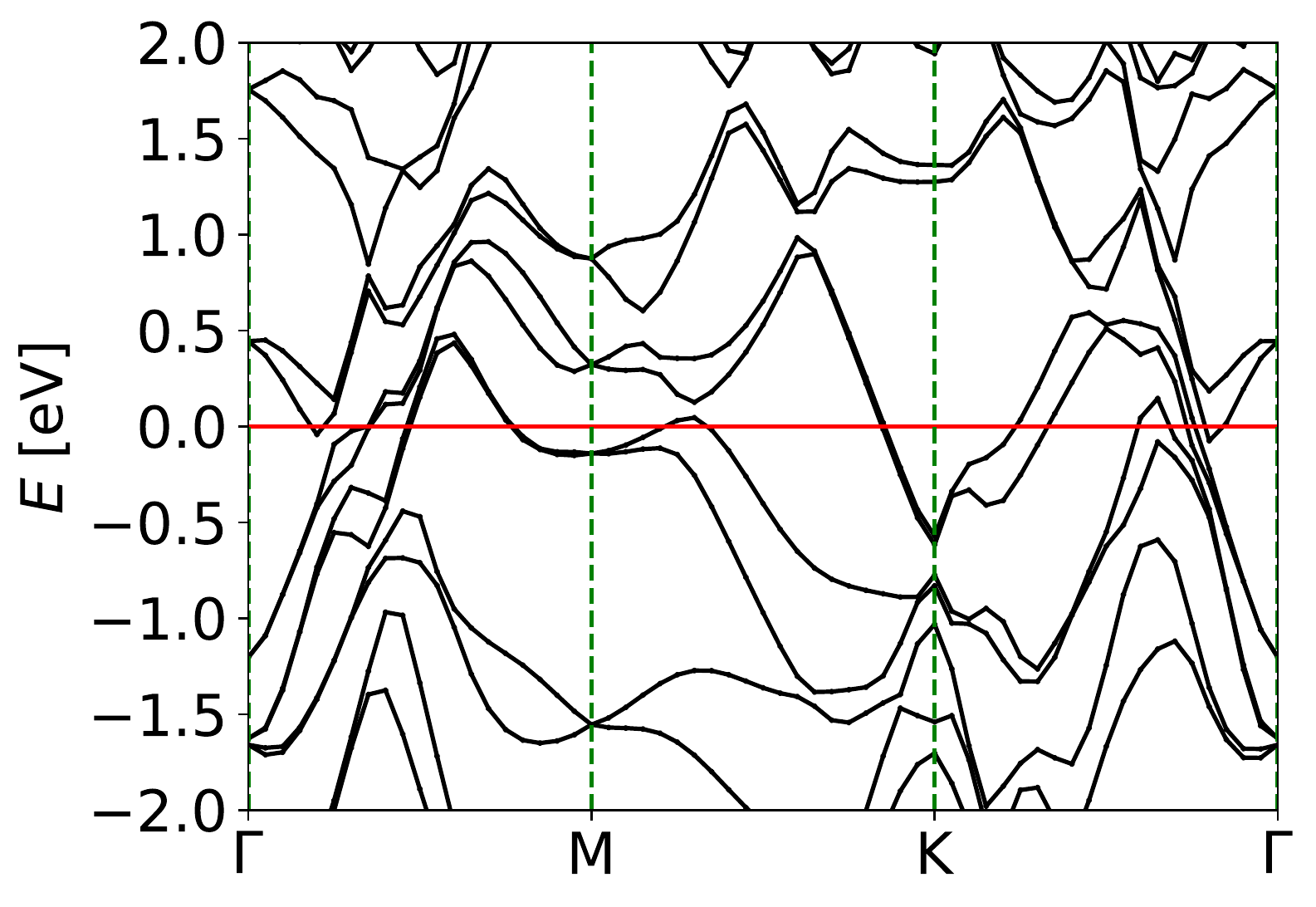}
\subcaption{2 Layer of Pb (111) grown on 1 quintuple layer of $\rm Sb_2Se_3$ substrate} \label{band_PbSb2Se3}
\end{subfigure}
\begin{subfigure}{0.45\linewidth}
\includegraphics[width=\linewidth,height=0.75\linewidth]{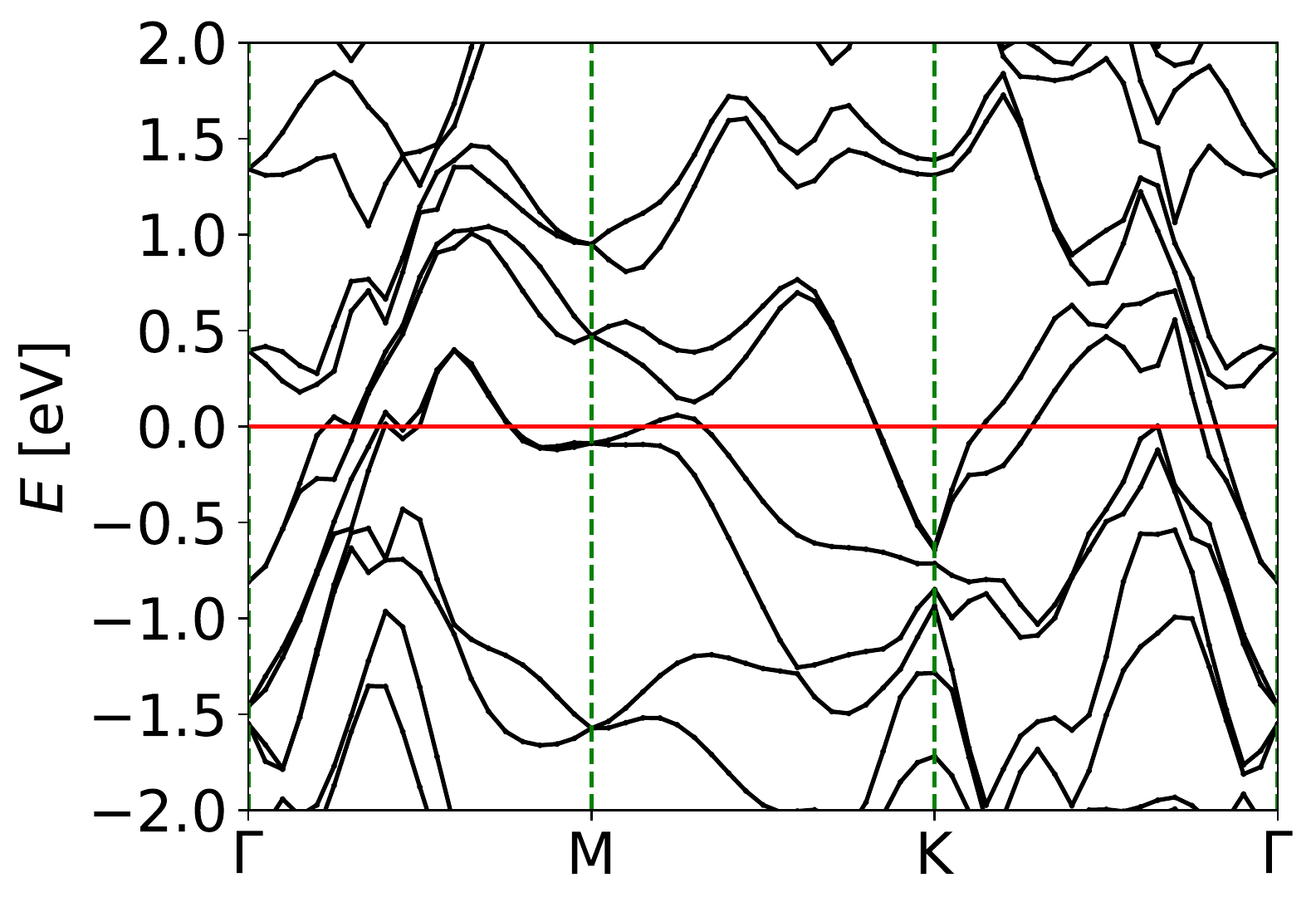}
\subcaption{2 Layer of Pb (111) grown on 1 quintuple layer of $\rm Bi_2Se_3$ substrate} \label{band_PbBi2Se3}
\end{subfigure}
\begin{subfigure}{0.45\linewidth}
\includegraphics[width=\linewidth,height=0.75\linewidth]{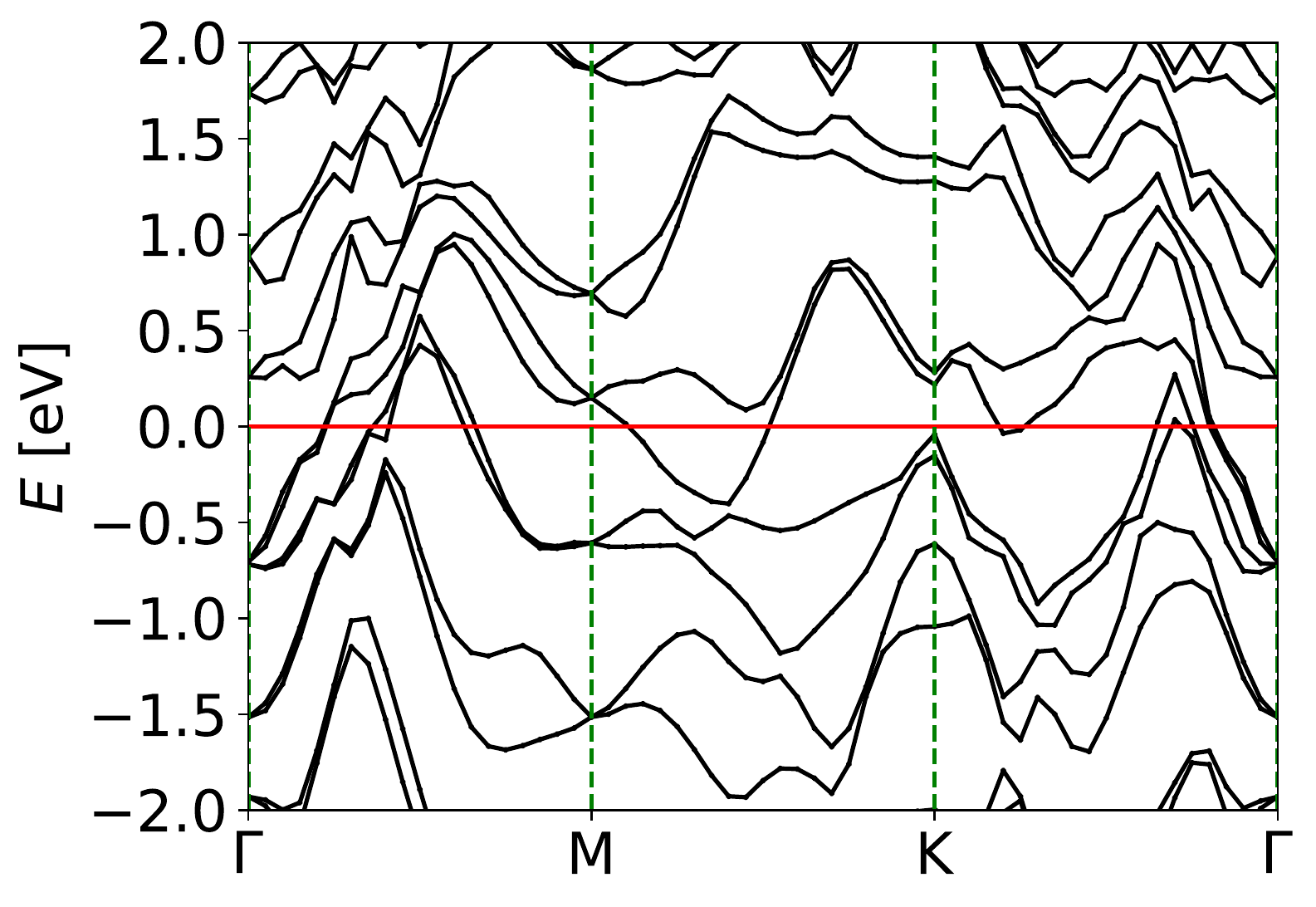}
\subcaption{2 Layer of Pb (111) grown on 1 quintuple layer of $\rm Bi_2Te_3$ substrate} \label{band_PbBi2Te3}
\end{subfigure}
\caption{
Band structure of a 2 layers Pb (111) grown on different substates}
\label{band_film2_substrate}
\end{figure*} \fi
All of the calculations are done for the lattice constant of $a = 4.9508 \AA$.

\subsection{Rashba Coupling}
To extract the Rashba splitting, the same methods are used as in Sn (001) thin film and Pb (111) thin film on $\rm As_2O_3$ substrate.
The Rashba coupling of 1 layer Pb (111) grown on different substrates are shown in Fig. \ref{rashba_pb1_sub} and \ref{rashba_pb1_sub_avg},
while the Rashba coupling of 2 layer Pb (111) are shown in Fig. \ref{rashba_pb2_sub} and \ref{rashba_pb2_sub_avg}. As the results show,
the substrates with larger atom numbers will give a larger Rashba splitting.
\ifpdf \begin{figure*}
\centering
\begin{subfigure}{0.45\linewidth}
\includegraphics[width=\linewidth,height=0.75\linewidth]{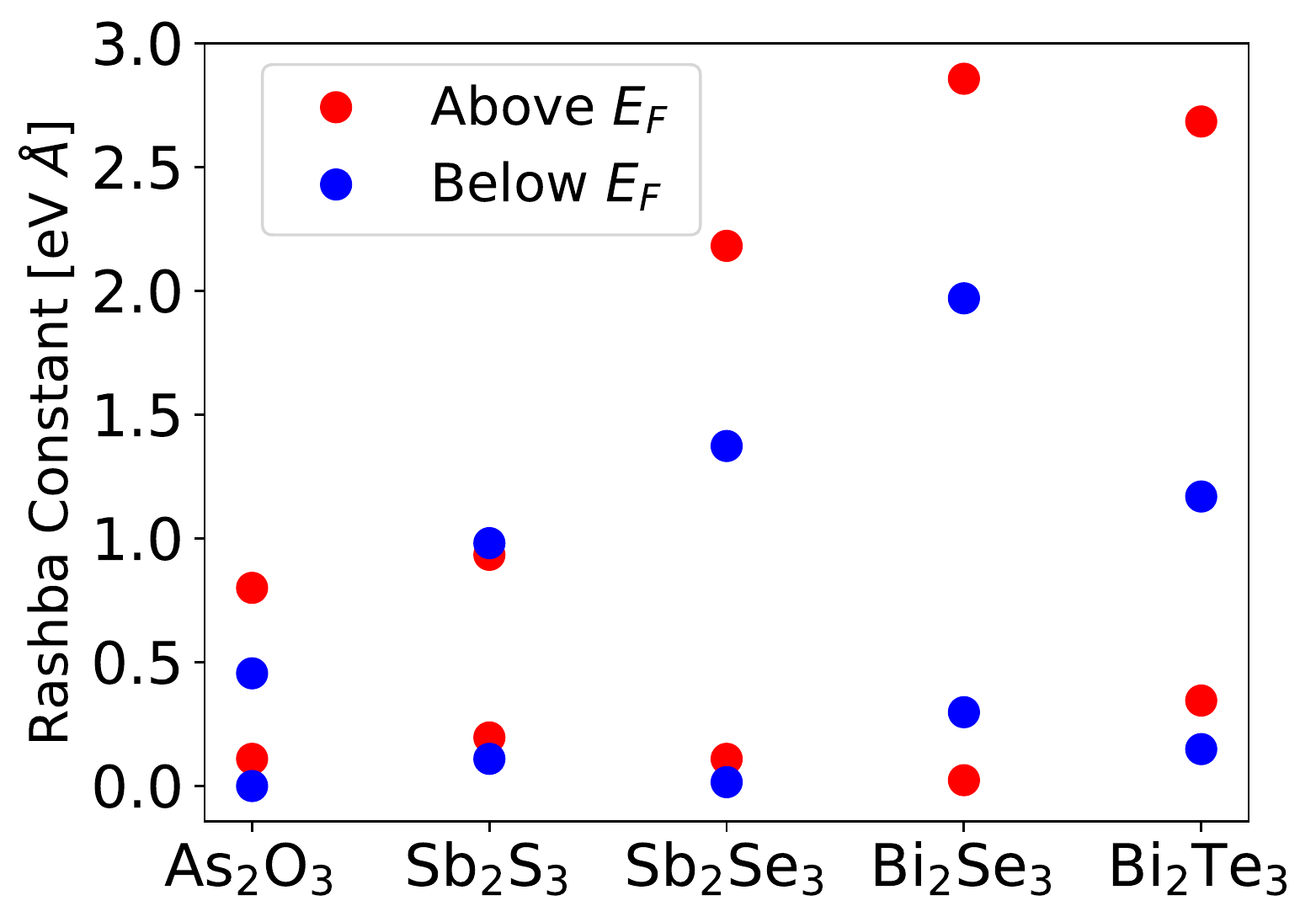}
\subcaption{Rashba coupling of 1 layer Pb (111) grown on different substrates} \label{rashba_pb1_sub}
\end{subfigure}
\begin{subfigure}{0.45\linewidth}
\includegraphics[width=\linewidth,height=0.75\linewidth]{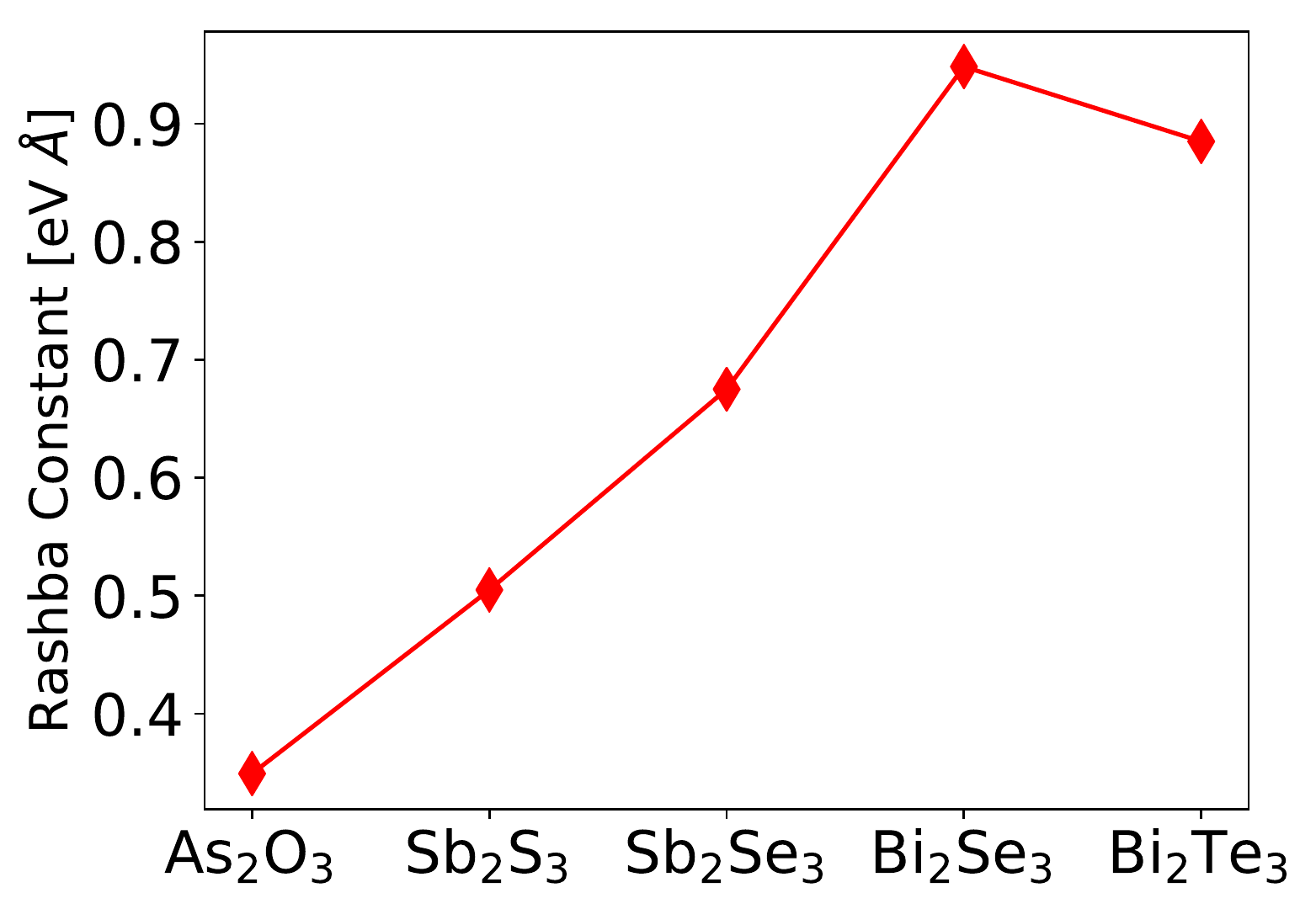}
\subcaption{Average Rashba coupling of 1 layer Pb (111) grown on different substrates} \label{rashba_pb1_sub_avg}
\end{subfigure}
\begin{subfigure}{0.45\linewidth}
\includegraphics[width=\linewidth,height=0.75\linewidth]{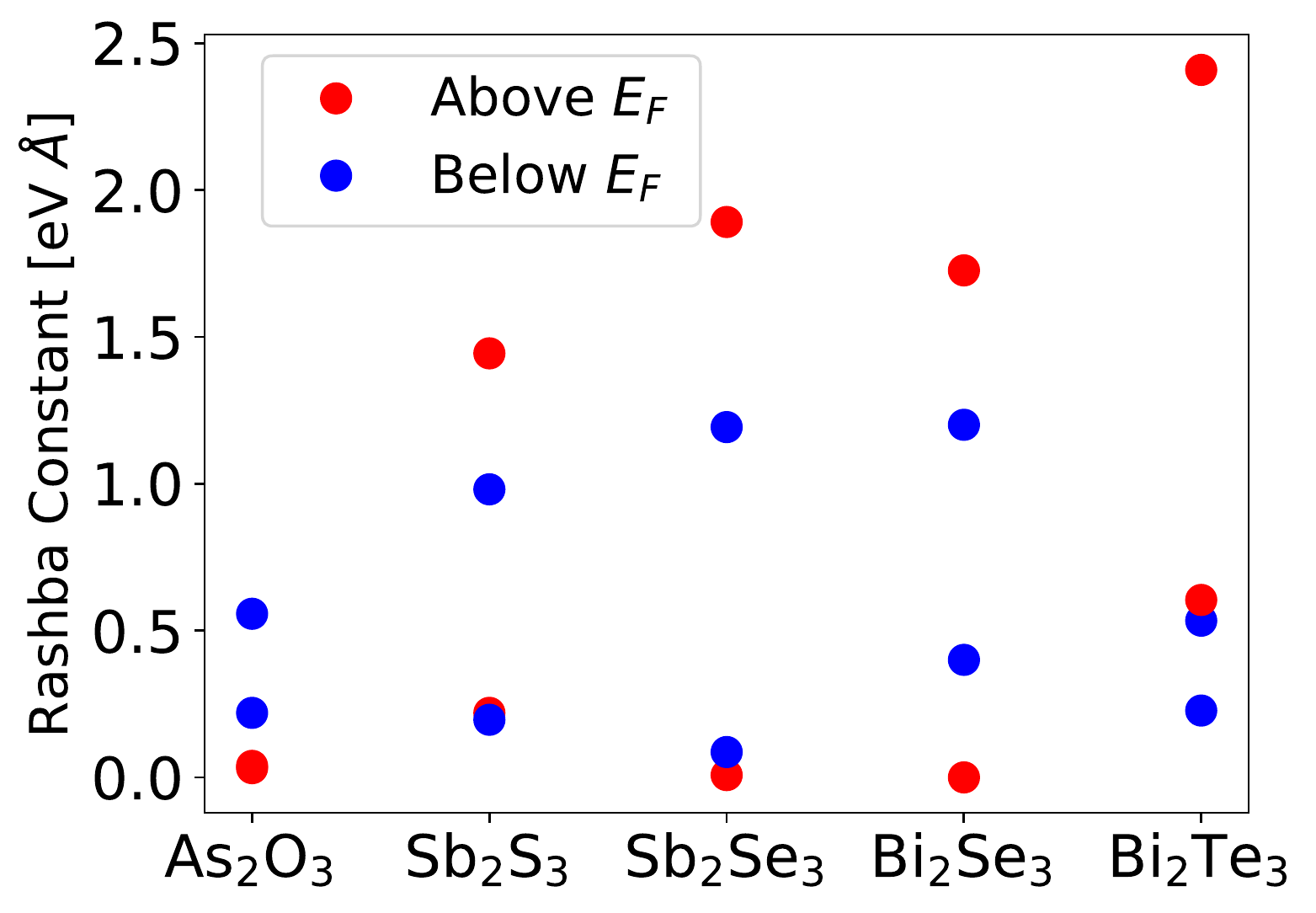}
\subcaption{Rashba coupling of 2 layer Pb (111) grown on different substrates} \label{rashba_pb2_sub}
\end{subfigure}
\begin{subfigure}{0.45\linewidth}
\includegraphics[width=\linewidth,height=0.75\linewidth]{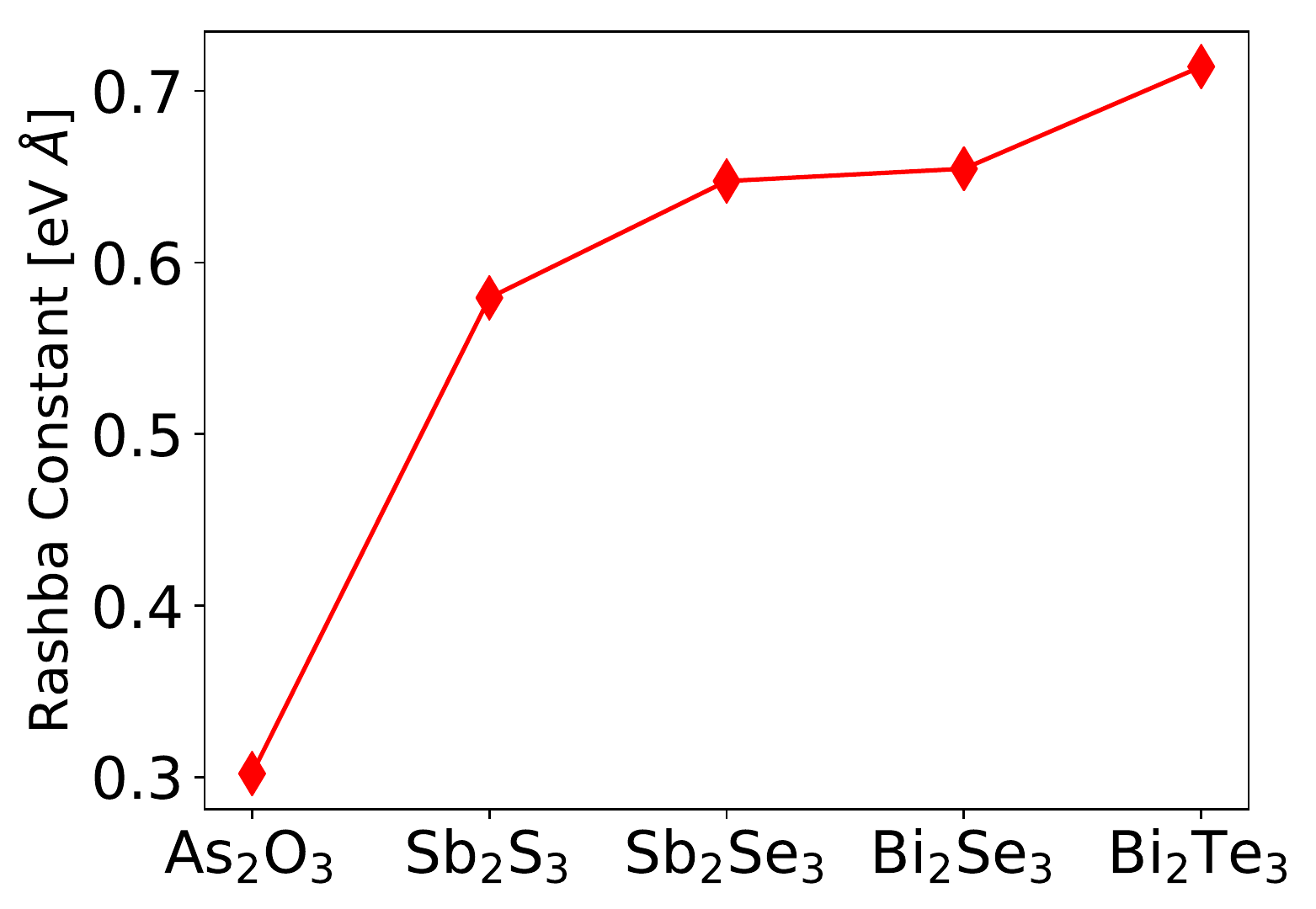}
\subcaption{Average Rashba coupling of 2 layer Pb (111) grown on different substrates} \label{rashba_pb2_sub_avg}
\end{subfigure}
\caption{
Rashba coupling of a 1 and 2 layers Pb (111) grown on different substates, the average is performed on the 12 subbands around Fermi level,
that is 6 above and 6 below.}
\label{rashba_sub_pb}
\end{figure*} \fi

\section{Mean Field Theory}
To analyse the class of topological phase for thin film metals, we construct a minimal mean field Hamiltonian.
For the subbands around Fermi level, the minimal mean field Hamiltonian of the system can be written as:
\begin{equation}\label{meanfield}
  H = \sum_{\bk} \psi_{\bk}^{\dagger} \mathcal{H}_{\bk} \psi_{\bk}
\end{equation}
with the Nambu basis $ \psi_{\bk}^{\dagger} = [c_{\bk\uparrow}^{\dagger}, c_{\bk\downarrow}^{\dagger}, c_{-\bk\uparrow}, c_{-\bk\downarrow}] $ and
\begin{equation}\label{Hamdens}
\begin{aligned}
  \mathcal{H}_{\bk} = & (\frac{\hbar^2}{2m^{\ast}} \bk^{2} -\mu +  g \mu_{B} B_x \sigma_x + g \mu_{B} B_z \sigma_z - \alpha_{_R} k_x \sigma_y) \tau_z \\
   & + \alpha_{_R}k_y \sigma_x + g \mu_{B} B_y \sigma_y + \Delta (i\sigma_y)(i\tau_y)
  \end{aligned}
\end{equation}
where $m^{\ast}$ is the effective mass,  $\mu$ is the chemical potential,$\alpha_{R}$ is the Rashba coupling constant, g is the g-factor, $ \mu_B $ is the Bohr magneton, $\bB$ is the magnetic field, $\bs$ is the Pauli matrices acting on spin , $\tau$ is the Pauli matrices acting on particle-hole space and $\Delta$ is the pairing potential of Cooper pairs.

For the time-reversal and particle-hole symmetry:
\begin{equation}\label{symmetry}
  T = U_T \mathcal{K}, C = U_C \mathcal{K}
\end{equation}
where $\mathcal{K}$ is the conjugate operator and
\begin{equation}\label{unitary}
  U_T = i\sigma_y, U_C = \tau_x
\end{equation}
it is obvious that:
\begin{equation}\label{hamsymmetry}
  T\mathcal{H}_{\bk}T^{-1} \ne \mathcal{H}_{-\bk}, C\mathcal{H}_{\bk}C^{-1} = -\mathcal{H}_{-\bk}
\end{equation}
which shows that the system is a class D topological superconductivity in Altland-Zirnbauer system,
for the reason of that $T^2 = -1$ and $C^2 = 1$.
Since this is a 2D system, the topological invariant can be calculated with Chern number.

In real space, the mean field Hamiltonian is :
\begin{equation}
 H = H_0 + H_R + H_{SC} + H_Z
\end{equation}
with
\begin{equation}
 H_0 = -\sum_{<ij>,\sigma}{t_{ij}c_{i\sigma}^{\dagger}c_{j\sigma}} - \sum_{i,\sigma} {\mu_{i}c_{i\sigma}^{\dagger}c_{j\sigma}}
\end{equation}
is the hopping term and chemical potential, where $ t \approx \frac{\hbar^2}{2m^{\ast} a^2} $, where a is the effective lattice constant and
\begin{equation}
 H_R = -\alpha \sum_{j}[c_{j-\hat{x},\downarrow}^{\dagger}c_{j,\uparrow} - c_{j+\hat{x},\downarrow}^{\dagger}c_{j,\uparrow}
 +i(c_{j-\hat{y},\downarrow}^{\dagger}c_{j,\uparrow} - c_{j+\hat{y},\downarrow}^{\dagger}c_{j,\uparrow}) + h.c. ]
\end{equation}
is the Rashba term with $\hat{x}$ and $\hat{y}$ the lattice vectors and $\alpha \approx \frac{\alpha_{_R}}{a}$,  and
\begin{equation}
 H_{SC} = \sum_{j} (\Delta_j c_{j,\downarrow}^{\dagger}c_{j,\uparrow}^{\dagger} + h.c.)
\end{equation}
is the s pairing superconductor and
\begin{equation}
 H_Z = g\mu_B\sum_{j} c_{j}^{\dagger} \bB \cdot \bs c_{j}
\end{equation}
is the Zeeman term where $c_{j}^{\dagger} = [c_{j,\uparrow}^{\dagger}, c_{j,\downarrow}^{\dagger}]$

\section{$\rm g$-factor of thin film}
The g-factors are obtained by evaluating the splitting between a Kramers pair at the $\Gamma$ point under a magnetic field, which is\cite{gfactor}:
\begin{equation}
 g(\bk) = \frac{4 m_e}{e\hbar} \frac{\partial \Delta E}{\partial \vec B} = \frac{4 m_e}{e\hbar} \Delta \mu(\bk)
\end{equation}
where $ \Delta \mu = \mu_1 - \mu_2$ and $ \mu_i $ is the eigenvalues of the matrix $ \hat{\mu} $, whose matrix elements are:
\begin{equation}
 \mu_{n,\alpha\alpha^{\prime}} = \mu_{n,\alpha\alpha^{\prime}}^s + \mu_{n,\alpha\alpha^{\prime}}^{or}
\end{equation}
where the total contributions of the orbital part as
\begin{widetext}
\begin{equation}\label{g_orbital}
\mu_{n,\alpha\alpha^{\prime}}^{or} = -\frac{e\hbar}{2i} \sum_{\alpha^{\dprime} = \alpha, \alpha^{\prime}} \vec{a} \cdot \vec{v}_{n\alpha,n\alpha^{\dprime}} \times \Omega_{n\alpha^{\dprime},n\alpha^{\prime}}(\bk)
 -\frac{e\hbar}{2i} \sum_{l\ne n,\alpha^{\dprime}} { \vec{a} \cdot \frac { \vec{v}_{n\alpha,l\alpha^{\dprime}} \times \vec{v}_{l\alpha^{\dprime},n\alpha^{\prime}} }{E_{n}(\bk) - E_{l}(\bk)}}
\end{equation}
\end{widetext}
where $ \Omega_{n\alpha^{\dprime},n\alpha^{\prime}}(\bk)  =  \bra{u_{n\alpha^{\dprime}}(\bk)}\frac{\partial}{\partial \bk}  \ket{u_{n\alpha^{\prime}}(\bk)} $,
and the velocity elements are:
\begin{equation}
\vec{v}_{n\alpha,m\alpha^{\prime}} = \frac{1}{\hbar} \bra{u_{n\alpha}(\bk)}\frac{\partial H}{\partial \bk}  \ket{u_{n\alpha^{\prime}}(\bk)}
\end{equation}
the pure spin parts is
\begin{equation}
\mu_{n,\alpha\alpha^{\prime}}^s = -\frac{e\hbar}{2m_e} \bra{u_{n\alpha}(\bk)} \vec a \cdot \vec{\sigma}  \ket{u_{n\alpha^{\prime}}(\bk)}
\end{equation}
where $ \vec a $ is the unit vector of direction of the magnetic field.
According to Eq. \ref{g_orbital}, the density of subbands and velocity determined the value of g-factor, for this reason if the large hybridization between
subbands will give a large g-factor. Also at $\Gamma$ point, the velocity can be approximately determined by the Rashba coupling constant $\alpha_R$(see Eq. \ref{Hamdens}),
thus a large $\alpha_R$ usually gives a large g-factor.

As shown in Fig.\ref{en_film}, as the film becomes thicker, the density of subbands becomes larger, while the Rashba coupling
(shown as in Fig.\ref{rashba_film_sn} and Fig. \ref{rashba_film_pb}) becomes smaller, it is thus hard to estimate the trends of g-factor vs layers.
From Fig.\ref{en_film} we can still get additional information about the subbands at $\Gamma$ point and a large difference between Sn and Pb exists:
that the hybridization of Sn thin film is larger, combined with the larger Rashba splitting of Sn thin film than Pb thin film, we can expect
a larger g-factor in Sn thin film.

\ifpdf \begin{figure*}
\centering
\begin{subfigure}{0.45\linewidth}
\includegraphics[width=\linewidth,height=0.75\linewidth]{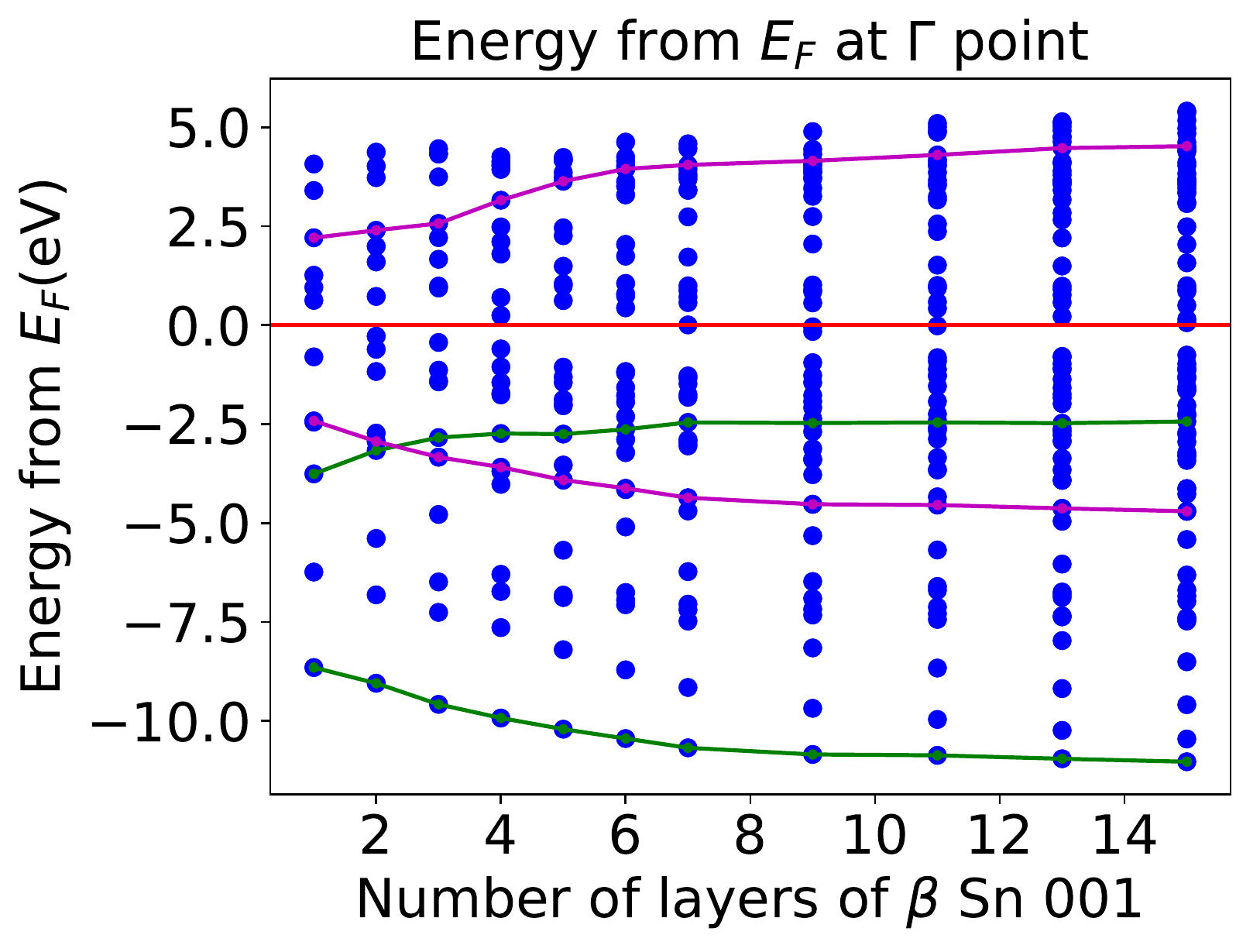}
\subcaption{Energy from Fermi level for Sn (001) thin film} \label{En_sn_lay}
\end{subfigure}
\begin{subfigure}{0.45\linewidth}
\includegraphics[width=\linewidth,height=0.75\linewidth]{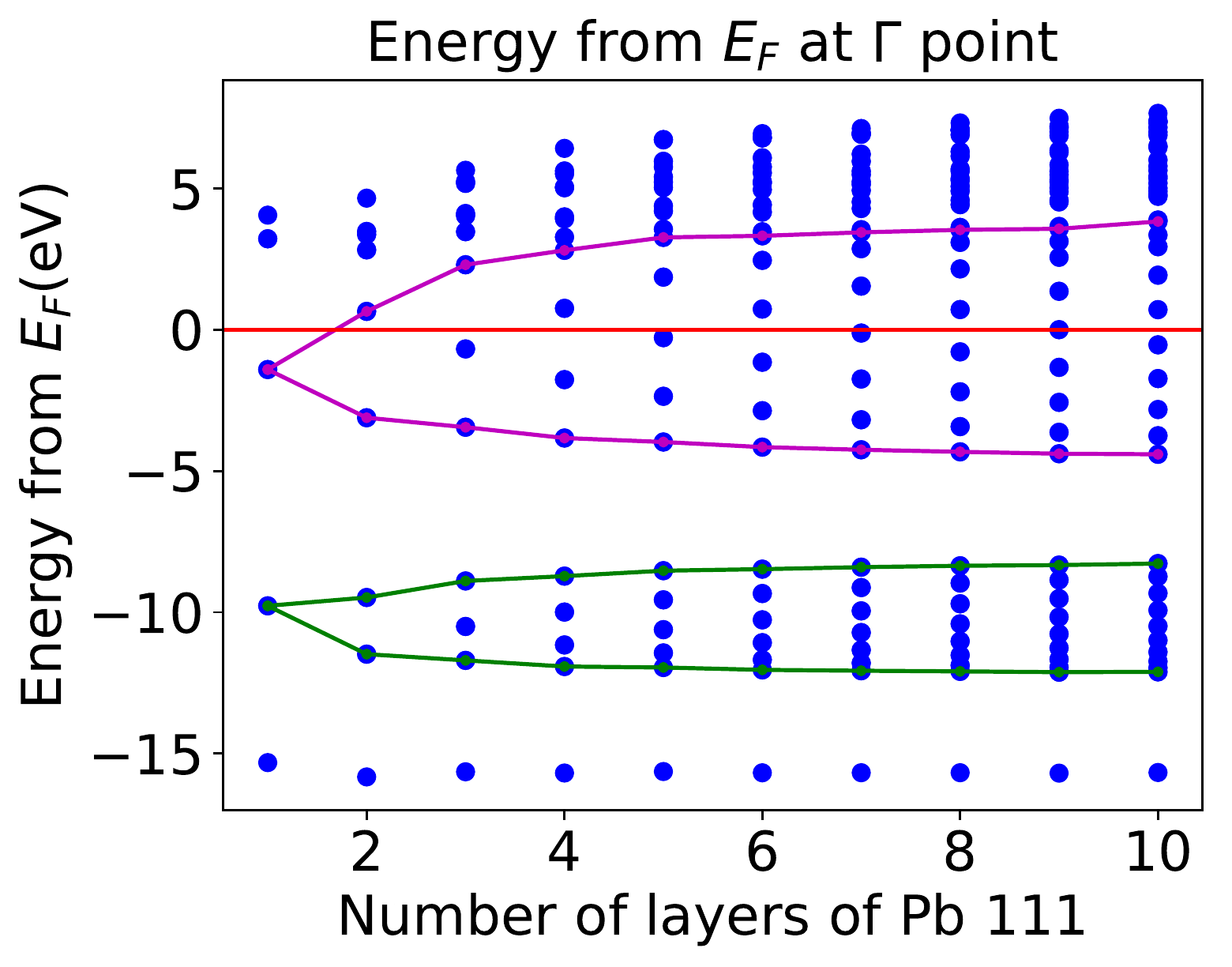}
\subcaption{Energy from Fermi level for Pb (111) thin film} \label{En_pb_lay}
\end{subfigure}
\caption{Energy from Fermi level for Sn (001) and Pb (111) thin film}
\label{en_film}
\end{figure*} \fi

\ifpdf \begin{figure*}
\centering
\begin{subfigure}{0.45\linewidth}
\includegraphics[width=\linewidth,height=0.75\linewidth]{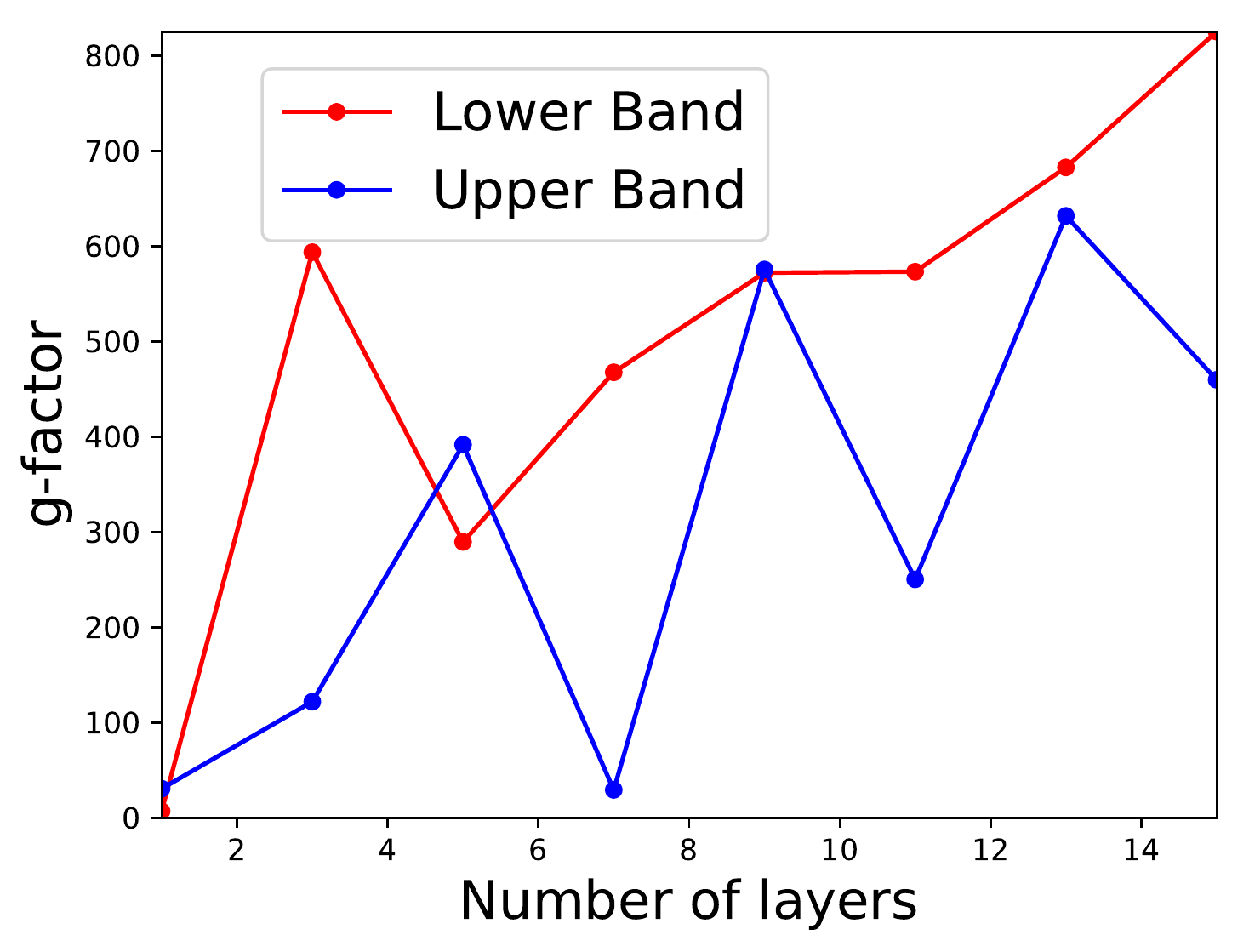}
\subcaption{g-factor of the bands around Fermi level at $\Gamma$ point} \label{g_sn_lay}
\end{subfigure}
\begin{subfigure}{0.45\linewidth}
\includegraphics[width=\linewidth,height=0.75\linewidth]{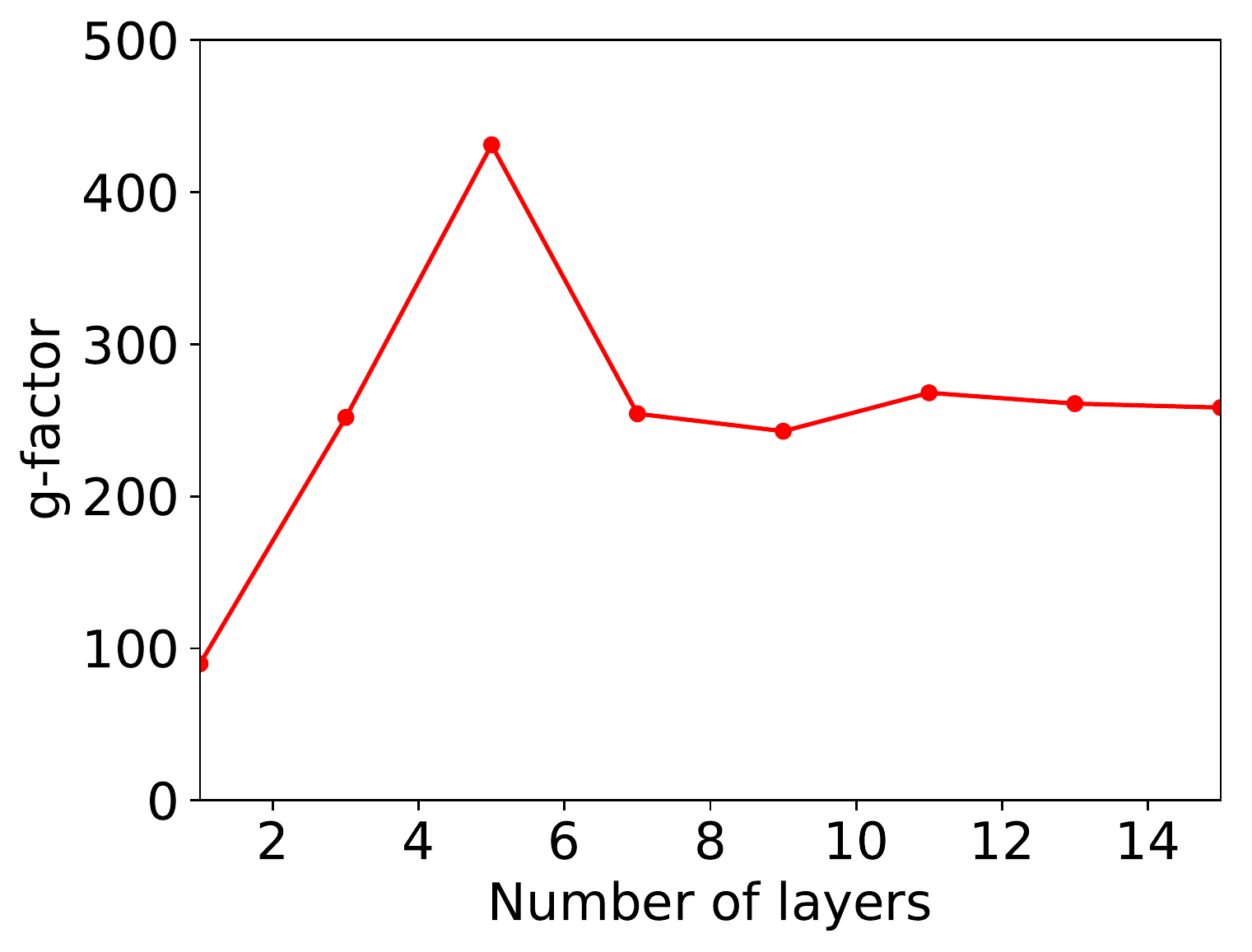}
\subcaption{Average g-factor of the bands around Fermi level at $\Gamma$ point} \label{g_sn_avg}
\end{subfigure}
\caption{g-factor vs number of layers for Sn (001) thin film}
\label{g_film_sn}
\end{figure*} \fi

\ifpdf \begin{figure*}
\centering
\begin{subfigure}{0.45\linewidth}
\includegraphics[width=\linewidth,height=0.75\linewidth]{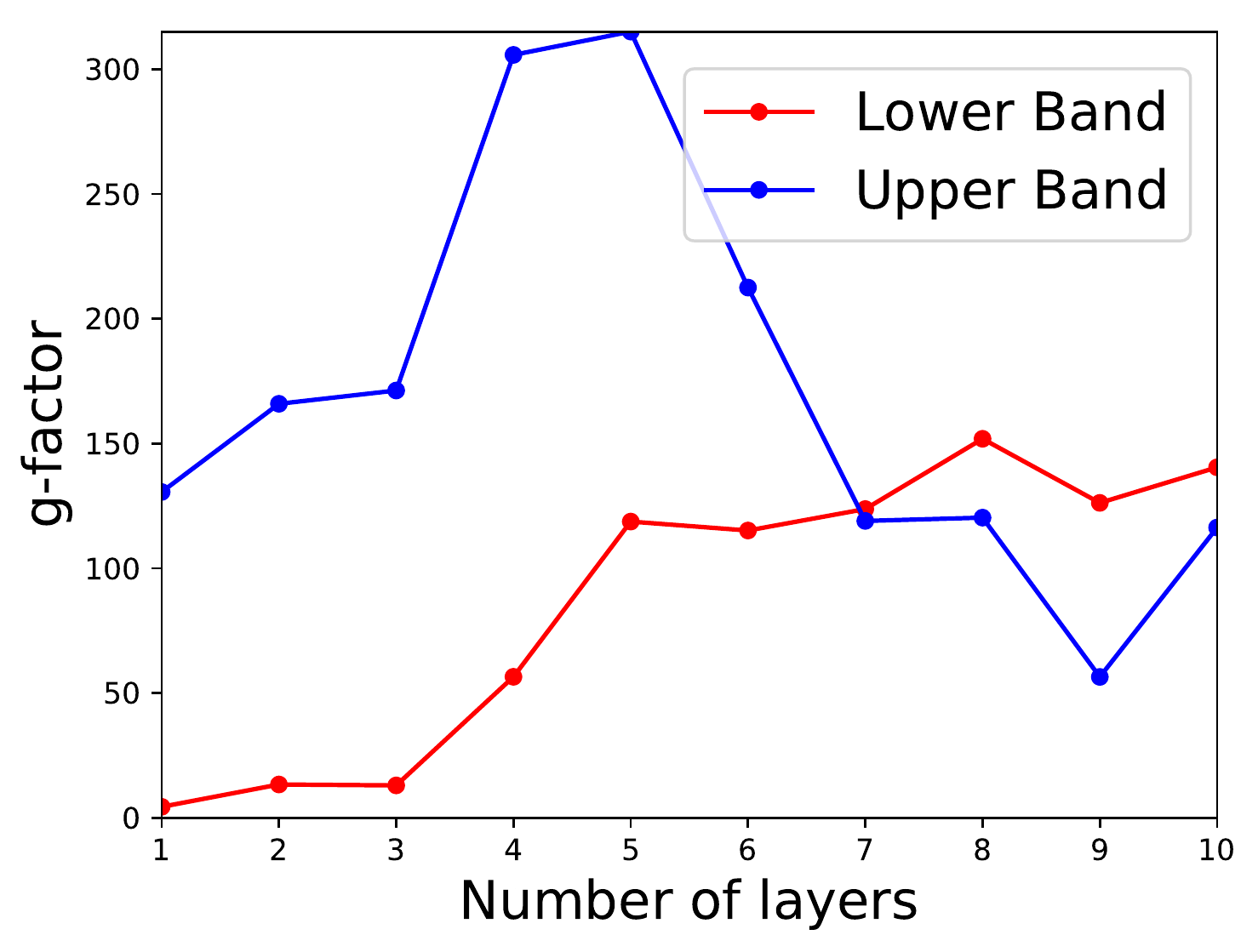}
\subcaption{g-factor of the bands around Fermi level at $\Gamma$ point} \label{g_pb_lay}
\end{subfigure}
\begin{subfigure}{0.45\linewidth}
\includegraphics[width=\linewidth,height=0.75\linewidth]{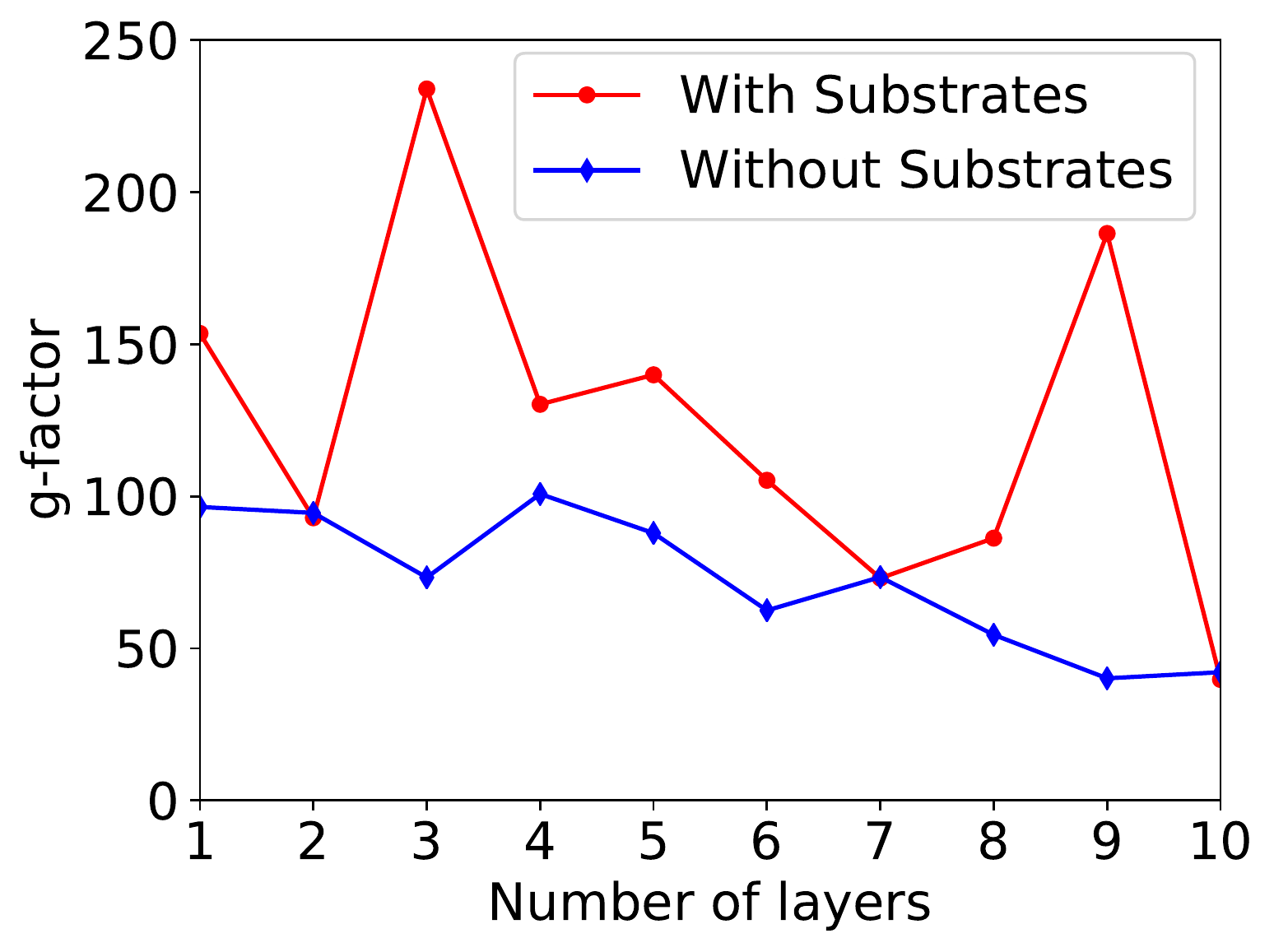}
\subcaption{Average g-factor of the bands around Fermi level at $\Gamma$ point} \label{g_pb_avg}
\end{subfigure}
\caption{g-factor vs number of layers for Pb (111) thin film}
\label{g_film_pb}
\end{figure*}
As Fig.\ref{g_film_sn} and \ref{g_film_pb} show, we can see that the g-factors in Sn thin film are several times larger than the g-factors in Pb thin film,
some oscillations happen in both Sn and Pb thin film, and the case of Sn thin film is even more complicated, although the quantum size effect has been studied
in Pb thin film\cite{Wei2002}, there is no study on Sn thin film as the best of our knowledge, not only in theory but also in experiments. And all of the calculations
show a greater prospective of Sn thin film than Pb thin film to be a topological superconductor, that with a g-factor several times larger.

From Fig. \ref{band_film_sn} and \ref{band_film_pb_s} we can also see that at the $\Gamma$ point, there are nearly degenerate subbands besides the Kramers pairs,
this will give a extremely large g-factor values due to the diverge, to avoid this we do not consider these subbands when calculate the average value of g-factors.
Although we did not calculate the g-factors under strain or with different substrates, we can give an approximate estimation according to the Rashba coupling,
since under the strain or with different substrates the density of subbands keep almost the same, thus a larger Rashba coupling will give a larger g-factor.

\bibliography{MFmetal}

\end{document}
%
% ****** End of file apstemplate.tex ******